%% file: main.tex
\documentclass[twoside, 11pt]{article}
\usepackage{german,amsmath,amssymb,fancyhdr,fancybox}
\usepackage{a4} 
\usepackage[dvips]{epsfig}
\usepackage{float,wrapfig}

\graphicspath{{bilder/}}
\renewcommand{\theequation}{\thesection.\arabic{equation}}

\renewcommand{\Pr}[1]{\text{Pr}\left\{#1\right\}}

\newcommand{\pd}{\mathcal{D}}     

\newcommand{\cstart}{---$\Downarrow$ \begin{itemize}}
\newcommand{\cend}{\end{itemize} ---$\Uparrow$}
\newcommand{\tic}{\rule{0pt}{0pt}}
\newcommand{\bi}{\begin{itemize}}
\newcommand{\ei}{\end{itemize}}

\newcommand{\str}[1]{#1^{\prime}}
\newcommand{\Od}[1]{{\cal O}\left(#1\right)}

\newcommand{\tmw}[1]{\left\langle #1\right\rangle} 
\newcommand{\kmw}[1]{\left\langle\left\langle #1\right\rangle\right\rangle} 
\newcommand{\umw}[1]{\overline{#1}}                
\newcommand{\gb}[1]{\left[#1\right]}               

\newcommand{\komm}[2]{\left[#1,#2\right]}

\renewcommand{\vec}[1]{\underline{#1}}
\newcommand{\mat}[1]{\underline{\mathbf{#1}}}

\newcommand{\csch} {\mathrm{csch}}

\newcommand{\nn}{\nonumber}

\newcommand{\trenner}{\medskip\hspace{.2\linewidth}\hrulefill\hspace{.2\linewidth}\medskip}

\newcommand{\onefigure}[4]
{
 \begin{figure}[hbt]
 \noindent
 \begin{center}
  \epsfig{figure=#1,width=#2}
  \caption{#3}\label{#4}
 \end{center}
 \end{figure}
}

\newcommand{\twofigures}[8]
{
  \begin{figure}[hbt]
   \noindent
   \begin{minipage}[t]{#2}
    \centering\epsfig{figure=#1,width=\linewidth}
    \caption{#3}\label{#4}
   \end{minipage}\hfill
   \begin{minipage}[t]{#6}
    \centering\epsfig{figure=#5,width=\linewidth}
    \caption{#7}\label{#8}
   \end{minipage}
 \end{figure}
}

\newcommand{\twovfigures}[8]
{
  \begin{figure}[hbt]
   \noindent
   \begin{center}
    \epsfig{figure=#1,width=#2}
    \caption{#3}\label{#4}
   \end{center}

   \begin{center}
    \epsfig{figure=#5,width=#6}
    \caption{#7}\label{#8}
   \end{center}
 \end{figure}
}

\newcommand{\threefigureso}[4]
{
  \begin{figure}[hbt]
   \noindent
   \begin{center}
    \epsfig{figure=#1,width=#2}
    \caption{#3}\label{#4}
   \end{center}
   \vspace{0.5cm}
}
\newcommand{\threefiguresu}[8]
{
   \noindent
   \begin{minipage}[b]{#2}
    \centering\epsfig{figure=#1,width=\linewidth}
    \caption{#3}\label{#4}
   \end{minipage}\hfill
   \begin{minipage}[b]{#6}
    \centering\epsfig{figure=#5,width=\linewidth}
    \caption{#7}\label{#8}
   \end{minipage}
 \end{figure}
}

\newcommand{\fourfigureso}[8]
{
  \begin{figure}[hbt]
   \noindent
   \begin{minipage}[b]{#2}
    \centering\epsfig{figure=#1,width=\linewidth}
    \caption{#3}\label{#4}
   \end{minipage}\hfill
   \begin{minipage}[b]{#6}
    \centering\epsfig{figure=#5,width=\linewidth}
    \caption{#7}\label{#8}
   \end{minipage}
   \vspace{1cm}
}
\newcommand{\fourfiguresu}[8]
{
  \noindent
   \begin{minipage}[b]{#2}
    \centering\epsfig{figure=#1,width=\linewidth}
    \caption{#3}\label{#4}
   \end{minipage}\hfill
   \begin{minipage}[b]{#6}
    \centering\epsfig{figure=#5,width=\linewidth}
    \caption{#7}\label{#8}
   \end{minipage}
 \end{figure}
}
\renewcommand{\sectionmark}[1]%
{\markboth{#1}{}}
\renewcommand{\subsectionmark}[1]%
{\markright{\thesubsection\ #1}}
\begin{document}

%

\pagestyle{empty}
\tic\vspace{2cm}
\begin{center}
\Large {Quasi-Eindimensionale
Ladungsdichtewellen \\bei endlichen Temperaturen}\\[5cm]
\large
Diplomarbeit\\
von\\
Andreas Glatz\\[3cm]
\vfill Institut f\"ur Theoretische Physik\\
Gruppe Prof. T. Nattermann\\
Juni 2001
\end{center}
\newpage

\vspace{2cm}
\begin{center}
{\bf Abstract}
\end{center}

\vspace{1cm}

Dynamic and static properties of the classical Fukuyama-Lee-Rice
model and the renormalization and phase diagrams of the related
quantum model with phase-slips are studied.

In the first part, the phase correlation function is calculated in
the weak pinning limit by an one-loop renormalization group
calculation and exactly in the strong pinning case. Further, the
creep dynamics of these quasi-one-dimensional systems is studied
by analytical and numerical approaches.

In the second part, the phase diagrams of the quantum model is
studied by an anisotropic, finite-temperature renormalization
group calculation.

\newpage
\pagestyle{fancy}
\setcounter{page}{1}
\pagenumbering{roman}
\tableofcontents
\newpage

\pagenumbering{arabic}
\setcounter{page}{1}

\setcounter{section}{-1}

\include{summary_en}
\include{intro}

\include{classham}

\include{weakpin}
\include{strongpin}
\include{dynamic}
\include{qmodel}
\include{RG}
\include{result}
\include{summary}
\include{appendix}

%
\newpage
\bibliographystyle{alpha}
\small
\bibliography{references}
\addcontentsline{toc}{section}{Literatur}
\newpage
\pagestyle{empty}
\tic\normalsize\newpage

\include{dank}
\newpage\tic\newpage

\include{erklaerung}

\end{document}

%% file: summary_en.tex
\section{Summary}

Charge density waves (CDWs) are possible low energy states of
quasi--one--dimensional metals, which appear below a certain
transition temperature $T_c$, typically below $200K$. The
underlying  mechanism for this {\it phase-transition} -- called
{\it Peierls transition} -- is the coupling of the band electrons
to the underlying lattice (electron--phonon coupling). At low
temperatures the system can reduce its energy by formation of a
periodic lattice distortion, which is of the order one percent of
the interatomic distance. This deformation leads to a gap in the
electronic dispersion and with that to a gain of electronic energy
larger than the loss of elastic energy.

CDWs can be described by a {\it Ginzburg--Landau model} as
elastic, disordered systems.

\vspace{0.5cm}

In the first part of this thesis (chapter 2) I study dynamic and
static properties of the classical {\it Fukuyama-Lee-Rice model}:
In the weak pinning limit the temperature dependency of the pair
correlation function is calculated using a one--loop approximation
of the related {\it Burgers equation}. This result is verified by
a {\it Monte--Carlo} simulation. For strong disorder, the
correlation function is calculated {\it exactly}, considering the
order statistics of the positions of the impurities, at zero
temperature.

As the central problem, the collective dynamics, especially the
{\it creep--dynamics}, is studied by analytical and numerical
approaches. Contrary to high dimensional systems an analytical
expression for the creep velocity or current is found, which is in
very good agreement with numerical simulation results. This result
extends the well known results for $D>2$.

\vspace{0.5cm}

In the second part (starting with chapter 3) the classical model
is extended by considering quantum fluctuations and space--time
vortices ({\it phase--slips}).

With an anisotropic, finite--temperature renormalization group
approach the phase diagrams for this model are studied. In former
works only the zero temperature behavior for either 1D disordered,
electronic systems or systems with dislocation is studied.

The full consideration of thermal fluctuations and phase--slips is
therefore an essential extension of these works and leads to a new
scenario for the well know delocalization transition at $T=0$.

\vspace{1.0cm}

{\bf partly published in}\\
Andreas Glatz and Mai Suan Li, {\it Collective dynamics of
one-dimensional charge density waves}, Phys. Rev. {\bf B} 64,
184301 (2001). (excerpts of chapter 2)\\
Andreas Glatz and Thomas Nattermann, {\it One-Dimensional
Disordered Density Waves and Superfluids: The Role of Quantum
Phase Slips and Thermal Fluctuations}, Phys. Rev. Lett. 88, 256401
(2002). (based on chapters 3 and 4)

%% file: intro.tex
\section{Einf"uhrung}

\subsection{Einleitung}

In dieser Arbeit werden einige neue Aspekte eindimensionaler
Ladungsdichtewellen behandelt. Es
werden statische und dynamische Eigenschaften mit Hilfe des
 klassischen Fukuyama--Lee-Rice--Modells, welches nur
thermische und Unordnungsfluktuationen ber"ucksichtigt,
und die Phasendiagramme eines
erweiterten, ph"anomenologischen Modells, welches
zus"atzlich noch Quantenfluktuationen und Dislokationen
beinhaltet, bei tiefen Temperaturen untersucht.

Doch zun"achst eine kurze Einf"uhrung zu Ladungsdichtewellen
(CDW) und warum diese
interessant sind:

Schon 1955 fand R.E. Peierls anhand von theoretischen "Uberlegungen,
dass eindimensionale Metalle welche an das unterliegende Gitter gekoppelt
sind, aufgrund der Elektron--Phonon--Wechselwirkung nicht stabil sind
und eine Ladungsdichtewelle ausbilden k"onnen. CDWs k"onnen auch
in h"oheren Dimensionen auftreten, werden aber vorwiegend als quasi-eindimensionales
Ph"anomen beschrieben.

\onefigure{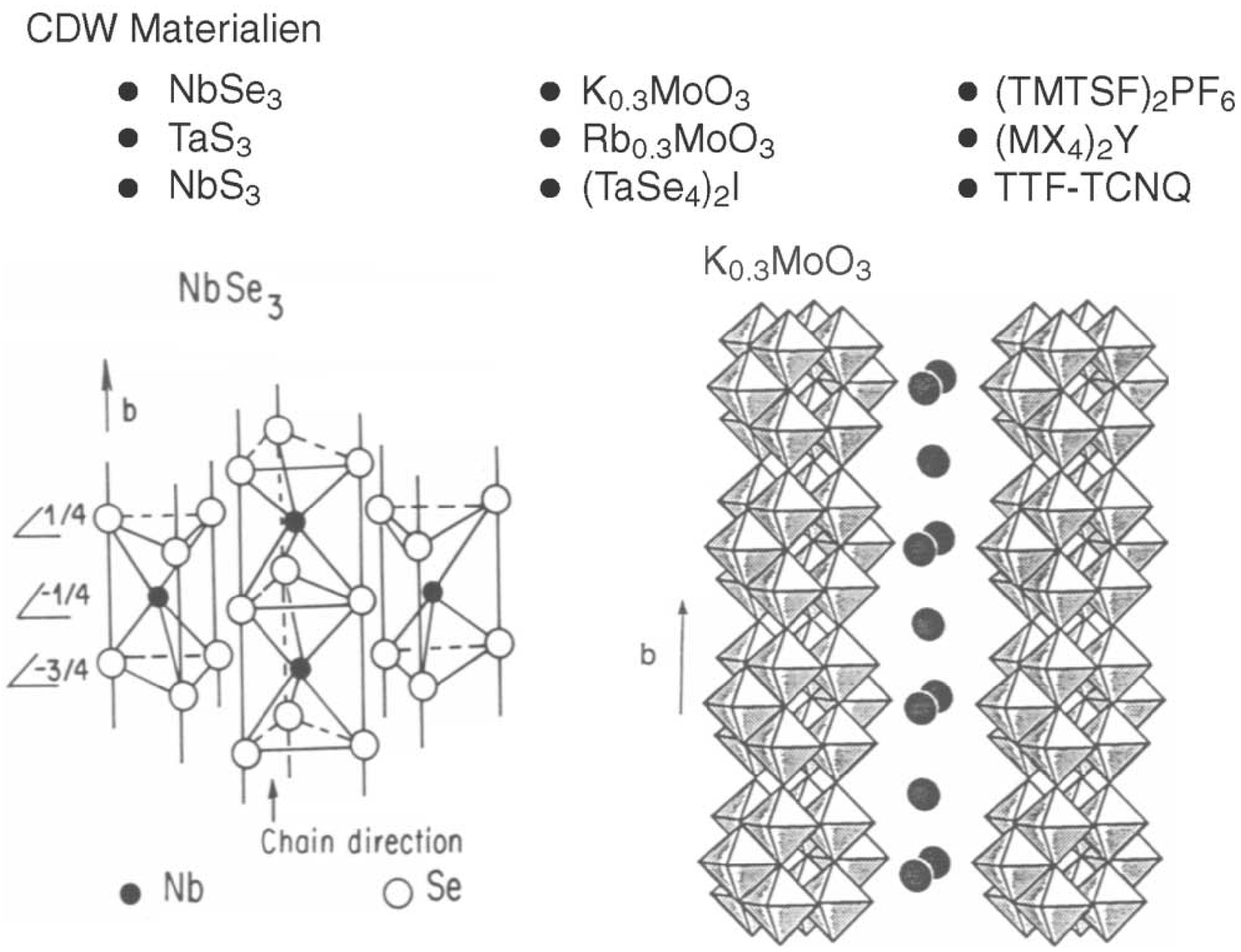}{0.7\linewidth}{Kristallstruktur einiger
CDW-Materialien. Diese bestehen aus schwach gekoppelten
Molek"ulketten.
Die "Ubergangstemperaturen $T_P$ zur CDW liegen f"ur
$\text{Nb}\text{Se}_3$ bei $145K$ und (!) $59K$ (diese Substanz
bildet zwei unabh"angige CDW Phasen aus) und f"ur
$\text{K}_{0.3}\text{Mo}\text{O}_3$ bei $180K$. }{fig.cdwmat}

\twofigures{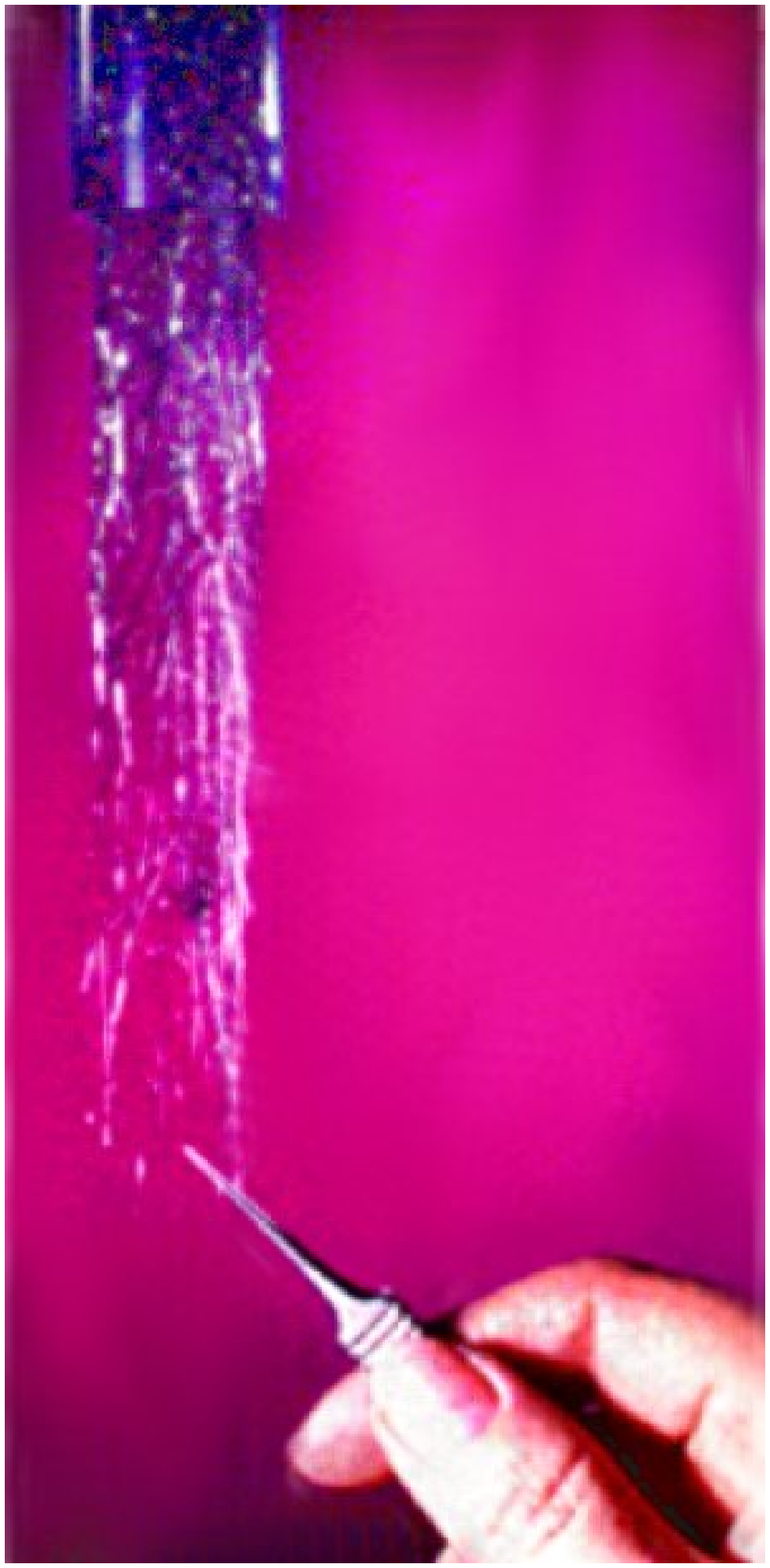}{0.35\linewidth}{$\text{Nb}\text{Se}_3$
Kristallf"aden
}{fig.nbse3}{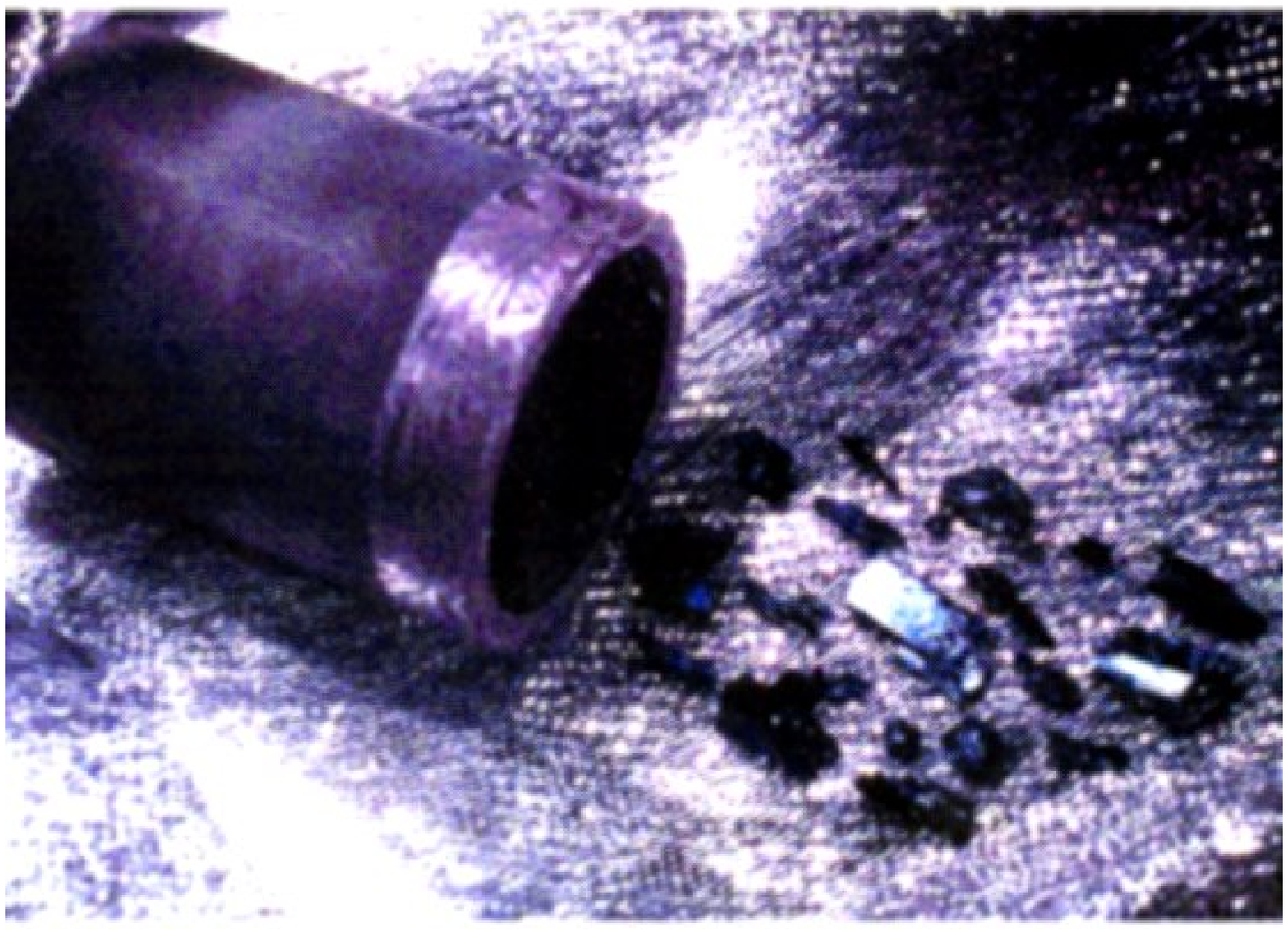}{0.55\linewidth}{$\text{K}_{0.3}\text{Mo}\text{O}_3$
Kristalle. (Aufgrund der tiefblauen Farbe auch Blau-Bronze genannt.)}{fig.kmoo}

Einige CDW Materialien sind in Abb. \ref{fig.cdwmat} mit Kristallstruktur
zusammengestellt. Aufgrund ihrer Struktur
k"onnen sie als eindimensionale Systeme behandelt werden.
In Abb. \ref{fig.nbse3} und \ref{fig.kmoo} sind Fotos von $\text{Nb}\text{Se}_3$
und $\text{K}_{0.3}\text{Mo}\text{O}_3$ zu sehen. Am Beispiel von $\text{Nb}\text{Se}_3$
kann man erkennen, dass die  Materialien als sehr feine F"aden (whiskers)
vorliegen k"onnen und somit deren transversale Ausdehnung, wenn diese
eine gewisse Ausdehnung ($\approx L_c$, s. Kap. \ref{sec.klassmod})
unterschreitet, vernachl"assigt werden kann.

\onefigure{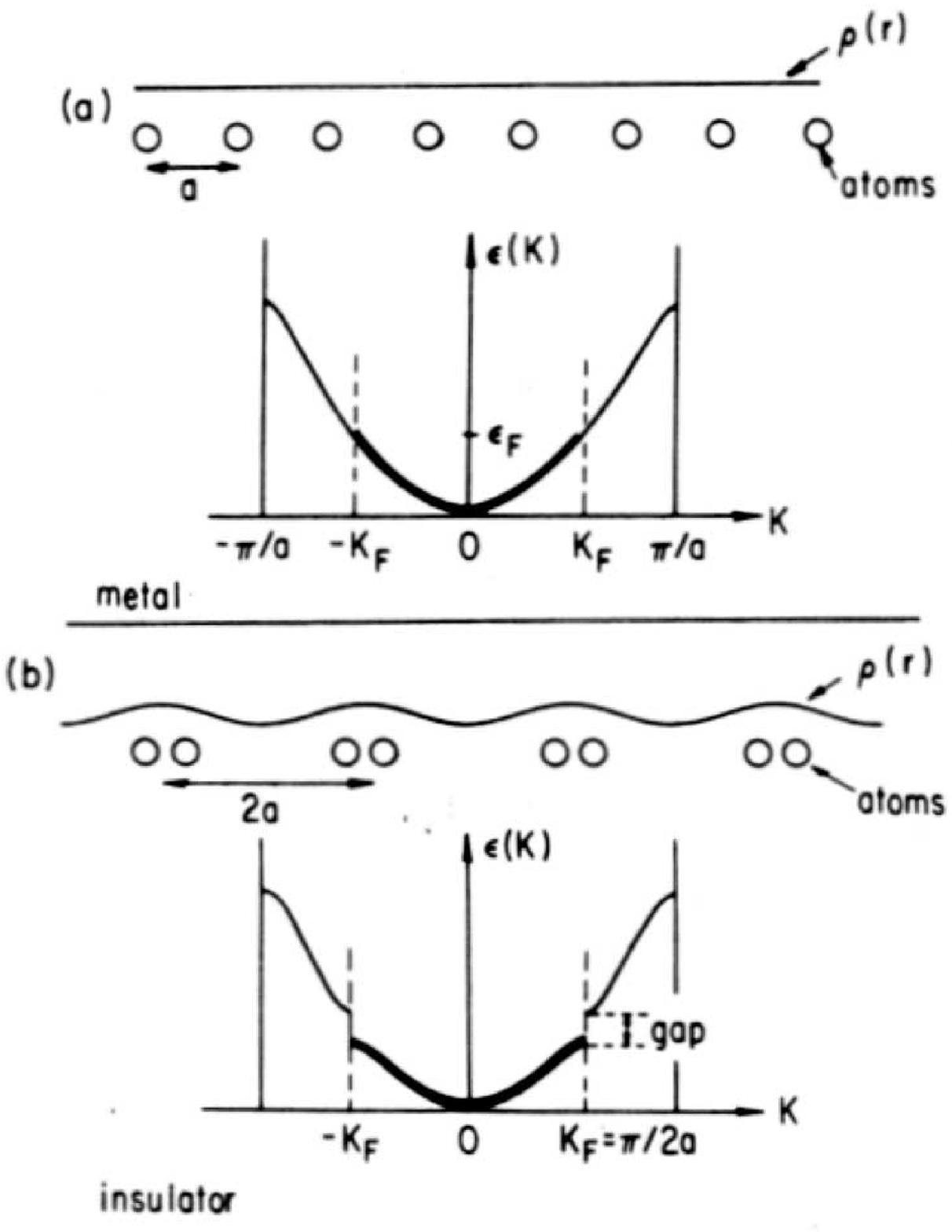}{0.55\linewidth}{Dispersion in einem ungest"orten
eindimensionalen Metall mit halber Bandf"ullung (a) und mit Peierls--St"orung (b).}{fig.espect}

Betrachten wir zun"achst den 1D Elektron--Phonon Hamiltonoperator
(Fr"ohlich Hamiltonian) f"ur spinlose Fermionen (In \cite{CDW:Gruener94,CDW:Gruener88}
ist eine ausf"uhrliche Darstellung der theoretischen und
experimentellen Grundlagen zu Ladungsdichtewellen zu finden.)
\begin{equation*}
 \hat H_{\textit{el-ph}}=\sum\limits_k \epsilon_k\hat c^{\dagger}_k \hat c_k+
\sum\limits_q \hbar\omega_q\hat b^{\dagger}_q \hat b_q+
\sum\limits_{k,q} g_q\hat c^{\dagger}_{k+q} \hat c_k(\hat b_q+\hat b^{\dagger}_q)\,,
\end{equation*}
wobei $\hat c^{(\dagger)}_k$ und $\hat b^{(\dagger)}_q$ Vernichtungs- (Erzeugungs-) Operatoren
f"ur Elektronen bzw. Phononen sind und $\epsilon_k=\frac{\hbar^2 k^2}{2m_e}$.
$g_q$ ist die Kopplungskonstante der Elektron--Phonon--WW.
Mit der Definition des komplexen Ordnungsparameters
\begin{equation}
 \Delta=|\Delta|e^{\imath\varphi}\equiv g_{2k_F}\tmw{\hat b_{2k_F}+\hat b^{\dagger}_{2k_F}}\,
\end{equation}
wobei $k_F$ der Fermi--Wellenvektor ist, und der Gitter--Verschiebung $u(x)=\sum_q Q_q e^{\imath qx}$ mit
$Q_q=\left(\frac{\hbar}{2M\omega_q}\right)^{1/2}(\hat b_{q}+\hat b^{\dagger}_{-q})$
($M$ ist die Ionenmasse) findet man f"ur $\tmw{u(x)}$ folgenden Ausdruck
\begin{equation*}
\tmw{u(x)}=\left(\frac{\hbar}{2M \omega_q}\right)^{1/2}
\left(\tmw{\hat b^{\dagger}_{2k_F}+\hat b^{\dagger}_{-2k_F}}e^{2\imath k_Fx}+c.c.\right)=2\Delta/g_{2k_F}\cos(2k_Fx+\varphi)\,.
\end{equation*}

Das bedeutet, dass das Gitter, welches ohne Elektron--Phonon--WW die Gitterkonstante $a$ hat,
aufgrund dieser WW periodisch gest"ort wird mit Periode $\lambda_0=2\pi/(2k_F)$,
was zur Ausbildung einer L"ucke ($\Delta$) im Elektronenspektrum bei $\pm k_F$ f"uhrt (s. Abb. \ref{fig.espect}).

Die elektronische Energie nimmt f"ur kleine Verschiebungen $\propto u^2\ln u$ ab \cite{CDW:RiceStr73},
die elastische Energie mit $\propto u^2$ zu. Insgesamt erh"alt man durch
die Ausbildung dieser periodischen Gitterst"orung einen Energiegewinn. Aus der Minimalbedingung
f"ur die Energie kann man den Wert von $|\Delta|$ bestimmen.
Mit dieser Gittermodulation wird auch die Ladungsdichte mit der Periode $\lambda_0$
moduliert, wobei pro Periode zwei Elektronen vorhanden sind.

F"ur das ph"anomenologische Modell, welches im n"achsten Kapitel behandelt wird, entwickelt man
einen Landau--Ginzburg--Hamiltonian f"ur den Ordnungsparameter $\Delta(x)$, welcher zun"achst
Fluktuationen der Amplitude $|\Delta|$ und der Phase $\varphi$ enth"alt. 
Da die Fluktuationen der Amplitude bei tiefen Temperaturen
keine Rolle spielen\footnote{Die Amplitudenfluktuationen k"onnen vernachl"assigt werden,
da diese im Landau--Ginzburg--Hamiltonian als quadratische Terme auftreten, 
und somit bei tiefen Temperaturen unterdr"uckt werden.}, 
werden diese im Rahmen dieser Arbeit nicht weiter (direkt) behandelt.
In Kapitel \ref{sec.qmodel} wird unser Modell erweitert und Dislokationen in der Phase
zugelassen. Da das Phasenfeld an den Orten der Dislokationen einen
Sprung hat, muss der Ordnungsparameter an diesen Stellen verschwinden und damit auch
die Amplitude.

Als einzig relevanter Beitrag bleibt noch der elastische Energieterm
\begin{equation}
 {\cal H}_{\textit{el}}=\int dx\,\frac{c}{2}(\partial_x\varphi)^2
\end{equation}
"ubrig (die Zeitabh"angigkeit des Ordnungsparameters bzw. ein daraus folgender kinetischer
Energieterm wurde vernachl"assigt, siehe dazu Kapitel \ref{sec.qmodel}/\ref{sec.results}). Folgt man den
Berechnungen in \cite{CDW:Gruener94} findet man $c=\hbar v_F/(2\pi)$.

\subsection{Bekannte Ergebnisse}

In diesem Abschnitt m"ochte ich kurz darstellen, was im Hinblick auf die
in dieser Arbeit behandelten Themen an Vorarbeiten geleistet wurde.

Viele Arbeiten, die die statischen Eigenschaften des klassischen Modells
behandeln, sind schon Ende der 1970er bis Anfang der 1980er Jahre geschrieben worden.
Hervorzuheben ist unter anderen die Arbeit von M.V. Feigel'man \cite{CDW:Feigel80}, 
in der Eigenschaften von klassischen, ungeordenten Systemen bei tiefen Temperaturen, 
wie z.B. Ladungsdichtewellen mit Unordnungspotential oder auch
Spinketten mit zuf"alligen Anisotropien, untersucht werden.
In dieser Arbeit wird, neben der Berechnung der kollektiven Leitf"ahigkeit von CDWs,
auch die Phasen--Paarkorrelationsfunktion im Fall von schwacher Unordnung
bei $T=0$ mittels einer Transfermatrixberechnung bestimmt.

J. Villain und J.F. Fernandez \cite{CDW:ViFer84} haben diese Korrelationsfunktion
f"ur $T\rightarrow 0$ f"ur ein harmonisches System mit Unordnungsfeld
in h"oheren Dimensionen ($2<d<4$) bestimmt, aber auch in $d=1$ im Fall eines
starken Unordnungspotentials.

Wir werden diese Ergebnisse auf endliche Temperaturen f"ur CDWs erweitern,
und im Fall von starkem Unordnungspotential bei $T=0$ ein exaktes Ergebnis
unter Ber"ucksichtigung der geordneten Statistik der Positionen der Verunreinigungen pr"asentieren.

Die Dynamik von elastischen Medien mit Unordnung bei einer extern angelegten
Kraft bzw. eines elektrischen Feldes ist ebenfalls schon in mehreren Arbeiten untersucht worden.

Wichtige Arbeiten in diesem Zusammenhang in Bezug auf Ladungsdichtewellen sind
u.a. von H. Fukuyama und P.A. Lee \cite{CDW:FuLee77} und
P.A. Lee und T.M. Rice \cite{CDW:LeeRice79} geschrieben worden, in denen
der Depinning-"Ubergang bei $T=0$ untersucht und eine Absch"atzung des Depinningfeldes
angegeben wird, aber auch Leitf"ahigkeiten mit Hilfe der Kubo Formel berechnet wurden.

Eine Sammlung von neueren Arbeiten zur Dynamik, auch experimentelle, ist in
\cite{CDW:Brazovski99} zu finden.

Der Beitrag dieser Arbeit zur Dynamik besch"aftigt sich im wesentlichen
mit der Kriechdynamik eindimensionaler CDWs bei tiefen, aber endlichen Temperaturen
und kleinen angelegten Feldern.
In h"oheren Dimensionen ist diese Dynamik wohl bekannt 
\cite{CDW:NatSchei} (f"ur $d=2$ auch \cite{CDW:TsaiShapir92})
und wir werden dies in Abschnitt \ref{sec.dynamic} f"ur $d=1$
studieren.

In einer Dimension erhalten wir f"ur die Kriechgeschwindigkeit einen analytischen Ausdruck,
wobei die Feldabh"angigkeit durch eine hyperbolische
Sinus-Funktion gegeben ist. Dieses Verhalten haben wir ausf"uhrlich numerisch
untersucht.
Bei $T=0$ haben wir au"serdem den Depinning--"Ubergang untersucht und den
kritischen Exponenten der CDW-Geschwindigkeit bestimmt. Unser Ergebnis
stimmt mit den von anderen Autoren \cite{CDW:SibaniLitt90,CDW:MyersSethna93},
ebemfalls numerisch, bestimmten Werten (innerhalb der Fehlergrenzen) gut "uberein.

Relevante Arbeiten die bzgl. der untersuchten Phasendiagramme
bei tiefen Temperaturen des erweiterten Modells schon geschrieben wurden,
werden in Kapitel \ref{sec.results} mit unseren Ergebnissen verglichen.


\subsection{"Uberblick}

In dieser Arbeit wird der Einfluss von Unordnung in Ladungsdichtewellen eine zentrale
Rolle spielen, welcher in einer Dimension besonders relevant ist und die Eigenschaften
des Systems stark bestimmt. Die Arbeit ist wie folgt aufgebaut:
\bi

\item In Kapitel \ref{sec.klassmod} wird zun"achst der Einfluss der Unordnung im klassischen Fall
(Fukuyama--Lee--Rice--Modell) auf die Phasen--Paarkorrelationsfunktion
$\umw{\tmw{(\varphi(x)-\varphi(\str{x}))^2}}$ im Fall von schwachem
und von starkem Unordnungspotential behandelt.
Anschlie"send wird die Dynamik des Systems bei einer von au"sen angelegten
treibenden Kraft bzw. einem elektrischen Feld untersucht.

\item In Kapitel \ref{sec.qmodel} wird dieses Modell um einen kinetischen Term,
da bei tiefen Temperaturen auch Quantenfluktuationen ber"ucksichtigt werden m"ussen,
und einen Term, welcher den Effekt von Dislokationen bzw. ''phase--slips''
ber"ucksichtigt, erweitert.

\item In Kapitel \ref{sec.RG} werden die Phasendiagramme dieses erweiterten Modells
ausf"uhrlich besprochen und
\item in Kapitel \ref{sec.results} deren physikalische Konsequenzen diskutiert und
mit verwandten Systemen verglichen.

\ei

%% file: classham.tex
\section{Klassisches Modell}\label{sec.klassmod}
\setcounter{equation}{0}

In diesem Kapitel behandele ich das klassische \emph{Fukuyama--Lee--Rice Modell}
~\cite{CDW:Fukuyama76,CDW:LeeRice79}
f"ur eindimensionale Ladunsdichtewellen. Zun"achst wird der Einflu"s der
Unordung auf die statischen Eigenschaften, insbesondere die Paarkorrelationsfunktion
$\umw{\tmw{(\varphi(x)-\varphi(0))^2}}$ bei niedrigen Temperaturen untersucht.
Dabei wird zwischen \emph{schwachem} (Abschnitt \ref{sec.weakpin}) und
\emph{starkem} (Abschnitt \ref{sec.strongpin}) Unordnungspotential unterschieden.
Mit Hilfe einer \emph{Monte--Carlo} Simulation werden die Resultate in Abschnitt \ref{sec.MC}
chlie"slich "uberpr"uft.

In Abschnitt \ref{sec.dynamic} wird das dynamische Verhalten des Systems
bei einer von au"sen an das System angelegten Kraft bzw. elektrischen Feld
analytisch und numerisch untersucht.

\subsection{Ladungsdichte}
Als fundamentale Observable wird ein Ausdruck f"ur die
Ladungsdichte ben"otigt.
Dazu wird einfach die Teilchendichte als $\rho(x)=n\sum_i\delta(x-x_i)$ geschrieben,
wobei $x_i=X_i-\frac{\varphi_i}{2\pi}a$ die Positionen der Elektronen sind und
$\frac{\varphi_i}{2\pi}a$ kleine Abweichungen von der mittleren Position $X_i=i a$ beschreibt
($\varphi_i$ hat demnach die Funktion eines Verschiebungsfeldes).
Dies l"a"st sich wie folgt umformen:
\begin{eqnarray}
 \rho(x)&=&n\sum_i\int dX\,\delta(X-X_i)\delta\left(x-X+\frac{\varphi(x)}{2\pi}a\right)\nn\\
 &\underset{\text{PSF}}{=}&n\int dX\,\sum_m \frac{1}{a}e^{\imath Q_mX}\delta\left(x-X+\frac{\varphi(x)}{2\pi}a\right)\nn
\end{eqnarray}
mit $\varphi(x_i)=\varphi_i$ und $Q_m=2\pi m/a$.
F"uhrt man noch die neue Variable $y=X-\frac{\varphi(x)}{2\pi}a$ ein,
folgt (vgl. mit Herleitung in \cite{CDW:Haldane81})
\begin{eqnarray}
 \rho(x)&=&\frac{n}{a}\int dy\,\left(1+\frac{1}{2\pi}\partial_x\varphi(x)a\right)
\sum_m e^{\imath 2\pi m/a (y+\varphi(x)/(2\pi) a)}\delta(x-y)\quad ,\quad 2k_F=\frac{2\pi}{a}\nn\\
&=& n\left(\frac{k_F}{\pi}+\frac{1}{2\pi}\partial_x\varphi(x)\right)\sum_m
  e^{\imath m(2k_F x+\varphi(x))}\,.
\end{eqnarray}
Im Fall der Ladungsdichtewelle ist $n=2$, 
da pro CDW--Periode zwei Elektronen vorhanden sind.

In niedrigster Ordnung erh"alt man folgenden Ausdruck
\begin{equation}
 \label{eq.halcd}
 \rho(x)=\rho_0+\frac{1}{\pi}\partial_x\varphi(x)+\rho_1\cos(\varphi(x)+2k_Fx)+\text{h.h.}\,,
\end{equation}
mit $\rho_0=2k_F/\pi$ und $\rho_1=2\rho_0$.


\subsection{Hamiltonian}\label{sec.klassham}

Die Hamiltonfunktion des Phasenfeldes
$\varphi$ setzt sich aus zwei Anteilen zusammen:


${\cal H}_{\textit{klass}}={\cal H}_{\textit{el}}+{\cal H}_{\textit{imp}}$, wobei ${\cal H}_{\textit{el}}=\int dx\,
\frac{1}{4\pi}\hbar v_F\left(\frac{\partial}{\partial x}\varphi\right)^2$ die
elastische Energie und ${\cal H}_{\textit{imp}}$ die ''Unordungsenergie'' des Systems beschreibt
(den Index ${}_{\textit{klass}}$ lasse ich im weiteren Verlauf dieses Abschnitts weg).
Das Unordungspotential ist durch $V(x)=-\sum\limits_{i=1}^{N}V_i\delta(x-x_i)$
gegeben, wobei $x_i$ die Position der i-ten Verunreinigung (\emph{imp}urity)
und $V_i>0$ dessen St"arke ist. Dieses Potential koppelt an die Ladungsdichte
$\rho(x)$ und liefert drei zus"atzliche Terme zum Hamiltonian: Einen konstanten
Beitrag durch $\rho_0$, einen Gradiententerm und einen Beitrag durch die
 Kopplung an den schnell variierenden
bzw. mikroskopischen Anteil der Ladungsdichte $\rho_1\cos{\big(\varphi(x)+Qx\big)}$
mit dem Wellenvektor $Q=2k_F$ der ungest"orten Welle. H"ohere Harmonische in
der Dichte (\ref{eq.halcd}) werden nicht ber"ucksichtigt.
Der konstante Beitrag kann vernachl"assigt werden und der Gradiententerm kann durch
eine geeignete Umdefinition des Phasenfeldes in den elastischen Term absorbiert werden
(siehe dazu auch im n"achsten Abschnitt).

Die Hamiltonfunktion des Systems der L"ange $L$ lautet daher~\cite{CDW:Gorkov77}:
\begin{eqnarray}\label{eq.cmodel}
   {\cal H}&=&\int\limits_0^L dx\,\left\{  \frac{\mathrm{\hbar} v_F}{4\pi}
   \left(\frac{\partial}{\partial x}\varphi\right)^2 \right.\nonumber\\
   &&-\left. \sum\limits_{i=1}^{N} V_i\rho_1\delta(x-x_i)\cos{(\varphi+Qx)}\right\}
\end{eqnarray}

Der mittlere Abstand der Verunreinigungen $c_{\textit{imp}}^{-1}=L/N$ ist gro"s
im Vergleich mit der CDW-Wellenl"ange, also
\begin{equation*}
 Q\gg c_{\textit{imp}}\,.
\end{equation*}

\onefigure{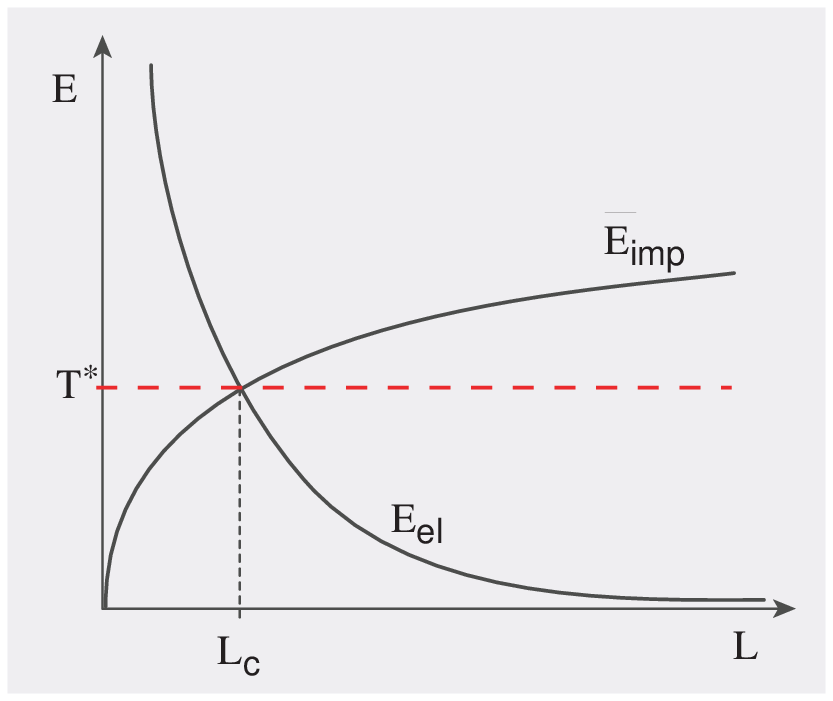}{0.4\linewidth}{Elastische Energie und mittlere
Unordnungsenergie als Funktion der L"ange}{fig.Lc}

Nun kann man eine L"angenskala $L_c$ definieren, die Fukuyama--Lee--L"ange~\cite{CDW:FuLee77},
auf der die elastische Energie gleich der mittleren Unordnungsenergie wird
($\varphi$ "andert sich um $\Od{1}$): $\hbar v_F/L_c=(\umw{V_i^2}\rho_1 c_{\textit{imp}}L_c)^{1/2}$
(siehe Abb. \ref{fig.Lc}).
Daraus folgt f"ur $L_c$
\begin{equation}
 L_c=(\hbar v_F^{\phantom{1}})^{2/3}(\umw{V_i^2}\rho_1 c_{\textit{imp}})^{-1/3}\,.
 \label{eq.Lc}
\end{equation}

Die durch $L_c$ definierte Energieskala bezeichne ich als
\begin{equation}
   T^{\ast}=\frac{\hbar v_F}{L_c}=\left(\hbar v_F^{\phantom{1}}\overline{V^2_i}\rho_1c_{\textit{imp}}\right)^{1/3}\,.
   \label{eq.ELc}
\end{equation}

Typische Werte f"ur $T^{\ast}/T$ liegen im Bereich $10^3-10^7$ \cite{CDW:MiddFi91}.

Im Folgenden unterscheiden wir die F"alle von \emph{schwacher} ($V_i\rho_1\ll\hbar v_F c_{\textit{imp}}$)
und \emph{starker} ($V_i\rho_1\gg\hbar v_F c_{\textit{imp}}$) Unordnung.

Die angegebenen Bedingungen f"ur starke und schwache Unordnung
 kann man sich folgenderma"sen "uberlegen:
Im Fall eines starken Unordnungspotentials wird sich die Phase an
den Positionen $x_i$ der Impurities so einstellen, dass
$H_{\textit{imp}}$ minimal wird.
Dabei gewinnnt das System die Energie $V_i\rho_1$ pro Verunreinigung.
Andererseits wird dabei die elastische Energie erh"oht, und zwar
um $\approx c_{\textit{imp}}^{-1} \hbar v_F(c_{\textit{imp}})^2=\hbar v_F c_{\textit{imp}}$
pro Verunreinigung. Die Unterscheidung von starker und schwacher Unordung
ist dadurch vorgegeben, ob der Quotient $V_i\rho_1/ (\hbar v_F c_{\textit{imp}})$
sehr viel gr"o"ser bzw. kleiner als eins ist.

Im Fall der schwachen Unordnung wird die Korrelationsfunktion
$\umw{\tmw{(\varphi(x)-\varphi(0))^2}}$ zuerst mit einem einfachen
Renormierungsgruppenargument und anschlie"send deren Temperaturabh"angigkeit
in \emph{one-loop}-N"aherung mit einer KPZ-artigen Gleichung berechnet.

%% file: weakpin.tex
\subsubsection{Schwaches Pinning - Burgers Gleichung}\label{sec.weakpin}

Aus der Bedingung f"ur die schwache Unordnung und der Definition  von $L_c$
folgt, dass die Fukuyama--Lee--L"ange gro"s im Vergleich
zum mittleren Abstand der Impurities ist, also
\begin{equation}
   L_c\gg c_{\textit{imp}}^{-1}\gg Q^{-1}\,.
   \label{eq.weakpinbed}
\end{equation}
Da wir uns f"ur die Physik auf L"angenskalen $\geq L_c$ interessieren,
kann der Hamiltonian (\ref{eq.cmodel}) in Form eines \emph{XY-Modells} mit
Zufallsfeld geschrieben werden:
\begin{equation}
   {\cal H}=\int dx\Bigg\{\frac{1}{4\pi}\hbar v_F
   \left(\frac{\partial}{\partial x}\varphi-g(x)\right)^2-V\cos{\big(\varphi+\alpha(x)\big)}\Bigg\}\,,
   \label{eq.xyrmodel}
\end{equation}

wobei $V^2_{\phantom{1}}\equiv\overline{V^2_i}\rho_1 c_{\textit{imp}}$ und $\alpha(x)$ eine
Zufallsphase ist, mit $\alpha(x_i)=2 k_F x_i \mid 2\pi$
( $\mid$ steht f"ur modulo). F"ur den Unordungsmittelwert
von $\alpha(x)$ gilt
\begin{equation}
   \overline{e^{\imath\big(\alpha(x)-\alpha(x^{\prime})\big)}}=
   \delta(x-x^{\prime})\,.
   \label{eq.phaseMW}
\end{equation}
In (\ref{eq.xyrmodel}) wurde au"serdem ber"ucksichtigt, dass unter der
Renormierungsgruppentransformation, welche von (\ref{eq.cmodel}) auf
(\ref{eq.xyrmodel}) f"uhrt, ein linearer Gradiententerm $\propto\int dx\, g(x)\partial_x\varphi$
generiert wird und die elastische Konstante $\hbar v_F$ nicht
renormiert wird.

F"ur $g(x)$ soll gelten
\begin{equation}
   \overline{g(x)}=0\quad \; , \; \quad \overline{g(x)\,g(x^{\prime})}=
   \sigma\delta(x-x^{\prime})\,.
   \label{eq.gMW}
\end{equation}
Dass die elastische Konstante nicht renormiert wird, folgt aus der Tatsache,
dass das Modell
(\ref{eq.cmodel}) eine sogenannte \emph{statistische Tilt--Symmetrie}
\cite{CDW:Schultz88} besitzt.
(In Anhang \ref{app.tiltsym} ist dies ausf"uhrlicher gezeigt.)

Es ist zweckm"a"sig zu reskalierten Gr"o"sen "uberzugehen:
\begin{eqnarray}
 y=L_c^{-1} x \quad &,& \quad \tilde L=L_c^{-1} L\nn\\
 \tilde\varphi(y)=\varphi(x) \quad &,& \quad \tilde\alpha(y)=\alpha(x)\nn\\
 \tilde g(y)=g(x) \quad &,& \quad \tilde\sigma=L_c\sigma\\
 \tilde T&=&T/T^{\ast}\nn
 \label{eq.rescale}
\end{eqnarray}

Damit l"a"st sich (\ref{eq.xyrmodel}) umschreiben zu
\begin{equation}
   \frac{\cal H}{T}=\tilde T^{-1}\int dy\left[\frac{1}{2}\left(
   \frac{\partial\tilde\varphi}{\partial y}-\tilde g(y)\right)^2-
   \cos{\big(\tilde\varphi-\tilde\alpha(y)\big)}\right]\,.
   \label{eq.rmodel}
\end{equation}


Au"serdem brauchen wir das Hilfsfeld $\tilde\phi$, definiert durch
\begin{equation}
 \frac{\partial\tilde\phi}{\partial y}\equiv\frac{\partial\tilde\varphi}{\partial y}-\tilde g(y)
\end{equation}
und damit
\begin{equation}\label{eq.defphi}
  \tilde\varphi(y)=\tilde\phi(y)+\int\limits_0^y dz\,\tilde g(z)+\tilde\varphi(0)-\tilde\phi(0)\,.
\end{equation}

In nullter Ordnung der St"orungstheorie kann der nicht--lineare Term in
(\ref{eq.rmodel}) ignoriert werden und die Paarkorrelationsfunktion $C(x)$ l"a"st
sich in diesem Fall unter Benutzung von (\ref{eq.defphi}) und (\ref{eq.gMW}) einfach
berechnen:
\begin{eqnarray}
   C(x) & = & \umw{\tmw{\big(\varphi(x)-\varphi(0)\big)^2}} = \umw{\tmw{\big(\tilde\varphi(y)-\tilde\varphi(0)\big)^2}}\nn\\
   &=&\umw{\tmw{\left( \tilde\phi(y)-\tilde\phi(0)+\int\limits_0^y dz\,\tilde g(z)\right)^2}}\nn\\
   &=&\tmw{\big(\tilde\phi(y)-\tilde\phi(0)\big)^2}+\iint\limits_0^y dzd\str{z}\,\umw{\tilde g(z)\tilde g(\str{z})}\nn\\
   &=&\tilde T|y|+\tilde \sigma|y|=\left(\frac{T}{\hbar v_F}+\sigma\right)|x|^{2\zeta}\,,
   \label{eq.cfwp}
\end{eqnarray}
wobei f"ur den Rauhigkeitsexponenten $\zeta=1/2$ gilt.

Die Berechnung der Korrelationsfunktion f"ur das einfache elastische
Modell (welche f"ur den vorletzten Schritt benutzt wurde) ist in Anhang
\ref{app.cfelmod} zu finden.

F"ur $T\rightarrow 0$ hat Feigel'man~\cite{CDW:Feigel80},
ausgehend von (\ref{eq.cmodel}), mit einer Transfermatrix--Methode
gezeigt, dass $\sigma$ von der Gr"o"senordnung $L_c^{-1}$ ist,
d.h. $\tilde\sigma$ ist von der Ordnung $1$.
Dies wird sehr gut durch die Monte--Carlo--Simulation
in Abschnitt \ref{sec.MC} best"atigt.
Ein "ahnliches Resultat wurde auch von Villain und Fernandez
~\cite{CDW:ViFer84} mit Hilfe einer Ortsraum--Renormierungsgruppenmethode
gefunden.
Mit (\ref{eq.cfwp}) haben wir ein Resultat bekommen, welches auch bei
endlichen Temperaturen g"ultig ist.


Bei endlichen Temperaturen kann der Einfu"s des nichtlinearen Terms der Unordnung mit einer \emph{KPZ-} oder \emph{Burgers-}"ahnlichen Gleichnung behandelt werden.
Dazu betrachtet man die eingeschr"ankte Zustandssumme
\begin{equation}
{\cal Z}(l,\varphi)\equiv\int\limits_{\varphi(0)=0}^{\varphi(l)=\varphi}\pd\varphi\, e^{-{\cal H}(l)/T}
\end{equation}
des Systems mit L"ange $l$ und den Randbedingungen $\varphi(0)=0,\;\varphi(l)=\varphi$.\\
Mit Hilfe der Transformation $F(l,\varphi)=-T\ln({\cal Z}(l,\varphi))$
(\emph{Cole-Hopf-Transformation}) erh"alt man aus der
''Transfermatix-Gleichung'' $-T\frac{\partial {\cal Z}(l,\varphi)}{\partial l}=
\left[-\frac{T^2}{2\hbar v_F}\frac{\partial^2}{\partial \varphi^2}+V(l,\varphi))\right]
{\cal Z}(l,\varphi)$ die folgende Gleichung f"ur die
eingeschr"ankte frei Energie $F(l,\varphi)$ \cite{CDW:HuseHenleyFi85}

\begin{equation}
   \frac{\partial F}{\partial l}=\frac{T}{2\hbar v_F}
   \frac{\partial^2F}{\partial\varphi^2}-\frac{1}{2\hbar v_F}
   \left(\frac{\partial F}{\partial\varphi}\right)^2-
   \underbrace{V\cos{\big(\varphi-\alpha(l)\big)}}_{\equiv V(l,\varphi)}\,.
   \label{eq.burgers}
\end{equation}

Nun geht man zur Fouriertransformierten $\hat F(k,\omega)=\iint_{\mathbb{R}^2}d\varphi dl\,
e^{\imath k\varphi+\imath\omega l}F(l,\phi)$ (analog f"ur $\hat V(k,\omega)$) "uber.
Setzt man die Umkehrtransformierte in Gleichung
(\ref{eq.burgers}) ein, erh"alt man \cite{CDW:Medina89}

\begin{eqnarray}
 -\imath \omega \hat F(k,\omega)&=&-\frac{T}{2\hbar v_F}k^2 \hat F(k,\omega)\nn\\
 &&-\frac{1}{2\hbar v_F} \iint\limits_{\{|\tilde\omega|<\Omega\}\times \mathbb{R}}\frac{d\tilde\omega d\tilde k}{(2\pi)^2}
       (k-\tilde k)\tilde k \hat F(k-\tilde k,\omega-\tilde\omega)\hat F(\tilde k,\tilde \omega)\nn\\
 &&+\hat V(k,\omega)\,.
 \label{eq.burgersFT}
\end{eqnarray}

Wird $\hat V(k,\omega)$ explizit aufgeschrieben, sieht man, dass das Zufallspotential
nur Fourierkomponenten mit $k=\pm 1$ enth"alt:
\begin{eqnarray}
 \hat V(k,\omega)&=&V\iint_{\mathbb{R}^2}d\varphi dl\cos(\varphi-\alpha(l))e^{\imath k\varphi+\imath\omega l}\nonumber\\
 &=&\frac{h_1(\omega)}{2}\int d\varphi \left(e^{\imath (k+1)\varphi}+e^{\imath (k-1)\varphi}\right)\nn\\
   &&+\frac{h_2(\omega)}{2\imath}\int d\varphi \left(e^{\imath (k+1)\varphi}-e^{\imath (k-1)\varphi}\right)\nonumber\\
 &=&\pi\delta(k+1)(h_1(\omega)-\imath h_2(\omega))\nn\\
    &&+\pi\delta(k-1)(h_1(\omega)+\imath h_2(\omega))
\end{eqnarray}
dabei ist $h_1(\omega)\equiv V\int dl \cos(\alpha(x)) e^{\imath\omega l}$ und $h_2(\omega)$ analog mit $\sin(\alpha)$ definiert.

Damit kann jetzt das Unordnungs-gemittelte Produkt $\umw{\hat V(\str{k},\str{\omega})\hat V(k,\omega)}$ unter
Ausnutzung von (\ref{eq.phaseMW}) berechnet werden:
\begin{eqnarray}
 \umw{\hat V(\str{k},\str{\omega})\hat V(k,\omega)}&=&2\pi^2
        \umw{h_1(\omega)h_1(\str{\omega})}\delta(k+\str{k})(\delta(k-1)+\delta(k+1))\nn\\
 &=&2V^2\pi^3\delta(\omega+\str{\omega})\delta(k+\str{k})(\delta(k-1)+\delta(k+1))\nn\\
 &\equiv&2\delta(\omega+\str{\omega})\delta(k+\str{k})D(k,\omega)
\label{eq.burgersUM}
\end{eqnarray}

Nun f"uhre ich folgende Bezeichnungen ein:
\begin{eqnarray}
 \nu&\equiv&\frac{T}{2\hbar v_F}\\
 \mathcal{G}_0(k,\omega)&\equiv&\frac{1}{\nu k^2-\imath\omega}\\
 \mathcal{G}(k,\omega)\hat V(k,\omega)&\equiv&\hat F(k,\omega)\\
 \epsilon&\equiv&\frac{1}{\hbar v_F}
\end{eqnarray}
$\nu$ ist im wesentlichen die inverse relativistische de Broglie-Wellenl"ange
eines Teilchens mit Geschwindigkeit $v_F$, $\mathcal{G}_0$ der ungest"orte und
$\mathcal{G}$ der effektive Propagator.

\onefigure{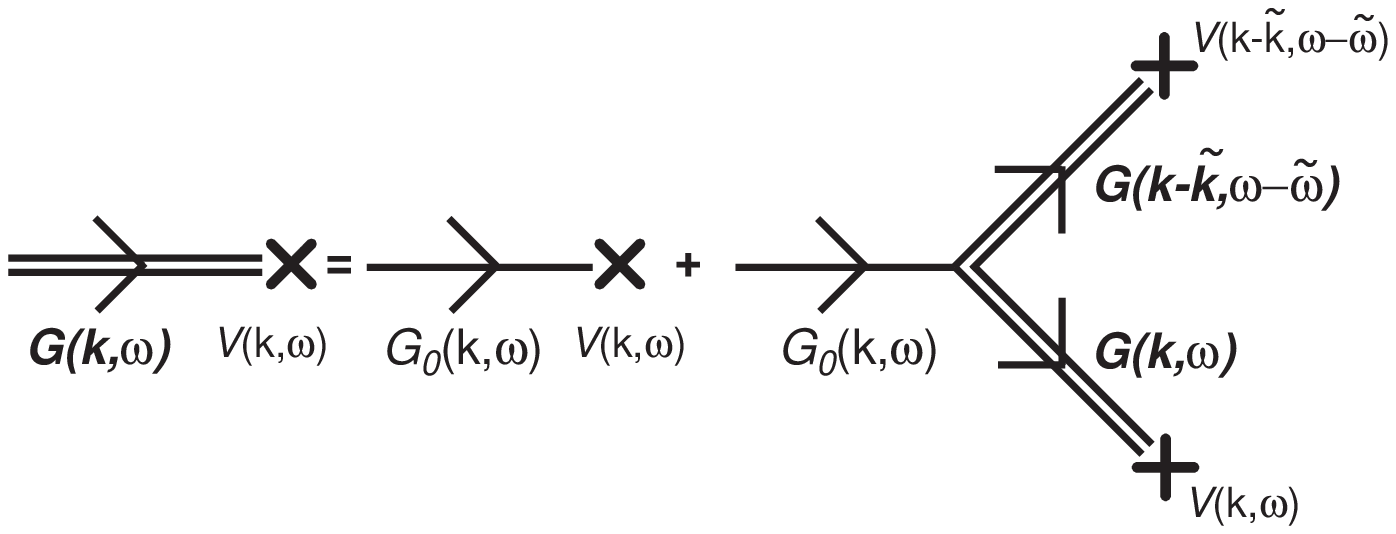}{.7\linewidth}{Diagrammatische Darstellung der
selbstkonsistenten Gleichung f"ur den effektiven Propagator (doppelte Linie).
Der ungest"orte Propagator ist durch eine einfache Linie
und der Unordnungsterm durch ein $\times$ gekennzeichnet.}{fig.burgers1}

Damit l"a"st sich Gl. (\ref{eq.burgersFT}) als eine selbstkonsistente Gleichung
f"ur $\mathcal{G}(k,\omega)$ schreiben:
\begin{eqnarray}
 \mathcal{G}(k,\omega)\hat V(k,\omega)&=&\mathcal{G}_0(k,\omega)\hat V(k,\omega)-
  \frac{\epsilon}{2}\mathcal{G}_0(k,\omega)\iint\limits_{\{|\tilde\omega|<\Omega\}\times \mathbb{R}}
     \frac{d\tilde\omega d\tilde k}{(2\pi)^2}(k-\tilde k)\tilde k\nn\\
       &&\times \hat V(k-\tilde k,\omega-\tilde\omega) \mathcal{G}(k-\tilde k,\omega-\tilde\omega)
         \hat V(\tilde k,\tilde \omega)\mathcal{G}(\tilde k,\tilde \omega)
 \label{eq.propagator}
\end{eqnarray}
\onefigure{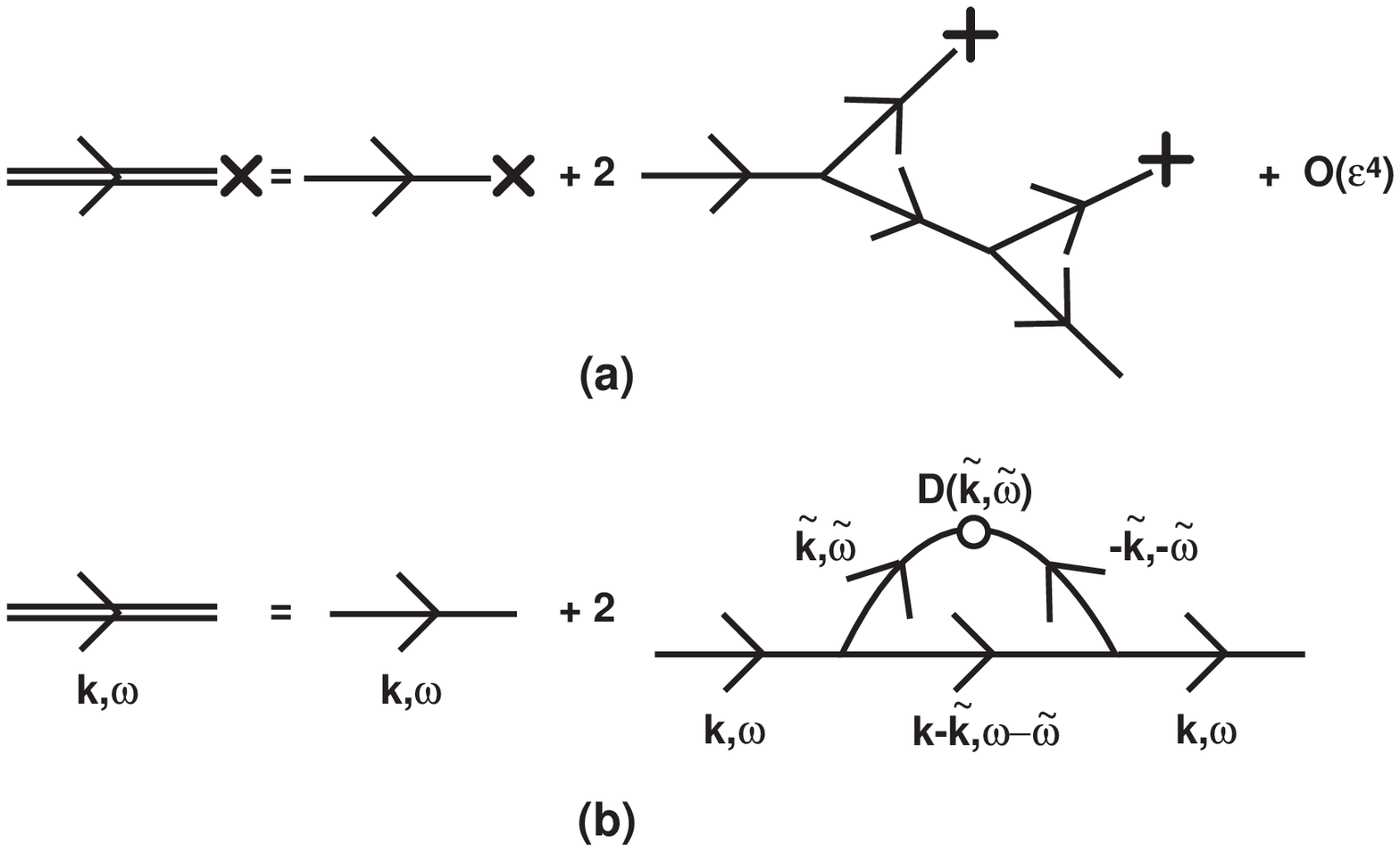}{.8\linewidth}{Propagator in \emph{one-loop} N"aherung vor (a) und nach (b)
Mittelung "uber die Unordnung, wobei $\bigcirc \hat = 2 D(k,\omega)$.}{fig.burgers2}

Diagrammatisch ist diese Gleichung in Abb. \ref{fig.burgers1} dargestellt.
Nun f"uhrt man die Picard-Iteration einmal durch, ersetzt $\mathcal{G}$
durch $\mathcal{G}_0$ und erh"alt so einen Korrekturterm
der niedrigsten, nicht verschwindenden Ordnung $\epsilon^2$ (\emph{one-loop}-N"aherung),
welcher in Abb. \ref{fig.burgers2}(a) dargestellt ist.
Nach Mittelung "uber die Unordnung, unter Benutzung von (\ref{eq.burgersUM}),
erh"alt man
\begin{eqnarray}
 \mathcal{G}(k,\omega)&=&\mathcal{G}_0(k,\omega)+4\left(-\frac{\epsilon}{2}\right)^2\mathcal{G}^2_0(k,\omega)
 \iint\limits_{\{|\tilde\omega|<\Omega\}\times \mathbb{R}}
     \frac{d\tilde\omega d\tilde k}{(2\pi)^2}(k-\tilde k)\cdot\tilde k (-\tilde k)\cdot k\nn\\
  &&\times\mathcal{G}_0(k-\tilde k,\omega-\tilde\omega)\mathcal{G}_0(\tilde k,\tilde\omega)
    \mathcal{G}_0(-\tilde k,-\tilde\omega)2D(\tilde k,\tilde \omega)+\mathcal{O}(\epsilon^4)\,,
    \label{eq.burgersOL}
\end{eqnarray}
siehe Abb. \ref{fig.burgers2}(b). Der rechte Vertex entspricht dem Faktor $(k-\tilde k)\cdot\tilde k$,
der linke $(-\tilde k)\cdot k$ und $\bigcirc$ entspricht $2 D(k,\omega)$.

F"ur $\omega=0$ kann das Integral auf der rechten Seite von (\ref{eq.burgersOL})
weiter vereinfacht werden und aufgrund der einfachen Struktur von
$D(\tilde k,\tilde \omega)$ ist das $\tilde k$-Integral
einfach zu berechnen
\footnote{Hat $D(k,\omega)$ eine kompliziertere $k$-Abh"anigkeit,
z.B. $D\propto k^{-2\rho}$ (r"aumlich korrelierte Unordnung), kann der Korrekturterm
divergieren. In diesem Fall muss der Einfluss der Unordnung schrittweise mit einer
Renormierungsgruppen-Rechnung behandelt werden~\cite{CDW:Medina89}.
F"uhrt man diese Rechnung im vorliegenden Fall durch, erh"alt man
letzendlich dasselbe Ergebnis (\ref{eq.burgersE}).}:
\begin{eqnarray}
 \mathcal{G}(k,0)&=&\mathcal{G}_0(k,\omega)\Big[1-\frac{V^2\pi\epsilon^2k}{2}\nn\\
   &&\int d\tilde\omega\left(\frac{k-1}{\nu(k-1)^2+\imath\tilde\omega}+
       \frac{k+1}{\nu(k+1)^2+\imath\tilde\omega}\right)\frac{1}{\nu^2+\tilde\omega^2}\Big]
 \label{eq.burgersR}
\end{eqnarray}

Um das $\tilde \omega$-Integral noch zu bestimmen, nutzt man aus, dass das physikalische
Verhalten des Systems bei tiefen Temperaturen durch kleine $k$ bestimmt wird
und entwickelt den Integranden in (\ref{eq.burgersR}) in eine Laurent-Reihe
um $k=0$. Als relevanter Term bleibt nur noch
$\frac{\nu-\imath\tilde\omega}{(\nu^2+\tilde\omega^2)^2}$ zu integrieren. F"ur
$\Omega\rightarrow\infty$ erh"alt man daher schlie"slich:
\begin{equation}
 \mathcal{G}(k,0)=\mathcal{G}_0(k,\omega)\left(1-\frac{\epsilon^2}{2}V^2k^2\pi^2\nu^{-2}\right)
 \label{eq.burgersE}
\end{equation}
Vergleicht man die rechte Seite von (\ref{eq.burgersE}) mit
$\mathcal{G}(k,0)\equiv\frac{2v_F}{T_{\textit{eff}} k^2}+\mathcal{O}(k^2)$,
findet man, dass die Temperatur oder besser die Diffusionskonstante
$2\hbar v_F/T$ aufgrund der Unordnung in niedrigster Ordnung durch
\begin{equation}
   \frac{2\hbar v_F}{T}\left(1-4\pi^2\frac{\hbar v_FV^2}{T^3}\right)\,,
   \label{eq.burgersT}
\end{equation}
 ersetzt wird. Der Rauhigkeitsexponent bleibt unver"andert
$\zeta=1/2$.

\onefigure{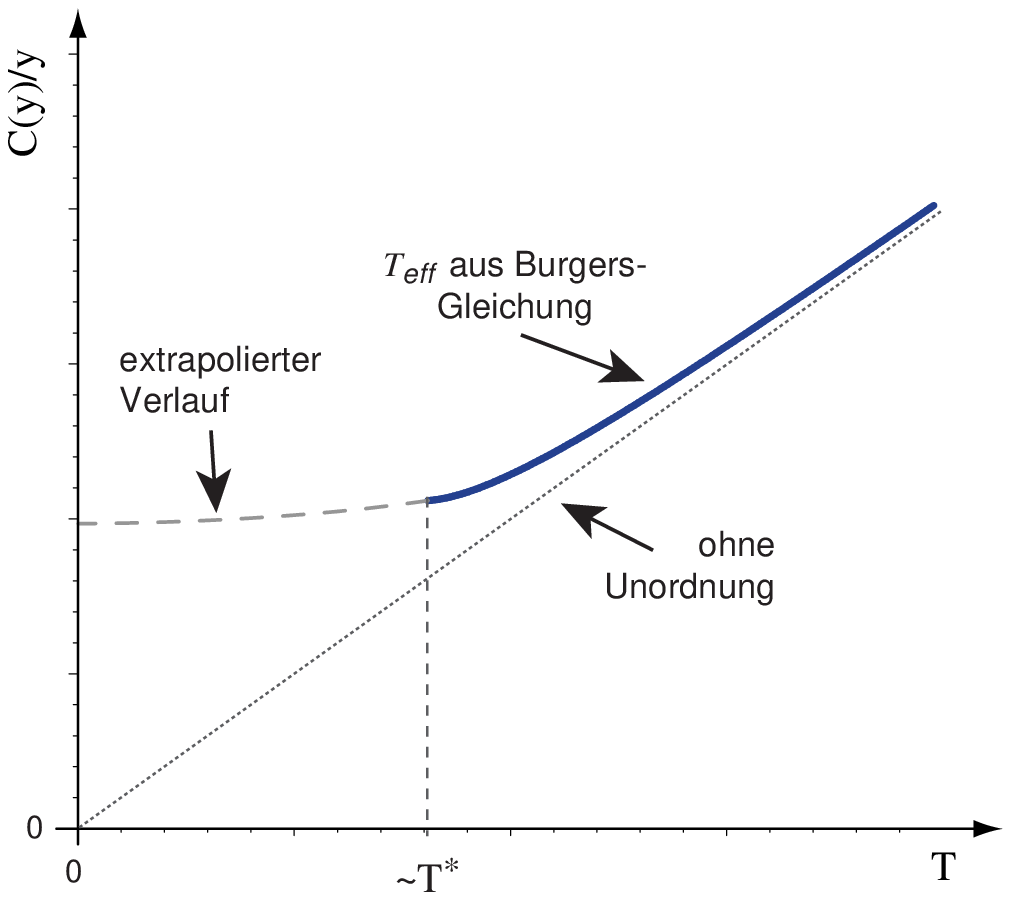}{.6\linewidth}{Temperaturabh"anigkeit der Korrelationsfunktion im
ungest"orten Fall (punktiert) und in \emph{one-loop}-N"aherung (durchgezogen).
Gestrichelt ist das Verhalten f"ur tiefe Temperaturen im Unordungsfall extrapoliert
dargestellt.}{fig.burgers3}

Die durch Gl. (\ref{eq.burgersT}) gegebene Abweichung der Temperaturabh"anigkeit
der Korrelationsfunktion vom ungest"orten Fall, erfa"st aufgrund der \emph{one-loop}-
N"aherung nur das Verhalten f"ur h"ohere Temperaturen $T>(4\pi^2)^{1/3}T^{\ast}$
bzw. kleine Korrekturen zur Diffusionskonstante. Daher liefert diese N"aherung f"ur
gro"se Temperaturen wieder das einfache lineare Verhalten,
aber auch eine Erh"ohung der effektiven Temperatur, aufgrund der Unordnung.
In Abb. \ref{fig.burgers3} ist die Temperaturabh"anigkeit der
Korrelationsfunktionen dargestellt.

Dass die Paarkorrelationsfunktion bei $T=0$ mit Unordnung bei
endlichen Abst"anden ($y>0$) nicht null ist, ist klar, da die Phase
an den Positionen der Verunreinigungen auf einen Wert fixiert wird,
und somit die perfekte Ladungsordnung zerst"ort wird und damit auch
die langreichweitige Ordnung.

%% file: strongpin.tex
\subsubsection{Starkes Pinning}\label{sec.strongpin}

F"ur starke Unordnung ($V_i\rho_i\gg\hbar v_F c_{\textit{imp}}$) geht man
zweckm"a"sigerweise zu einem
effektiven, diskreten Modell "uber, indem man das Phasenfeld $\varphi(x)$
jeweils zwischen den Orten der Verunreinigungen in der
Zustandssumme $\int\pd\varphi\, e^{-{\cal H}_{\textit{klass}}/T}$ ausintegriert
($L_c$ ist jetzt nicht mehr gro"s gegen"uber $c_{\textit{imp}}^{-1}$)~\cite{CDW:Feigel80};
hier exemplarisch zwischen $\varphi_i=\varphi(x_i)$ und $\varphi_{i+1}$
($\epsilon=(x_{i+1}-x_i)/M$):
\begin{eqnarray}
 &&\int\limits_{\varphi=\varphi_i}^{\varphi=\varphi_{i+1}}\pd\varphi\,
    e^{-\frac{\hbar v_F}{4\pi T}\int\limits_{x_i}^{x_{i+1}} dx\, (\partial_x\varphi)^2}\,\nn\\
 &=&\lim_{M\rightarrow\infty}\int\limits_{\mathbb{R}^M}\frac{1}{\mathcal{N}}\prod\limits_{k=1}^{M-1}
    d\varphi_{i,k}e^{-\frac{\hbar v_F}{4\pi T}\sum\limits_{k=1}^{M-1}\epsilon
     \left(\frac{\varphi_{i,k+1}-\varphi_{i,k-1}}{2\epsilon}\right)^2}\nn\\
 &=&\lim_{M\rightarrow\infty}\int\limits_{\mathbb{R}^M}\frac{d\vec{\tilde \varphi}_i}{\mathcal{N}}
   e^{-\frac{\hbar v_F}{2\pi 8T\epsilon}(\vec{\tilde \varphi}_i^T\cdot\mat{\tilde A}_i\cdot\vec{\tilde \varphi}_i
+\varphi_{i,0}^2+\varphi_{i,M}^2-2(\varphi_{i,0}\varphi_{i,2}+\varphi_{i,M-2}\varphi_{i,M}))}
\end{eqnarray}
Dabei wurden folgende Bezeichnungen eingef"uhrt:
\bi
 \item $\varphi_{i,k}\equiv\varphi(x_i+k\epsilon)$
 \item $\vec{\tilde \varphi}_i^T\equiv (\varphi_{i,1},\ldots,\varphi_{i,M-1})$
 \item $\mat{\tilde A}_i\equiv \begin{pmatrix}
  2 &  0 & -1 &  0 &  \ldots &  0\\
  0 &  2 &  0 & -1 & \ddots & \vdots \\
  -1 &  0 & 2 & \ddots & \ddots & 0\\
  0 & -1 & \ddots & \ddots & 0 & -1 \\
  \vdots & \ddots & \ddots & 0 & 2 & 0\\
  0 & \ldots & 0 & -1 & 0 & 2
  \end{pmatrix}\in \mathcal{M}_{M-1}(\mathbb{R})$
\ei

Das $M$-dimensionale Gau"sintegral kann leicht berechnet werden (siehe (\ref{eq.nGI})) und man erh"alt:
\begin{equation}
 \int\limits_{\varphi=\varphi_i}^{\varphi=\varphi_{i+1}}\pd\varphi\,
   e^{-\frac{\hbar v_F}{4\pi T}\int\limits_{x_i}^{x_{i+1}} dx\, (\partial_x\varphi)^2}
   =Re^{-\frac{\hbar v_F}{4\pi T}\frac{(\varphi_{i+1}-\varphi_i)^2}{x_{i+1}-x_i}}
\end{equation}
mit einer Konstanten $R$.

Daher hat der effektive Hamiltonian die Form
\begin{equation}
   {\cal H}_{\textit{eff}}=\sum_{i=1}^{N} \left\{ \frac{\hbar v_F}{4\pi}\frac{(\varphi_i-\varphi_{i+1})^2}{x_i-x_{i+1}}
    - V_i\rho_1\cos(\varphi_i+Qx_i)\right\}\,.
   \label{eq.effmod}
\end{equation}

Da wir das Verhalten bei tiefen Temperaturen untersuchen wollen,
nehmen wir weiter an, dass $V_i\gg T$. $\varphi_i$ wird, um die
Energie zu minimieren, nur Werte annehmen, f"ur die gilt: $\varphi_i+Qx_i=2\pi n_i$
mit $n_i\in \mathbb{Z}$.
Nun setze ich $n_{i+1}-n_i=h_i+\gb{\frac{Q}{2\pi c_{\textit{imp}}}}$ und
$x_{i+1}-x_i=\frac{1}{c_{\textit{imp}}}+\epsilon_i$, wobei f"ur die $x_i$ gilt:
$0\le x_1\le x_2\le\ldots\le x_{N+1}\le L$ ($L$ war die Systeml"ange).
($\gb{x}=k$ ist die Gau"sklammerfunktion, welche ein $k\in\mathbb{Z}$
liefert, so da"s $|\gb{x}-k|$ minimal wird, insbesondere $\leq 1/2$.)
Der effektive Hamiltonian lautet damit
\begin{equation}
   {\cal H}_{\textit{eff}}=\frac{\hbar v_F}{4\pi}\sum_{i=1}^{N}
     \frac{4\pi^2\left(h_i-\frac{Q\epsilon_i}{2\pi}-\gamma\right)^2}{\frac{1}{c_{\textit{imp}}}+\epsilon_i}\,,
   \label{eq.effham}
\end{equation}
wobei $\gamma=\frac{Q}{2\pi c_{\textit{imp}}}-\gb{\frac{Q}{2\pi c_{\textit{imp}}}}$,
so da"s $|\gamma|\le\frac{1}{2}$ ist.

Au"serdem kann man schreiben:
\begin{eqnarray}
 \left(\varphi_n-\varphi_1\right)^2&=&\left((2\pi n_n- Q x_n)+\sum\limits_{i=2}^{n-1}(\varphi_i-\varphi_i)-(2\pi n_1-Qx_1)\right)^2\nn\\
  &=&4\pi^2\left(\sum_{i=1}^{n-1}\left(h_i+\gb{\frac{Q}{2\pi c_{\textit{imp}}}}\right)
      -\frac{Q}{2\pi}\sum_{i=1}^{n-1}\left(\epsilon_i+\frac{1}{c_{\textit{imp}}}\right)\right)^2\nonumber\\
  &=&4\pi^2\left(\sum_{i=1}^{n-1}\left(h_i-\frac{Q\epsilon_i}{2\pi}-\gamma\right)\right)^2
 \label{eq.phidiff}
\end{eqnarray}
Im folgenden ist es n"otig, die Statistik geordneter System zu benutzen - hier
genauer gesagt die Statistik der Abst"ande der Verunreinigungen $\epsilon_i+c_{\textit{imp}}^{-1}$.
In Anhang \ref{app.ostat} ist die Herleitung der Verteilungsfunktion der Abst"ande
allgemein dargestellt (Grundlage ist das Buch von David~\cite{STAT:David70}).
Sind die Positionen $x_i$ der Impurities gleichwahrscheinlich zwischen $0$ und $L$
verteilt, erh"alt man die \emph{Wahrscheinlichkeitsdichtefunktion} (pdf) oder
kurz Verteilungsfunktion der $\epsilon_i$, $p(\epsilon_i)$:
\begin{equation}
  p(\epsilon_i)=c_{\textit{imp}}(1-\frac{1}{L}(\epsilon_i+\frac{1}{c_{\textit{imp}}}))^{N-1}
\end{equation}
mit $-\frac{1}{c_{\textit{imp}}}\le\epsilon_i\le L-\frac{1}{c_{\textit{imp}}}$ (siehe Def. von $\epsilon_i$).
Im thermodynamischen Limes kann diese pdf gut durch
\begin{equation}
 p(\epsilon_i)\approx\frac{c_{\textit{imp}}}{e}e^{-c_{\textit{imp}}\epsilon_i}\,,\quad -\frac{1}{c_{\textit{imp}}}\le\epsilon_i<\infty
 \label{eq.pdfe}
\end{equation}
gen"ahert werden.

Damit k"onnen die n-ten Momente $\overline{\epsilon_i^n}$ f"ur $n\ge 1$ berechnet werden:
\begin{eqnarray}
 \overline{\epsilon_i^n}&=&\int_{-\frac{1}{c}}^{\infty}\frac{c}{e}e^{-c\epsilon_i}\epsilon_i^n\,d\epsilon_i\nonumber\\
 &=&\frac{c}{e}(-1)^n\left.\frac{\partial^n}{\partial \lambda^n}\right|_{\lambda=c}\int_{-\frac{1}{c}}^{\infty}e^{-\lambda\epsilon_i}\,d\epsilon_i\nonumber\\
 &=&\frac{c}{e}(-1)^n\left.\frac{\partial^n}{\partial \lambda^n}\right|_{\lambda=c}\frac{e^{\lambda/c}}{\lambda}
\end{eqnarray}
Substituiert man $x=\lambda/c$ und benutzt (\ref{eq.expdiff}), erh"alt man
\begin{equation}
 \overline{\epsilon_i^n}=\frac{n!}{c_{\textit{imp}}^n}\sum_{k=2}^{n}\frac{(-1)^k}{k!}
\end{equation}
f"ur $n>1$, $\overline{\epsilon_i}=0$ und f"ur den Korrelator
$\overline{\epsilon_i\epsilon_j}=\frac{1}{c_{\textit{imp}}^2}\delta_{ij}$.

Nutzt man aus, dass $\overline{h_i}=0$ und $\overline{h_i h_j}\propto\delta_{ij}$,
folgt f"ur die Korrelationsfunktion mit (\ref{eq.phidiff}):
\begin{equation}
 \overline{\left\langle\left(\varphi_n-\varphi_1\right)^2\right\rangle}=
 4\pi^2\overline{\left\langle\left(h_i-\frac{Q\epsilon_i}{2\pi}-\gamma\right)^2\right\rangle}\cdot n\,.
\end{equation}
Bei niedrigen Temperaturen $T\ll\hbar v_F c_{\textit{imp}}$,  nimmt $(h_i-\frac{Q\epsilon_i}{2\pi}-\gamma)$
den minimalen Wert an, was genau der Fall ist, wenn $h_i=\gb{\frac{Q\epsilon_i}{2\pi}+\gamma}$,
also
\begin{eqnarray}
 \underbrace{\overline{\left\langle\left(h_i-\frac{Q\epsilon_i}{2\pi}-\gamma\right)^2\right\rangle}}_{\equiv \textit{Cf}}&=&
 \overline{\left(\frac{Q\epsilon_i}{2\pi}+\gamma-\gb{\frac{Q\epsilon_i}{2\pi}+\gamma}\right)^2}\nonumber\\
 &=&\int_{-1/c_{\textit{imp}}}^{\infty}d\epsilon_i\,\frac{c_{\textit{imp}}}{e}e^{-c_{\textit{imp}}\epsilon_i}\left(\frac{Q\epsilon_i}{2\pi}+\gamma-\gb{\frac{Q\epsilon_i}{2\pi}+\gamma}\right)^2
\end{eqnarray}
Da $Q\gg c_{\textit{imp}}$ k"onnen wir den {\bf kleinen Parameter} $\alpha=\frac{\pi c_{\textit{imp}}}{Q}$
einf"uhren. Unter Beachtung dass $\gb{x+n}=\gb{x}+n$ f"ur $n\in \mathbb{Z}$ und
der Substitution $x=c_{\textit{imp}}\epsilon_i+1$ l"a"st sich $\textit{Cf}$ vereinfachen zu
\begin{equation}
 \textit{Cf}=\int_{0}^{\infty}dx\,e^{-x}\left(\frac{x}{2\alpha}-\gb{\frac{x}{2\alpha}}\right)^2\,.
\end{equation}
Dieses Integral l"a"st sich exakt berechnen (siehe Anhang \ref{app.cfint}) und man
erh"alt folgenden Ausdruck:
\begin{eqnarray}
 \overline{\left\langle\left(h_i-\frac{Q\epsilon_i}{2\pi}-\gamma\right)^2\right\rangle}&=&
 \frac{1}{2\alpha^2}-\frac{1}{2\alpha\sinh(\alpha)}\nn\\
 &=&\frac{1}{12}-\frac{7}{720}\alpha^2+\mathcal{O}(\alpha^4)
\end{eqnarray}
wobei $(\sinh \alpha)^{-1}$ im letzten Schritt f"ur kleine $\alpha$ entwickelt wurde
(Reihe f"ur $\csch$ ist in (\ref{eq.csch}) aufgeschrieben).

Zu bemerken ist, dass man unter der Annahme, dass
$\frac{Q\epsilon_i}{2\pi}+\gamma-\gb{\frac{Q\epsilon_i}{2\pi}+\gamma}$ im Intervall
$[-1/2,1/2]$ gleichwahrscheinlich verteilt ist, das einfache Resultat
$\textit{Cf}=\int_{-1/2}^{1/2}dx\,x^2=\frac{1}{12}$ bekommen w"urde.

Bei hohen Temperaturen $T\gg\hbar v_F c_{\textit{imp}}$ kann man vernachl"assigen, dass $h_i$ eine diskrete Variable ist,
und erh"alt in diesem Fall
$\overline{\left\langle\left(2\pi h_i-Q\epsilon_i\right)^2\right\rangle}=\frac{T}{c_{\textit{imp}}\hbar v_F}$.

Eine sinnvolle Interpolationsformel dieser beiden Ergebnisse ist durch
   \begin{equation}
    \overline{\left\langle\left(\varphi(x)-\varphi(0)\right)^2\right\rangle}
     \approx\left(\frac{\pi^2}{3}c_{\textit{imp}}-\frac{7\pi^4}{180}\frac{c_{\textit{imp}}^3}{Q^2}+\frac{T}{\hbar v_F}\right)|x|
    \label{eq.cfsp}
   \end{equation}
gegebeb.
Vergleicht man dies mit (\ref{eq.cfwp}) sieht man, dass auch hier der Rauhigkeitsexponent
unver"andert $\zeta=1/2$ ist.


\subsubsection{numerische Untersuchungen}\label{sec.MC}

Um das Resultat (\ref{eq.cfwp}) f"ur schwache Unordung numerisch zu "uberpr"ufen,
wurde ein \emph{Monte-Carlo} Algorithmus verwendet.
Ausgangspunkt war die diskrete Version des Hamiltonians
(\ref{eq.rmodel}) mit $\tilde g(y)=0$, d.h. nicht renormiert, mit
welchem die Paarkorrelationsfunktion berechnet wurde.

\emph{Bemerkung zum Algorithmus:} F"ur die Phase wurden freie Randbedingungen
benutzt und das Erreichen des thermischen Gleichgewichtes
wurde "uberpr"uft, indem die Daten von Programmdurchl"aufen
mit mindestens dreimal l"angerer Equilibierungszeit verglichen
wurden. F"ur den Mittelungsprozess wurde die erste H"alfe
der Monte-Carlo-Schritte nicht ber"ucksichtigt.

\twofigures{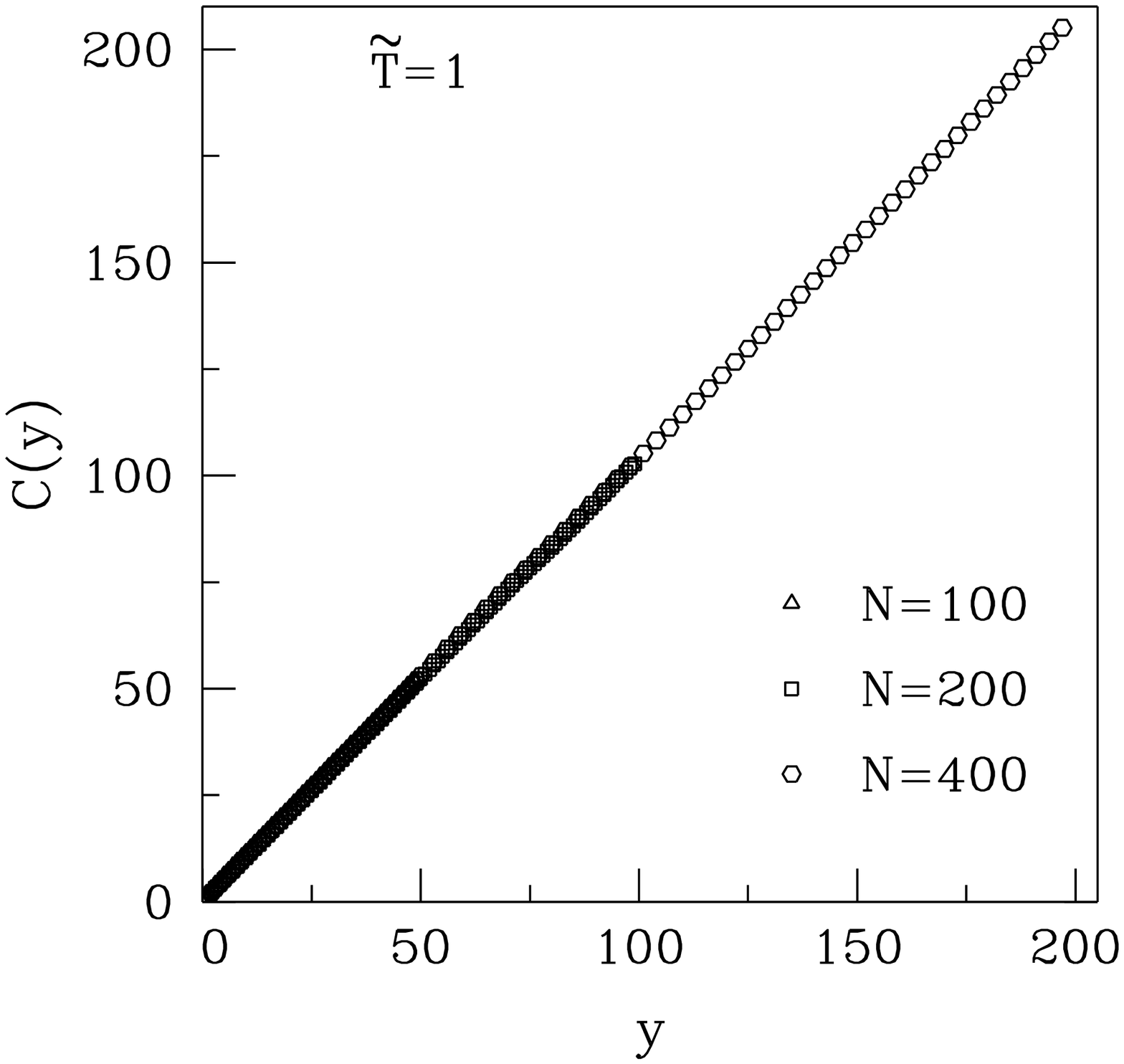}{0.45\linewidth}{Die Abstandsabh"angigkeit
der Korrelationsfunktion $C(y)$ f"ur verschiedene Systemgr"o"sen:
$N=100, 200$ und 400; $\tilde T=1$. Die Daten sind "uber 50--200
Unordnungskonfigurationen gemittelt.}{fig.mc1}{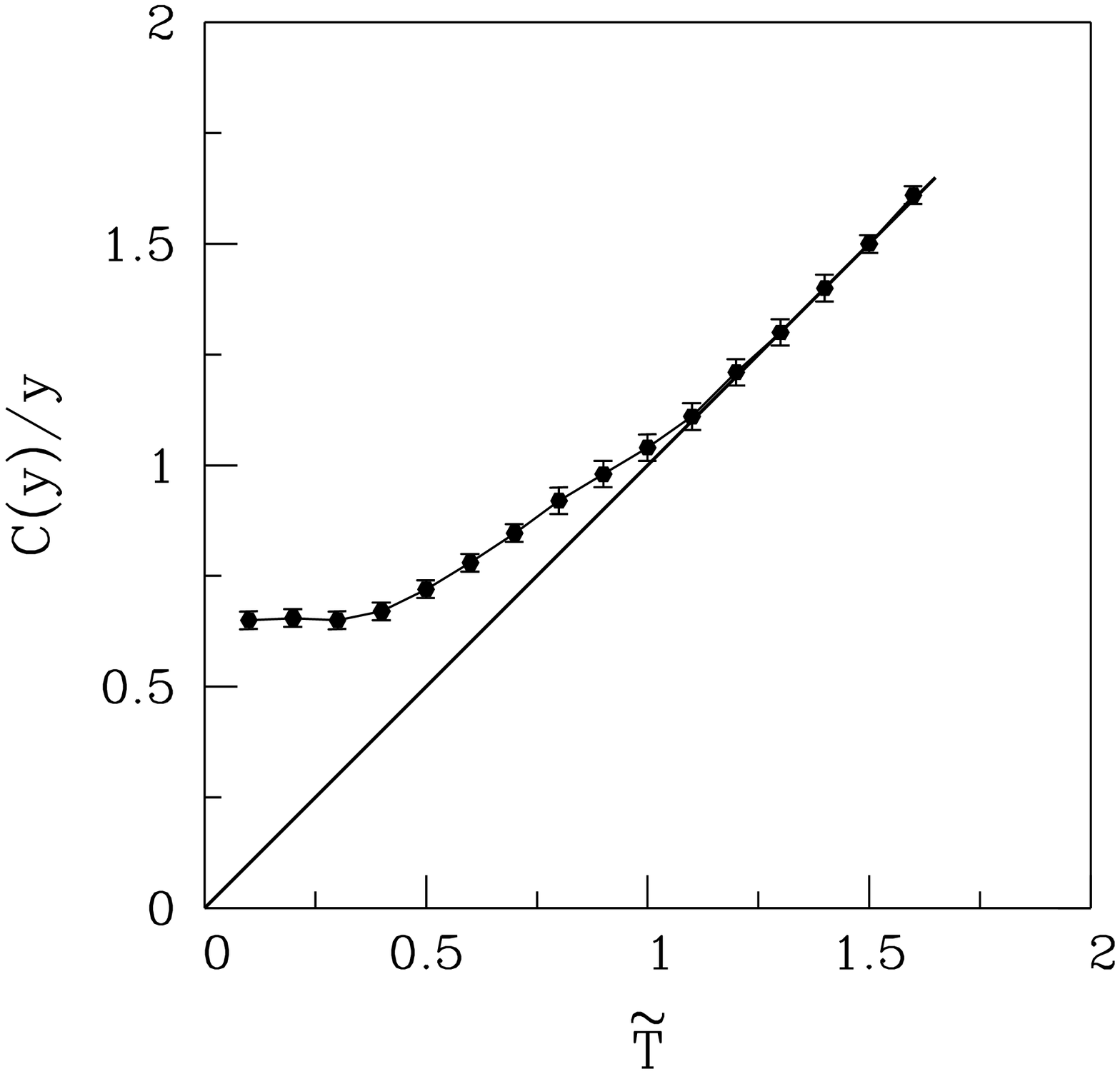}{0.45\linewidth}{Die Temperaturabh"angigkeit
der Funktion $C(y)/y$ - auch "uber 50--200 Unordnungskonfigurationen
gemittelt. Die Gerade zeigt die Temperaturabh"angigkeit der
Korrelationsfunktion ohne Unordnung.}{fig.mc2}

Abb. \ref{fig.mc1} zeigt die Korrelationsfunktion $C(y)$ als
Funktion des Abstandes $y$ bei fixierter Temperatur f"ur
verschiedene Systemgr"o"sen. Wie man sieht, hat man f"ur alle
Systemgr"o"sen eine lineare Abh"angigkeit vom Abstand $y$ 
und best"atigt somit den
Rauhigkeitsexponenten $\zeta=1/2$.

In Abb. \ref{fig.mc2} ist die (interessantere) Temperaturabh"angigkeit
der Paarkorrelationsfunktion abbgebildet. Die Berechnung
wurde mit einem System der Gr"o"se $N=200$ durchgef"uhrt.
F"ur $y\rightarrow 0$ sieht man, dass $C(y)/y$ gegen
einen Wert $\Od{1}$, genauer gegen $\tilde\sigma\approx0.65$, geht,
was wiederum das Ergebnis (\ref{eq.cfwp}) plausibel erscheinen l"a"st.
Au"serdem sieht man, dass f"ur $\tilde T\lesssim 0.3$ die Unordnung
dominiert und die Korrelationsfunktion temperaturunabh"angig wird,
wo hingegen f"ur $\tilde T\gtrsim 1.3 $ die Unordnung irrelevant
wird (siehe auch Abschnitt \ref{sec.flussdia}).

%% file: dynamic.tex
\subsection{Dynamik}\label{sec.dynamic}

In diesem Abschnitt wird das CDW-System im Fall eines externen
elektischen Feldes untersucht. Insbesondere wird das Kriechverhalten
bei endlichen Temperaturen betrachtet.

\onefigure{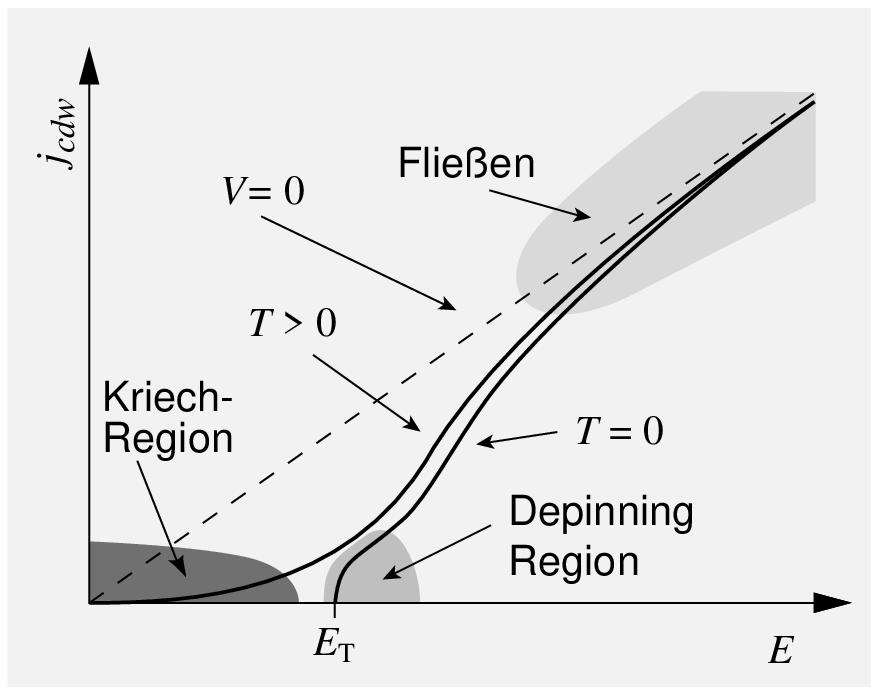}{0.7\linewidth}{''dynamische Regionen'' der
kollektiven CDW--Geschwindigkeit bzw. des CDW--Stroms:
Depinning Region bei $T=0$, Kriech-Region f"ur
$T^{\ast}>T>0$ und Flie"s--Regime f"ur $E\gg E_T$.}{fig.dyndia}

\onefigure{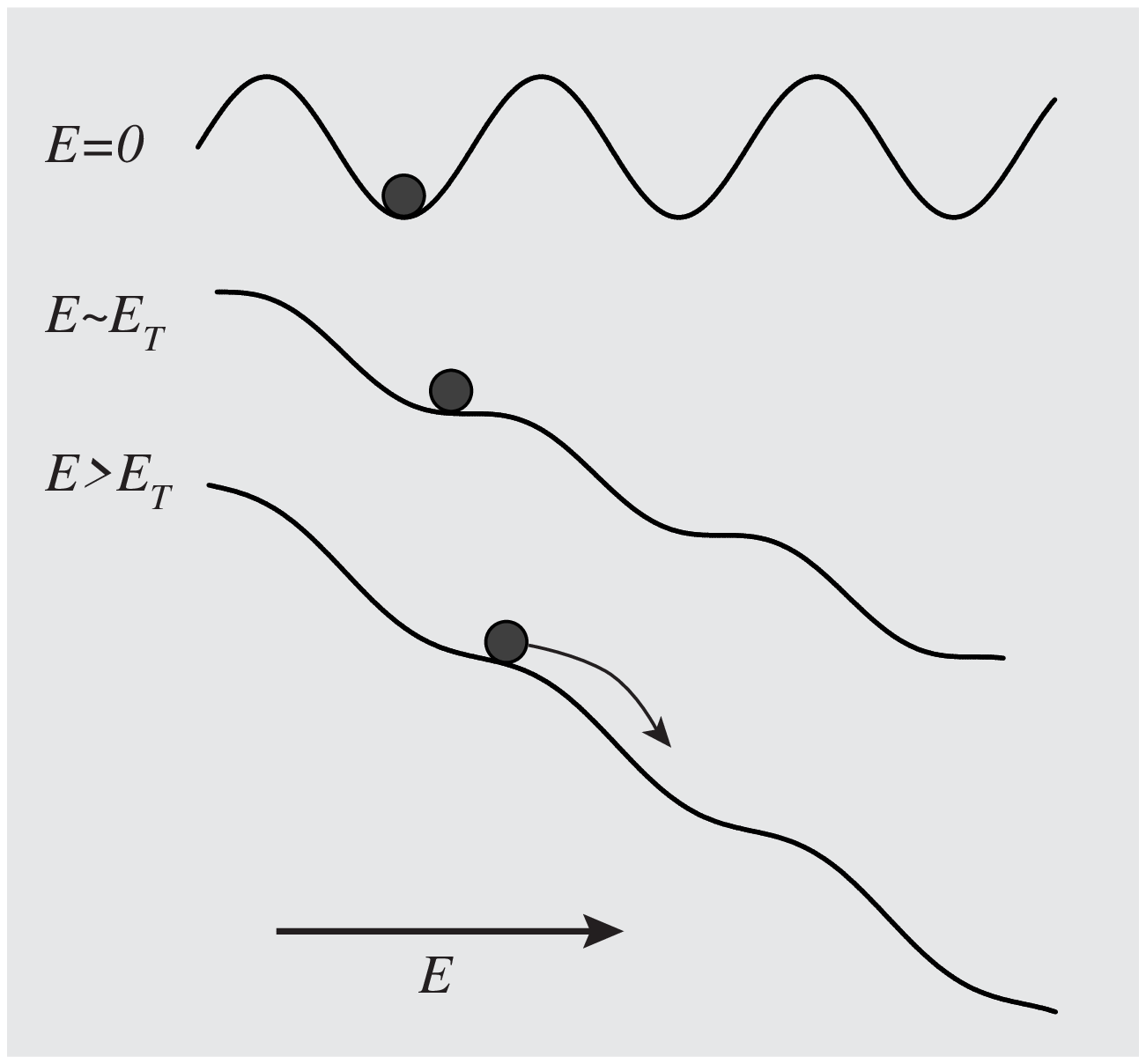}{0.5\linewidth}{Energielandschaft bei $T=0$
f"ur ein Teilchen bei verschiedenen Werten von $E$.}{fig.einteil}

Wie schon fr"uher gesehen, erzeugt die Unordnung auf L"angenskalen
$\geq L_c$ Energiebarrieren der H"ohe $T^{\ast}$. Bei $T=0$ und ohne
externes Feld bleibt das Phasenfeld an den Verunreinigungen gepinnt.
Legt man nun ein elektrisches Feld an das System an, bleibt dieser
Zustand auch bis zu einem kritischen Schwellenwert des elektrischen
Feldes $E_T$, welches durch die Barrierenh"ohe
vorgegeben ist, (bei $T=0$ !) erhalten. Wird die Feldst"arke weiter
erh"oht, k"onnen die Barrieren "uberwunden werden,
sodass das Phasenfeld eine endliche mittlere Geschwindigkeit
hat, d.h. ein Strom flie"st (\emph{Depinning "Ubergang},
siehe Abschnitt \ref{sec.depin}). In Abb. \ref{fig.einteil} ist dies vereinfacht
am Beispiel eines Teilchens veranschaulicht.

Bei endlichen Temperaturen, insbesondere im Bereich $0<T\lesssim T^{\ast}$,
wird man f"ur $E>0$ immer einen Strom messen, da die thermischen Fluktuation der
Ladungsdichtewelle erlauben die Barrieren zu "uberwinden (\emph{Kriech-Bewegung},
siehe Abschnitt \ref{sec.creep}). In Abb. \ref{fig.dyndia} sind die Regionen
der m"oglichen dynamische Prozesse dargestellt.

F"ur weitere Untersuchungen wird der Beitrag des externen Feldes zum Hamiltonian ben"otigt:

Das elektrische Feld koppelt an den langsam variierenden Teil der Ladungsdichte
$\propto \partial_x\varphi$. Der entsprechende Hamiltonian hat die Form
\footnote{Bisher war die Wahl des Hilbertraumes bzw. welche Werte $\varphi$
annehmen konnte nicht festgelegt, d.h. man konnte $\varphi$ als Winkelvariabel $\in [0,2\pi]$
oder $\varphi\in\mathbb{R}$ w"ahlen. In Anwesenheit eines elektrischen Feldes
ist der Hamiltonian allerdings nicht mehr invariant unter der Transformation
$\varphi\rightarrow\varphi+2n\pi$, daher mu"s $\varphi$  alle Werte annehmen k"onnen
(siehe \cite{CDW:Schoen88}).}:
\begin{equation}
 {\cal H}_{\textit{ext}}=-\int dx\, E x \partial_x\varphi(x)=\text{const.}+\int dx\, E \varphi(x)
\label{eq.hext}
\end{equation}

Im n"achsten Abschnitt wird die Bewegungsgleichung hergeleitet, welche
Ausgangspunkt f"ur die numerischen Untersuchunungen ist.


\subsubsection{Bewegungsgleichung und Kriechgesetz}\label{sec.bewgl}

Als Ansatz f"ur die Bewegungsgleichung benutzen wir
 eine \emph{Langevin-Gleichung}, gegeben durch
\begin{equation}
   \frac{\partial\varphi}{\partial t}=-\gamma
   \frac{\delta {\cal H}}{\delta\varphi}+\eta(x,t)\,,
   \label{eq.langevin}
\end{equation}
wobei $\gamma$ ein kinetischer Koeffizient und $\eta(x,t)$
ein gau"sverteilter thermischer Rauschterm mit
\begin{eqnarray}
  \big<\eta\big>&=&0\,\qquad\text{und}\nn\\
  \big<\eta(x,t)\,\eta(x^{\prime},t^{\prime})\big>&=&
   2T\gamma\,\delta(x-x^{\prime})\,\delta(t-t^{\prime})
   \label{eq.rausch}
\end{eqnarray}
ist.

In Gleichung (\ref{eq.langevin}) ist ${\cal H}$ der volle Hamiltonian
inklusive ${\cal H}_{\textit{ext}}$.

F"uhrt man zus"atzlich zu (\ref{eq.rescale}) noch folgende Reskalierung
der Zeit $t$ ein
\begin{equation}
 \tau=\gamma\frac{T^{\ast}}{L_c}t\,,
\end{equation}
erh"alt man folgende reskalierte ("uberd"ampfte) Bewegungsgleichung
\begin{equation}
 \frac{\partial\tilde\varphi(y,\tau)}{\partial \tau}=
   \Delta \tilde\varphi(y,\tau)+\sin(\tilde\varphi(y,\tau)-\tilde\alpha(y))+\tilde E+\tilde\eta(y,\tau)\,.
   \label{eq.bewgl}
\end{equation}
Diese Gleichung enth"alt als einzige Parameter $\tilde T=T/T^{\ast}$
und $\tilde E=E/E^{\ast}$, wobei $E^{\ast}=T^{\ast}/L_c$ ($E^{\ast}$
ist von der Ordnung des $T=0$ Depinning-Schwellenfeldes $E_T)$.

Zur numerischen Untersuchung dieser Gleichung wird eine diskretisierte
Version der Bewegungsgleichung (\ref{eq.bewgl}) ben"otigt:
\begin{eqnarray}
  \frac{\Delta \tilde{\varphi}_i}{\Delta\tau}=&&(\tilde{\varphi}_{i+1}-2\tilde{\varphi}_i+\tilde{\varphi}_{i-1})\nonumber\\
  &&+\sin(\tilde{\varphi_i}-\tilde{\alpha_i})+\tilde E+\tilde\eta(i,\tau)\,,\quad i=1\ldots N
\end{eqnarray}
Der erste Term ist der eindimensionale \emph{Gitter-Laplaceoperator} und
$\tilde{\alpha_i}$ ist eine Zufallsphase, welche im Intervall $[0,2\pi[$
gleichwahrscheinlich verteilt ist. Der letzte Term $\tilde\eta(i,\tau)$,
das thermische Rauschen, ist durch (\ref{eq.rausch}) definiert.

Die mittlere kollektive Kriechgeschwindigkeit ist durch
$v(\tilde E,\tilde T)=\left\langle \tilde L^{-1}\int dy\,\partial_{\tau}\tilde\varphi(y,\tau)\right\rangle_{\tau}$
gegeben \cite{CDW:Gruener94}, bzw. in diskreter Version:
 \begin{equation}
  v(\tilde E,\tilde T)=\left\langle\frac{1}{N}\sum_{i=1}^{N}
     {\frac{\Delta \tilde{\varphi_i}}{\Delta\tau}}\right\rangle_{\tau}\,.
 \end{equation}
Zu bemerken ist, dass die Kriechgeschwindigkeit proportional zum CDW-Strom
$j_{\textit{cdw}}$ ist.

Die Bewegungsgleichung wurde numerisch mit einem speziellen
\emph{Runge--Kutta} Algorithmus, welcher f"ur
stochastische System geeignet ist \cite{NUM:Greenside81}, integriert, wobei
periodische Randbedingungen benutzt wurden. Der Hauptteil des
benutzten C-Programms ist in Anhang \ref{app.simprog} zu finden.

\twofigures{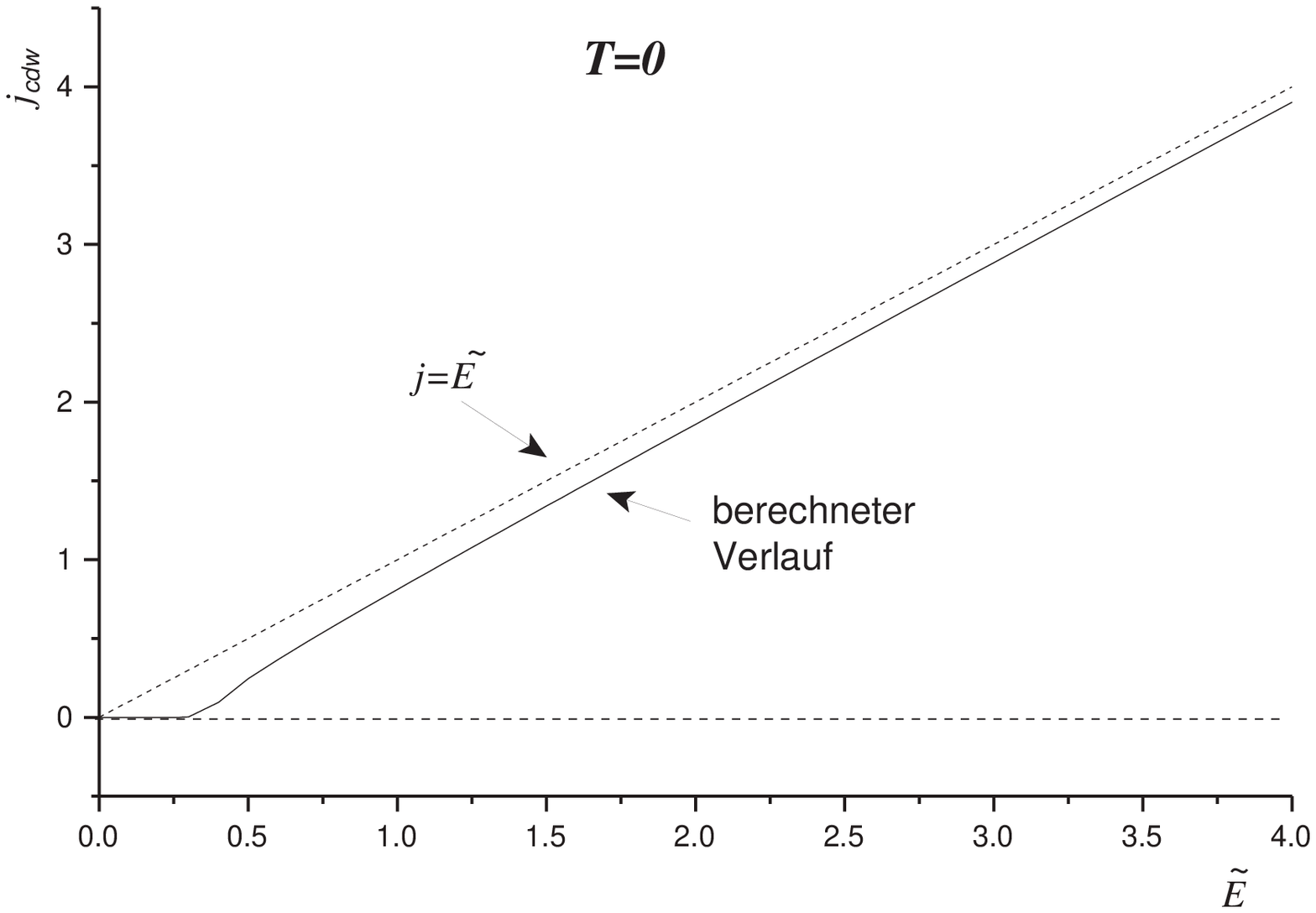}{0.45\linewidth}{Stromverlauf f"ur gro"se $E$ mit und ohne Unordnung bei $T=0$.}{fig.highe}
{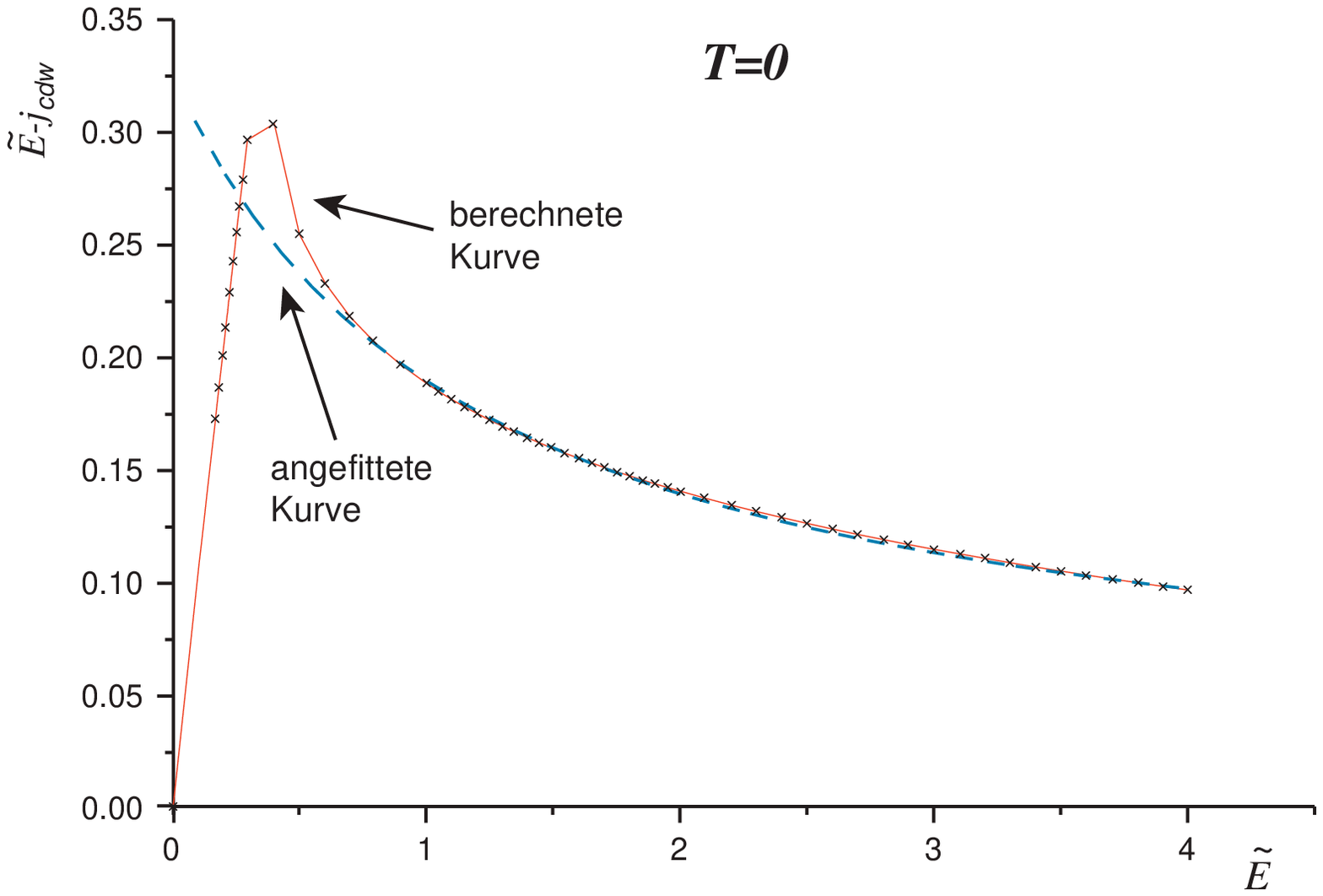}{0.45\linewidth}{Differenz der Str"ome aus Abb. \ref{fig.highe}.
Eine Funktion $\propto \tilde E^{-1}$ ist angepasst worden (gestrichelte Linie).}{fig.diffhighe}

F"ur Temperaturen $\tilde T\gg 1$ sind die Energiebarrieren
nicht mehr relevant, d.h. die ''Energielandschaft'' ist flach,
und die Ladungdichtewelle f"uhrt gem"a"s  $\dot\varphi\approx\gamma E$
eine ged"ampfte Bewegung aus. Dasselbe gilt nat"urlich auch f"ur
$\tilde E\gg 1$. In Abb. \ref{fig.highe} ist der Verlauf
der Geschwindigkeit mit dem angelegten Feld f"ur hohe Felder
mit und ohne Unordnung aufgetragen (bei $T=0$). In Abb. \ref{fig.diffhighe}
ist die Differenz dieser beiden Kurven aufgetragen und f"ur
gro"se $\tilde E$ mit einer Funktion $\propto \tilde E^{-1}$
angefittet, d.h. $v$ n"ahert sich mit $1/\tilde E$ dem Verlauf
ohne Unordnung bzw. bei hohen Temperaturen, was von S. Scheidl
und V.M. Vinokur gezeigt wurde \cite{CM:Scheidl98}.

Im umgekehrten Fall $T\ll T^{\ast}$ "andern sich die Energiebarrieren
auf der Skala $L>L_c$ um Betr"age von der Gr"o"senordnung $E$. Wenn man
von (\ref{eq.hext}) ausgeht, sieht man, dass aufgrund des
externen Feldes die Energiebarrieren (\ref{eq.ELc}) einen
zus"atzlichen Term proportional zu
\begin{equation}
 E L \tmw{\varphi^2}^{1/2}\sim E L \left(\frac{L}{L_c}\right)^{1/2}
\end{equation}
erhalten. Hierbei wurde benutzt, dass $\sigma\in\Od{L_c^{-1}}$
und der Rauhigkeitsexponent $\zeta=1/2$ ist.
Die Energiebarrieren $E_B^{\pm}$ haben infolgedessen die Form:
\begin{equation}
 E_B^{\pm}(L)\approx c_B T^{\ast}\mp c_E E \left(\frac{L}{L_c}\right)^{1/2} L\,,
\end{equation}
wobei $\pm$ sich jeweils auf die Bewegung parallel und entgegengesetzt dem
angelegten Feld bezieht und $c_B$ und $c_E$ Konstanten von der Ordnung $1$ sind.
Ohne externes Feld sind die Barrieren $E_B^{\pm}$, wie schon gesehen, von der
Ordnung $T^{\ast}$.

Da f"ur $L\approx L_c$ die Barrieren (parallel zum Feld) maximal werden,
kann mit Hilfe des \emph{Arrhenius Gesetzes} folgender Ausdruck f"ur die Kriechgeschwindigkeit
aufgeschrieben werden:
\begin{eqnarray}
   v(E,T)&\approx&\frac{\gamma}{2}\frac{T}{L_c}\left(e^{-E_B^{+}/T}-e^{-E_B^{-}/T}\right)\nn\\
    &=&\gamma\frac{T}{L_c}e^{-c_B(T^{\ast}/T)}\sinh\left({\frac{c_EEL_c}{T}}\right)\,,
   \label{eq.creeplaw}
\end{eqnarray}
d.h. $v(E,T)$ ist im wesentlichen die Differenz zwischen den Wahrscheinlichkeiten f"ur
eine Bewegung parallel bzw. antiparallel zum Feld.
Der Vorfaktor in (\ref{eq.creeplaw}) wurde so gew"ahlt, dass man f"ur
$T\gg (T^{\ast}+|E|L_c)$ das lineare Verhalten $v(\tilde E)\approx\gamma\tilde E$ erh"alt.

Man hat in einer Dimension demnach ein \emph{Kim--Anderson} Verhalten.
Dieses analytische \emph{Kriechgesetz} ist v"ollig
verschieden von dem nicht-analytischen in $d> 2$~\cite{CDW:Natter90}:
\begin{equation}
 v(E)\propto \exp\left(-\frac{T_{\xi}}{T}\left(\frac{E_{\xi}}{E}\right)^{\mu}\right)\,,\quad \mu=\frac{d-2}{2}
\end{equation}
($T_{\xi}$ und $E_{\xi}$ sind Parameter, die von der effektiven Barrierenh"ohe
und der typischen L"angenskala des Problems abh"angen.)
In $d=2$ folgt die Kriechgeschwindigkeit einem Potenz-Gesetz~\cite{CDW:TsaiShapir92}.

In den n"achsten zwei Abschnitten werden die Ergebnisse der numerischen Untersuchung
behandelt.


\subsubsection{Depinning bei $T=0$}\label{sec.depin}

\twofigures{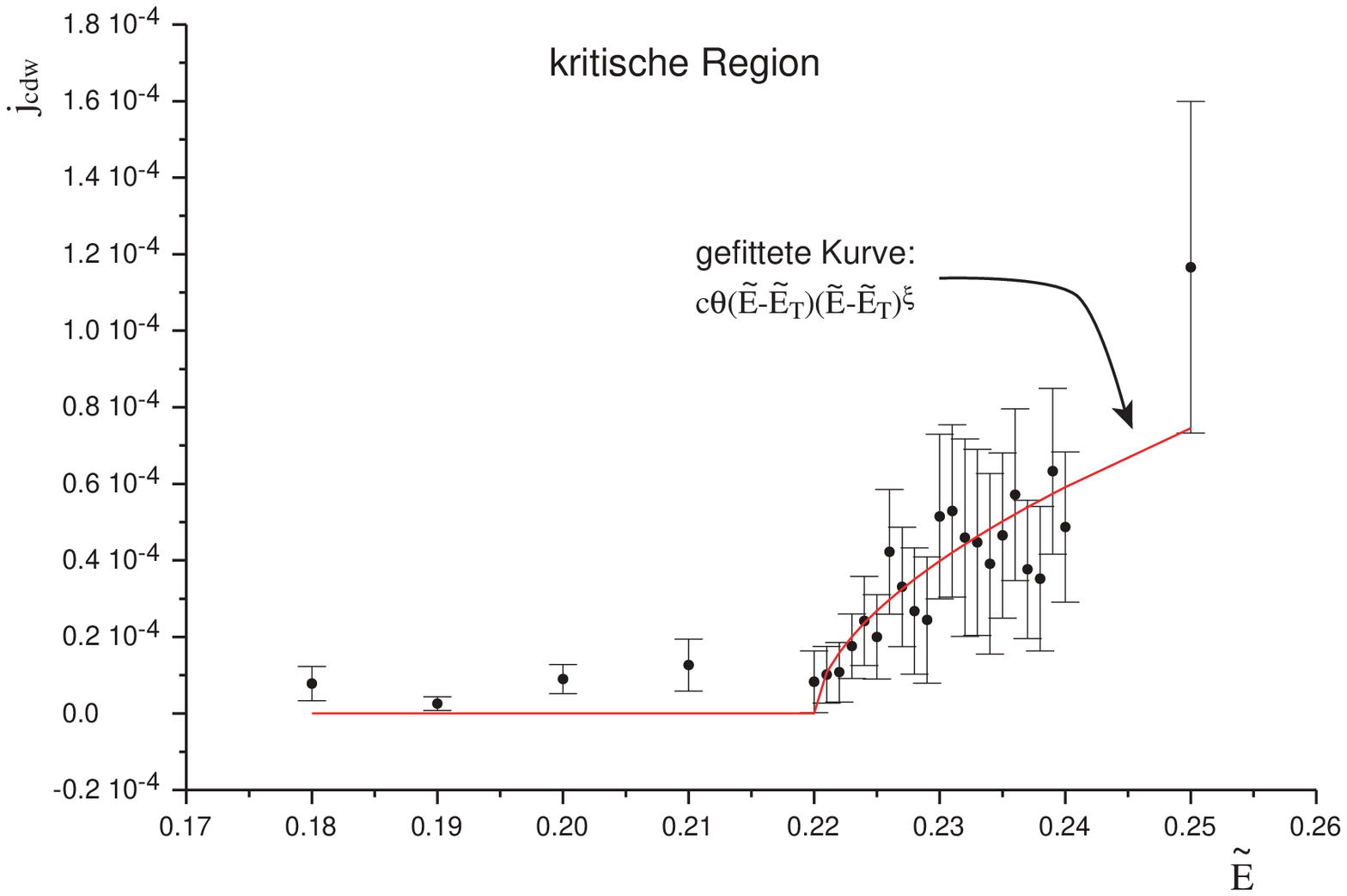}{0.45\linewidth}{Depinning Bereich. Die durchgezogene
Linie wurde mit dem in der doppelt-logarithmischen Auftragung ermittelten Wert f"ur $\xi$
erstellt.}{fig.critreg}
{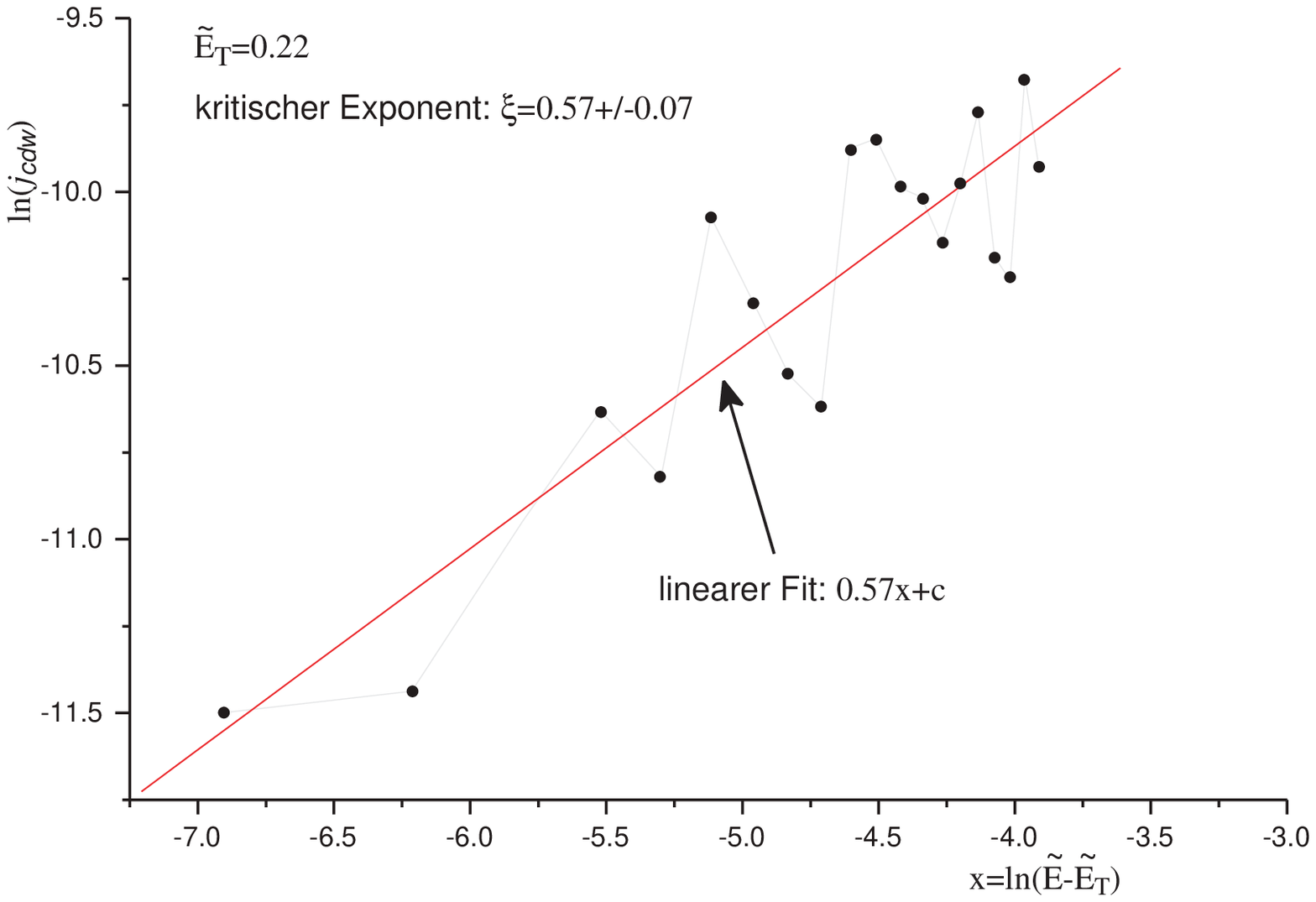}{0.45\linewidth}{doppelt-logarithmische Auftragung des Depinning Bereiches
mit angepasster Gerade.}{fig.logcrit}

Zur "Uberpr"ufung des Algorithmus wurde zun"achst der Depinning-"Ubergang
bei $T=0$ untersucht. F"ur den Schwellenwert des externen Feldes
wurde ein Wert von $\tilde E_T\approx 0.22$ gefunden
($E^{\ast}$ ist tats"achlich von der Gr"o"senordnung $E_T$).
Der kritische Exponent $\xi$, definiert durch
$v(E)\propto \Theta(\tilde E-\tilde E_T)(\tilde E-\tilde E_T)^{\xi}$ (im kritischen Bereich),
wurde durch eine Geradenanpassung in der doppelt-logarithmischen
Auftragung ermittelt und hat den Wert $\xi=0.57\pm0.07$.
Dieser Wert stimmt mit dem in anderen Arbeiten von
u.a. Sibani und Littelwood~\cite{CDW:SibaniLitt90} oder
auch Myers und Sethna~\cite{CDW:MyersSethna93} ermittelten gut "uberein.
In Abb. \ref{fig.critreg} und \ref{fig.logcrit} sind die ermittelten
Kurven aufgetragen.


\subsubsection{Kriechen}\label{sec.creep}

In diesem Abschnitt werden wir die CDW-Geschwindigkeit bei
endlichen Temperaturen untersuchen und Gleichung (\ref{eq.creeplaw})
f"ur die Kriechgeschwindigkeit "uberpr"ufen.

\twovfigures{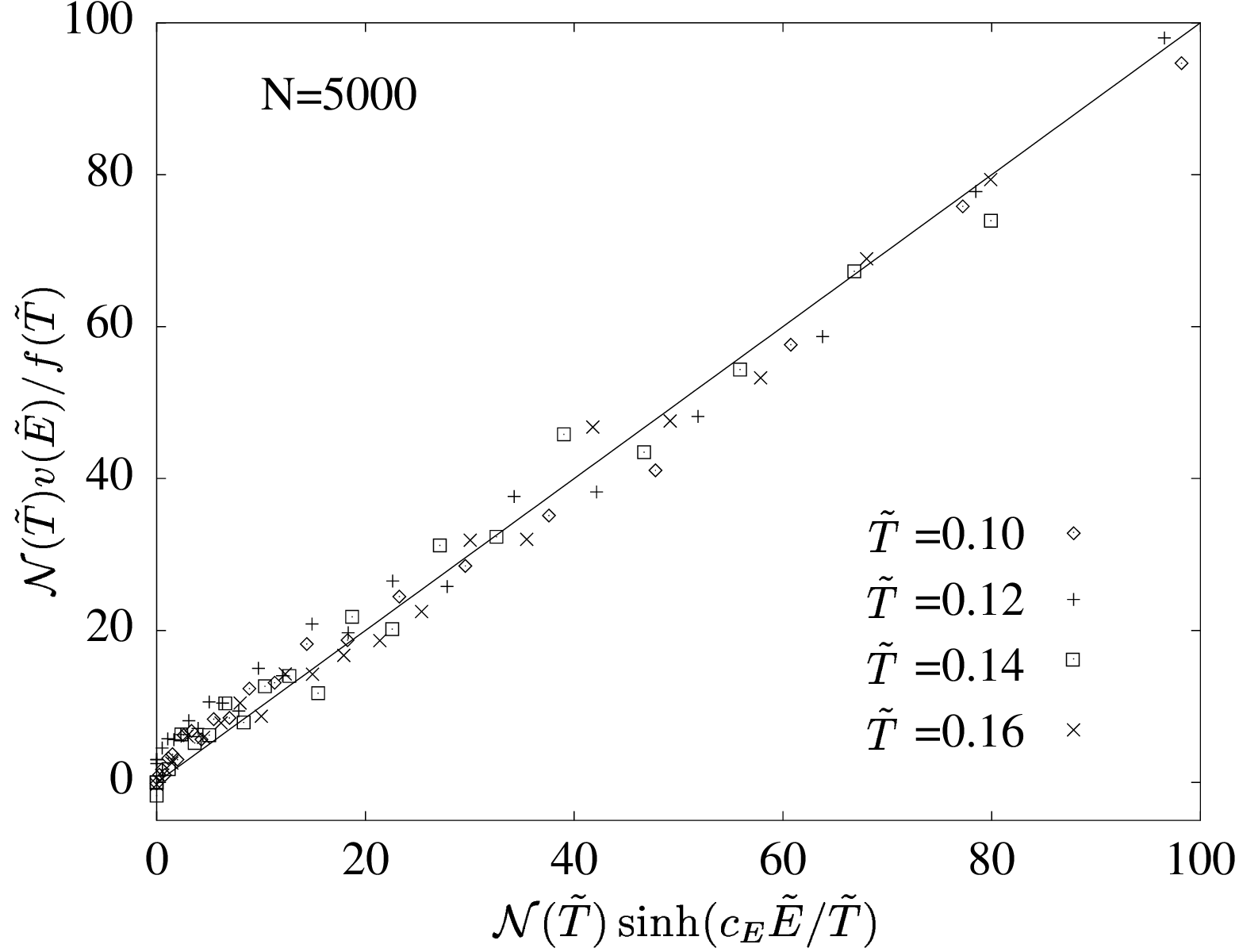}{0.7\linewidth}{Anh"angigkeit der Kriechgeschwindigkeit
vom externen Feld bei fixierter Temperatur $\tilde T$.
$f(\tilde T)=\gamma\frac{T}{L_c} e^{-\frac{c_B}{\tilde T}}$
und ${\cal N} (\tilde T)$ ist eine Skalierungs-Funktion,
um das lineare Verhalten besser darzustellen, mit folgenden Werte:
${\cal N} (0.10)=1.0$, ${\cal N} (0.12)=2.5$, ${\cal N} (0.14)=6.5$ und ${\cal N}(0.16)=9.0$.
F"ur $c_E$ wurde $2.5$ benutzt (siehe  Text).}{fig.3}{fig4.eps}{0.7\linewidth}
{Die Temperaturabh"angigkeit der Kriechgeschwindigkeit
f"ur verschiedene jeweils fixierte Werte von $\tilde E$, welche an den Kurven stehen.
F"ur eine "ubersichtlichere Darstellung wurden die Daten skaliert oder verschoben.
F"ur $\tilde E=0.08$ ist $3\tilde v(x)$ dargestellt, f"ur $\tilde E=0.10$ und $0.15$
wurden die Daten um $6\cdot 10^{-6}$ bzw. $10^{-5}$ nach oben verschoben. Typische
Fehlerbalken sind eingezeichnet.}{fig.4}

\threefigureso{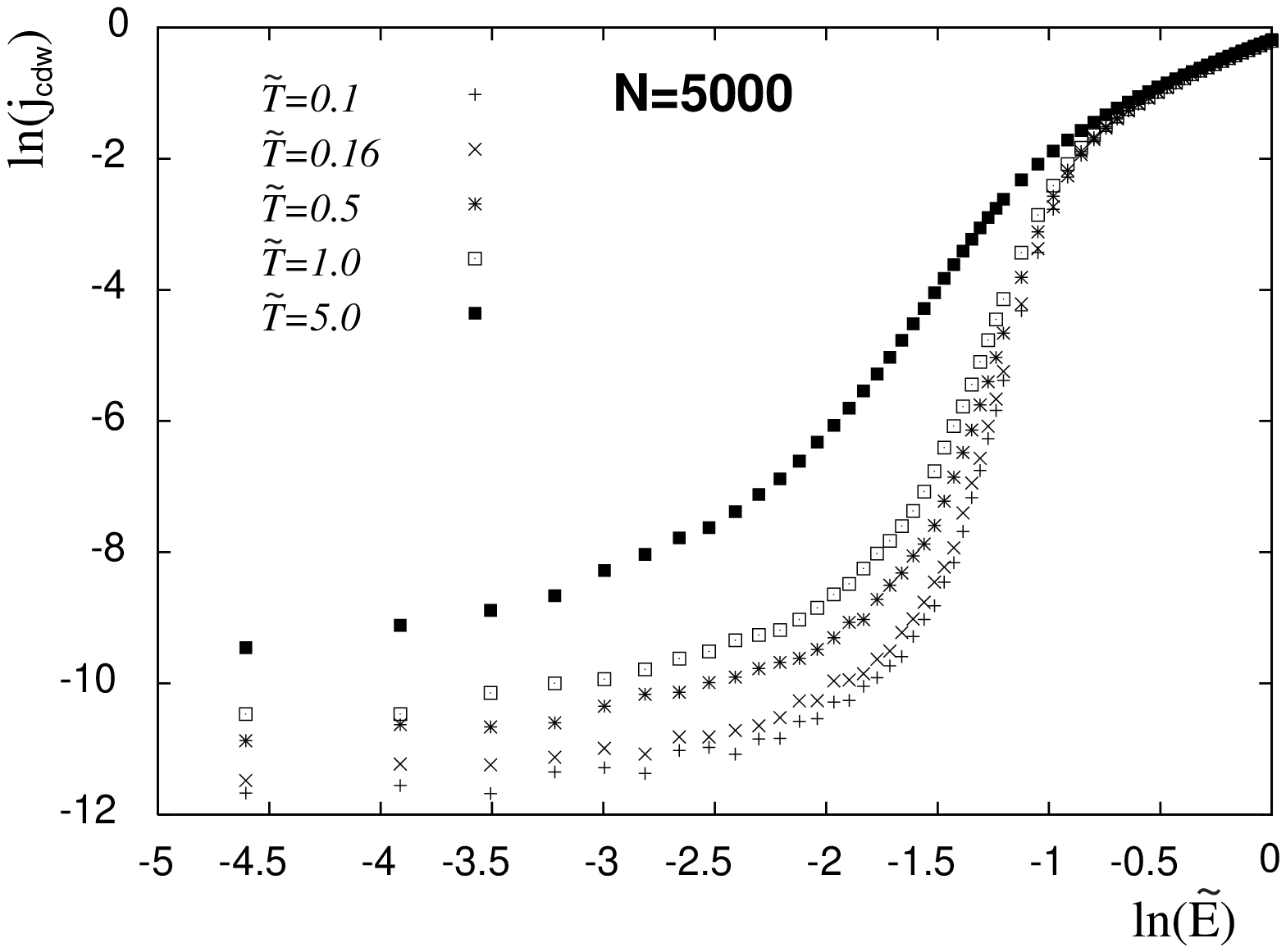}{0.7\linewidth}{Grafik analog zu Fig 1 in \cite{CDW:ZZ93}.}{fig.zz1}
\threefiguresu{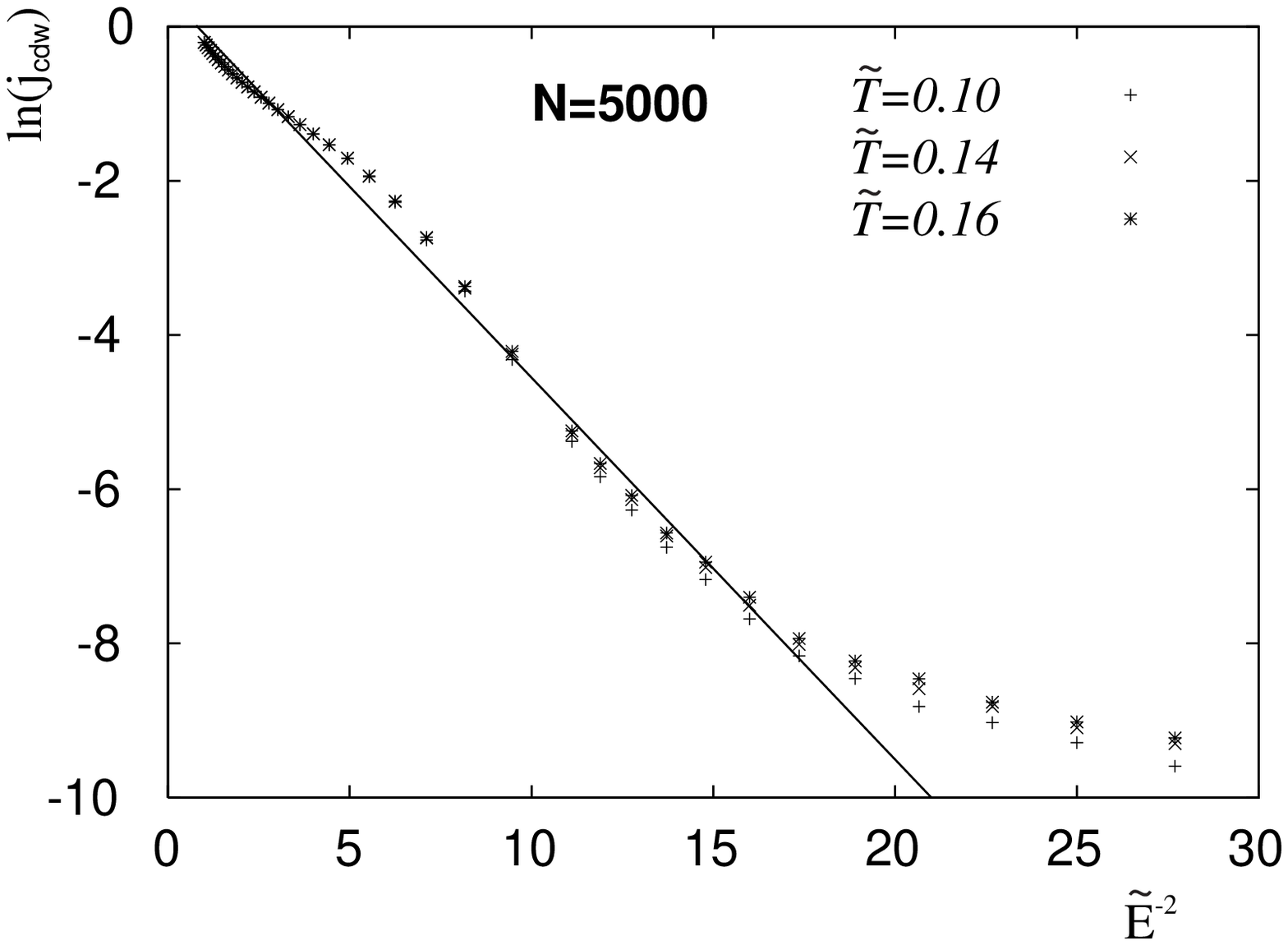}{0.5\linewidth}{Grafik analog zu Fig 3 in \cite{CDW:ZZ93}. (Erkl"arung siehe Text)}{fig.zz3}
{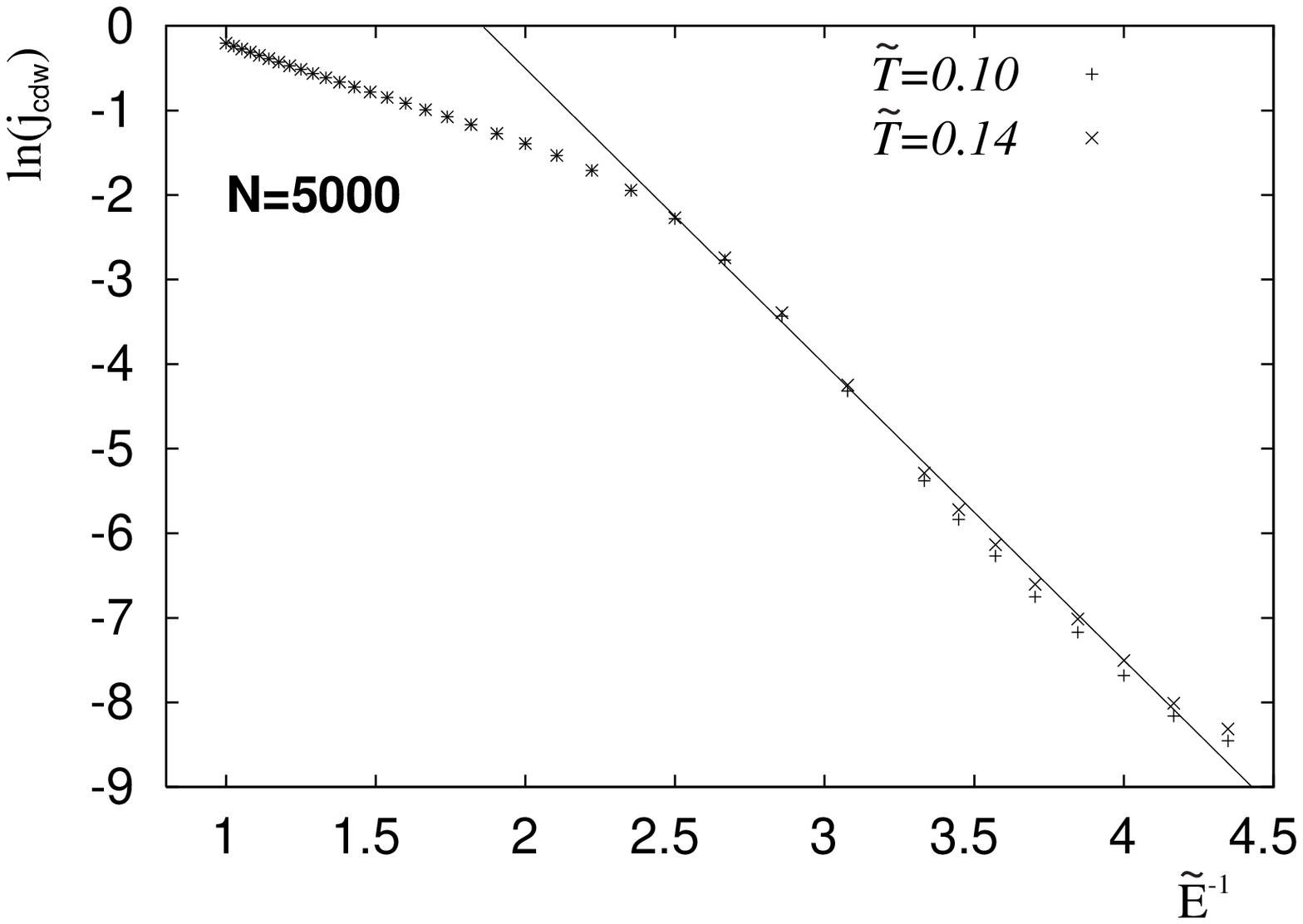}{0.4\linewidth}{Grafik analog zu Fig 3 (inset) in \cite{CDW:ZZ93}.}{fig.zz3i}

F"ur die Simulation des Kriechverhaltens wurde eine Simulationszeit von
$1000\tau_0$ und $\Delta\tau=0.05\tau_0$ benutzt.
Dreimal l"anger Zeiten haben das Resultat im wesentlichen
nicht ver"andert. Um (\ref{eq.creeplaw}) nachzuweisen und die Parameter
$c_E$ und $c_B$ zu bestimmen, wurde zun"achst $\tilde T$ festgehalten
und $\tilde E$ variiert. In diesem Fall wurde eine Systemgr"o"se von
$N=5000$ benutzt und "uber $50$ Unordnungskonfigurationen $\{\tilde\alpha_i;i=1\ldots N\}$
gemittelt. Das Resultat war wiederum stabil gegen"uber gr"o"seren
Systemenl"angen.
In Abb. \ref{fig.3} sind die Resultate dargestellt. Die gute Anpassung
an eine Gerade unterst"utzt, dass durch (\ref{eq.creeplaw}) die
Abh"angigkeit vom externen Feld richtig erfa"st wird.
Um den Wert von $c_E$ zu erhalten, wurde ein \emph{iteratives Minimal-Quadrat}-
Verfahren verwendet und damit $c_E=2.5\pm0.2$ ermittelt. Damit
wird auch die erwartete Gr"o"senordnung dieser Konstante als
richtig best"atigt.
Um den berechneten Stromverlauf mit Experimenten von S. Zaitsev-Zotov zu vergleichen,
tr"agt man die Daten doppelt-logarithmisch auf (Abb. \ref{fig.zz1})
und vergleicht dies mit dem in Fig. 1 in \cite{CDW:ZZ93} dargestellten
gemessenen Verlauf (Die Experimente wurden mit TaS$_3$ durchgef"uhrt).
Wie man sieht, stimmt das Verhalten qualitativ
gut "uberein, insbesondere im G"ultigkeitsbereich von (\ref{eq.creeplaw})
($\tilde T<1$, $\tilde E<1$ bzw. im Experiment $T\lesssim 20 K$, $E\lesssim 100 V/cm$).
In Abb. \ref{fig.zz1} wurden auch Kurven f"ur den "Ubergangsbereich $\tilde T\gtrsim 1$ vom
Kriechverhalten zum Flie"sen eingezeichnet, welche im Experiment
f"ur Temperaturen $T\gtrsim 20 K$ zu finden sind.
Au"serdem sieht man, wenn man den Stromverlauf mit Fig. 3 in \cite{CDW:ZZ93} vergleicht
(und beachtet, dass $\tilde E=1$ in etwa $E=100 V/cm$ entspricht),
dass (\ref{eq.creeplaw}) das Kriechverhalten sehr viel besser erfa"st, insbesondere f"ur
 kleine $\tilde E$, als das dort vorgeschlagene $\propto e^{-E_0/E}$-Verhalten
f"ur sehr kleine $E$ (siehe Abb. \ref{fig.zz3i} bzw. inset in Fig. 3 in \cite{CDW:ZZ93})
oder das $\propto e^{-c/(E^2 T)}$-Verhalten f"ur gr"o"sere $E$
(siehe Abb. \ref{fig.zz3} bzw. Fig. 3 in \cite{CDW:ZZ93}).

Um $c_B$ zu bestimmen, wurde diesmal die Temperaturabh"angigkeit
von $v(\tilde E,\tilde T)$ f"ur fixiertes $\tilde E$ untersucht.
Da hier eine Reduktion der Fehlerbalken sehr schwer zu erreichen
war, wurde diesmal eine Systemgr"o"se von $N=2000$ gew"ahlt
und "uber mindestens $500$ Unordnungskonfigurationen gemittelt.
Die Resultate sind in Abb. \ref{fig.4} zu sehen.
Mit dem ermittelten Wert f"ur $c_E$, wurde die Funktion
$\tilde v(x)=\gamma e^{-c_Bx}\sinh(Bx)$ mit $x=1/\tilde T$ und
$B=c_E\tilde E$ an die berechneten Werte angefittet und
damit $c_B=0.35\pm0.10$ gefunden.

Insgesamt wird das durch (\ref{eq.creeplaw}) bestimmte
Kriechverhalten gut durch die numerischen Simulationen
gest"utzt.

An Abb. \ref{fig.4} ist auch zu erkennen, dass der Strom $j_{\textit{cdw}}$
f"ur tiefe Temperaturen bei fixiertem $\tilde E$ saturiert. Dies stimmt
mit dem in Experimenten gemessenen Verhalten~\cite{CDW:ZZ93,CDW:ZZ94} "uberein
(siehe z.B. Abb. 2 in \cite{CDW:ZZ93}).


%% file: qmodel.tex
\newpage

\section{Erweiterung des Modells}\label{sec.qmodel}
\setcounter{equation}{0}

In diesem Kapitel wird das einfache klassische Modell aus dem
vorherigen Kapitel zu einem quantenmechanischen erweitert. Hinzu
kommen ein kinetischer Term und ein \emph{phase--slip}--Term.

In den folgenden Abschnitten wird zun"achst der vollst"andige
Hamiltonoperator aufgeschrieben, die Wirkung
aus der Pfadintegraldarstellung berechnet und schlie"slich
der ''Mechanismus'' der phase-slips diskutiert.
Zus"atzlich wird die Paarkorrelationsfunktion
des $1+1$-dimensionalen harmonischen Modells, d.h. ohne
Unordnung und phase slips, berechnet, da diese
in Kapitel \ref{sec.RG} wichtig sein wird.

\subsection{quantenmechanischer Hamiltonian}\label{sec.qham}

Das quantenmechanisch erweiterte Modell hat die Form
\begin{equation}
 {\cal \hat H}={\cal \hat H}_{\textit{kin}}+{\cal H}_{\textit{klass}}+{\cal \hat H}_{\textit{disl}}
\end{equation}
wobei ${\cal \hat H}_{\textit{kin}}$ der kinetische Anteil ist, mit
\begin{equation}
 {\cal \hat H}_{\textit{kin}}=\int\limits_{0}^{L} dx\, \frac{v^2}{2 c}\hat P^2\,,
\qquad \komm{\hat P(x)}{\hat \varphi(\str{x})}=-\imath\hbar\delta(x-\str{x})\,.
\end{equation}
$\hat P$ ist der zu $\varphi$ konjugierte Impulsoperator und $\frac{c}{v^2}=\rho$
entspricht einer Massendichte.

${\cal \hat H}_{\textit{disl}}$ ist der neu hinzugekommene \emph{phase--slip}--Term,
definiert durch
\begin{equation}
 {\cal \hat H}_{\textit{disl}}= \lambda\int\limits_0^L dx\,
  \cos\left(\frac{q\pi}{\hbar}\int\limits^{x} dy\, \hat P(y)\right)\,,
 \label{eq.hdisl}
\end{equation}
wobei $\lambda$ proportional zur Wahrscheinlichkeit der Ausbildung einer Dislokation
und $q$ eine zun"achst beliebige
Zahl ist. In Abschnitt \ref{sec.phaseslip} wird dieser Term ausf"uhrlicher motiviert
und diskutiert.

Der gesamte, erweiterte Hamiltonian lautet also:
\begin{equation}
 {\cal\hat H}=\int\limits_{0}^{L}\left[\frac{v^2}{2c}\hat P^2+\frac{c}{2}(\partial_x\hat \varphi)^2
 -V\cos(p\hat\varphi(x)+\alpha(x))+\lambda\cos\left(\frac{q\pi}{\hbar}\int\limits^{x} dy\,
\hat P(y)\right)\right]\,.
 \label{eq.qham}
\end{equation}
$p$ ist ebenfalls eine zun"achst beliebige Konstante.
Die elastische Konstante $c$ wird aufgrund der Quantenfluktuationen
i.a. verschieden von dem Wert $\hbar v_F/(2\pi)$ im klassischen Fall sein.
(Da im folgenden nur der Fall der schwachen Unordnung ben"otigt wird,
ist der Unordnungsterm analog zu (\ref{eq.xyrmodel}) aufgeschrieben.)

\subsection{Pfadintegraldarstellung}\label{sec.pathint}

Nun gehen wir "uber zur Wirkung; daf"ur f"uhre ich folgendes ''duales''
Feld $\theta$ ein, gegeben durch:
\begin{equation}
 \partial_x\hat\theta\equiv-\frac{\pi}{\hbar}\hat P
\end{equation}
(siehe auch Kapitel \ref{sec.results}).
Die Kommutatorrelation lautet damit: $\komm{\hat \theta(x)}{\hat\varphi(\str{x})}=\pi\Theta(x-\str{x})$.
Damit vereinfacht sich ${\cal \hat H}_{\textit{disl}}$ zu $\lambda\int dx\, \cos(q\hat\theta)$.
Da im folgenden der Unordnungsterm und der Dislokationsterm getrennt behandelt werden, wird
die Wirkung auch getrennt f"ur beide Terme berechnet. Beide Terme zusammen
lassen sich i.a. nicht als geschlossener Ausdruck in einem der beiden Felder
 aufschreiben.

Beginnen wir mit dem bekannten Unordnungsterm ($\lambda=0$):
Die Wirkung ist in diesem Fall durch
\begin{equation}
 \int\pd\varphi\pd\theta\, e^{-\int d\tau\,\tilde {\cal L}_{\varphi}}\equiv
  \int\pd\varphi\, e^{-{\cal S}_{\varphi}/\hbar}
\end{equation}
definiert, d.h. im Funktionalintegral wird das $\theta$-Feld ausintegriert.
$\tilde {\cal L}_{\varphi}$ ist gegeben durch
\begin{equation}
 \tilde {\cal L}_{\varphi}\equiv\hbar^{-1}\int dx\, \left[\frac{v^2\hbar^2}{2\pi^2 c}
  (\partial_x\theta)^2+\frac{c}{2}(\partial_x\varphi)^2-V\cos(p\varphi(x)+\alpha(x))+
   \frac{\imath\hbar}{\pi}\partial_x\theta\partial_{\tau}\varphi\right]\,.
\end{equation}
Da ${\cal \hat H}$ (f"ur $V=0$ oder $\lambda=0$) in einer einfachen (normal geordneten)
Form vorliegt und ${\cal \hat H}_0$ im Impulsraum im wesentlichen
in eine Summe von harmonischen Oszillatoren zerf"allt, l"a"st sich diese Form von
$\tilde {\cal L}_{\varphi}$ (oder auch $\tilde {\cal L}_{\theta}$) einfach
mit Hilfe des Pfadintegrals herleiten (siehe z.B. \cite{CM:Negele88}).

Um das $\theta$-Integral einfach berechnen zu k"onnen, bringt man  $\tilde {\cal L}_{\varphi}$
auf folgende quadratische Form
\begin{eqnarray}
 \tilde {\cal L}_{\varphi}\equiv\hbar^{-1}\int dx\,\bigg[&&
  \frac{v^2\hbar^2}{2\pi^2 c}\left(\Big(\partial_x\theta+
  \frac{\imath c \pi}{\hbar v^2}\partial_{\tau}\varphi\Big)^2+
  \frac{c^2\pi^2}{\hbar^2 v^4}(\partial_{\tau}\varphi)^2\right)\nn\\
  &&+\frac{c}{2}(\partial_x\varphi)^2-
  V\cos(p\varphi(x)+\alpha(x))\bigg]\,.
\end{eqnarray}
und kann mit der Substitution
$\partial_x\tilde\theta\equiv\partial_x\theta+\frac{\imath c \pi}{\hbar v^2}\partial_{\tau}\varphi$
das Gau"sintegral im Impulsraum einfach berechnen.

Letzendlich erh"alt man:
\begin{equation}\label{eq.actphi}
 {\cal S}_{\varphi}=\int\limits_0^{\hbar\beta}d\tau\,\int\limits_0^L dx\,
 \left[\frac{c}{2}\left\{(\partial_x\varphi(x,\tau))^2+v^{-2}(\partial_{\tau}\varphi(x,\tau))^2\right\}
-V\cos(p\varphi(x,\tau)+\alpha(x))\right]
\end{equation}
($\beta=((k_B)T)^{-1}$).

Auch hier kann man (f"ur $T>0$) zu reskalierten Gr"o"sen $\tilde x=x\lambda_T^{-1}$ und
$\tilde \tau=\tau(\hbar\beta)^{-1}$ "ubergehen, wobei $\lambda_T=\hbar\beta v$ die relativistische
\emph{de Broglie}-Wellenl"ange eines Teilchens mit der Geschwindighkeit $v$ ist:
\begin{eqnarray}
 \tilde {\cal S}_{\varphi}\equiv\frac{\mathcal{S}_{\varphi}}{\hbar}=
  \int_0^{1}d\tilde\tau\,\int_0^{L/\lambda_T}d\tilde x\,
 \Big[&&\frac{c}{2\hbar v}\left((\partial_{\tilde x}\tilde\varphi)^2+(\partial_{\tilde\tau}\tilde\varphi)^2\right)\nonumber\\
 &&-\beta V\lambda_T\cos(p\tilde\varphi(\tilde x,\tilde\tau)+\tilde\alpha(\tilde x))\Big]
\end{eqnarray}

F"ur den Dislokationsterm ($V=0$) geht man analog vor. Die Wirkung ist in diesem Fall
durch
\begin{equation}
 \int\pd\varphi\pd\theta\, e^{-\int d\tau\,\tilde {\cal L}_{\theta}}\equiv
  \int\pd\theta\, e^{-{\cal S}_{\theta}/\hbar}
\end{equation}
definiert, also wird jetzt, im Gegensatz zum Unordnungsterm das $\varphi$-Feld ausintegriert.
$ \tilde {\cal L}_{\theta}$ ist durch
\begin{equation}
 \tilde {\cal L}_{\theta}\equiv\int dx\, \hbar^{-1}\int dx\, \left[\frac{v^2\hbar^2}{2\pi^2 c}
  (\partial_x\theta)^2+\frac{c}{2}(\partial_x\varphi)^2+\lambda\cos(q\theta(x))+
   \frac{\imath\hbar}{\pi}\partial_{\tau}\theta\partial_{x}\varphi\right]\,.
\end{equation}

gegeben. Die entsprechende Wirkung lautet
\begin{equation}\label{eq.acttheta}
 {\cal S}_{\theta}=\int_0^{\hbar\beta}d\tau\,\int_0^L dx\, \left[\frac{\hbar^2}{2\pi^2 c}\left(v^2(\partial_x\theta)^2+
 (\partial_{\tau}\theta)^2\right)+\lambda\cos(q\theta)\right]
\end{equation}
oder in reskalierten Einheiten:
\begin{equation}
 \tilde{\cal S}_{\theta}= \frac{{\cal S}_{\theta}}{\hbar}=\int\limits_0^1 d\tilde \tau\,\int\limits_0^{L/\lambda_T}d\tilde x\,
 \left(\frac{\hbar v}{2\pi^2 c}\left[(\partial_{\tilde x}\tilde\theta)^2+(\partial_{\tilde\tau}\tilde\theta)^2\right]+\beta
 \lambda\lambda_T\cos(q\tilde\theta)\right)
\end{equation}
D.h. der Dislokationsterm wird durch ein $1+1$-dimensionales \emph{Sinus-Gordon}-Modell im Feld
$\theta$ beschrieben. Die Bedeutung des Feldes $\theta$ im Zusammenhang mit dem \emph{
Tomonaga--Luttinger}--Modell wird kurz in Kapitel \ref{sec.results} erkl"art.

Nachdem nun die Wirkungen hergeleitet sind, werde ich den zus"atzlichen
\emph{phase--slip}--Term etwas besser motivieren und herleiten.

\subsection{Phase--slips}\label{sec.phaseslip}
Der Term ${\cal \hat H}_{\textit{disl}}$ wurde von Pokrovskii
\cite{CDW:Pokro82} f"ur ein Modell benutzt, mit dem ein Phasen"ubergang
von einer inkommensurablen zu einer kommensurablen Phase in einem anisotopen
System mit Dislokationen untersucht wurde.
Dieser Term kann auf zwei verschiedene Weisen verstanden werden.
Zun"achst kann der Term in einer diskreten Version aufgeschrieben werden, also z.B.
\begin{eqnarray}
  \cos\left(\frac{q\pi}{\hbar}\int\limits^{x} dy\, \hat P(y)\right)&=&
   \cos\left(\frac{q\pi}{\hbar}\sum\limits_{k=1}^{N_x} \hat P_k\right)\nn\\
&=&\frac{1}{2}\left(\prod_{k=1}^{N_x} e^{\imath\frac{q\pi}{\hbar}\hat P_k}+\prod_{k=1}^{N_x}
   e^{-\imath \frac{q\pi}{\hbar}\hat P_k}\right)\,.
\end{eqnarray}
Man sieht, dass ${\cal \hat H}_{\textit{disl}}$ eine Superposition (und damit hermitesch) von
Produkten von Translationsoperatoren $e^{\pm\imath\frac{q\pi}{\hbar}\hat P_k}$ ist,
d.h. bis zum Teilchen $N_x$ an der Positon $x$ wird die Phase $\varphi$ (!) um
$\pm q\pi$ verschoben. Diese Verschiebung kann auch als Einf"uhrung eines $q\pi$\emph{--Vortex} bei $x$ verstanden werden\cite{CM:Sachdev99}.
Am Ort des Vortex mu"s das Phasenfeld eine Singularit"at haben, d.h.
das Phasenfeld undefiniert sein und damit der Ordnungsparameter verschwinden,
was zur Folge hat, dass die Amplitude auf null gedr"uckt wird. Man ber"ucksichtigt
somit indirekt Amplitudenfluktuationen.

\onefigure{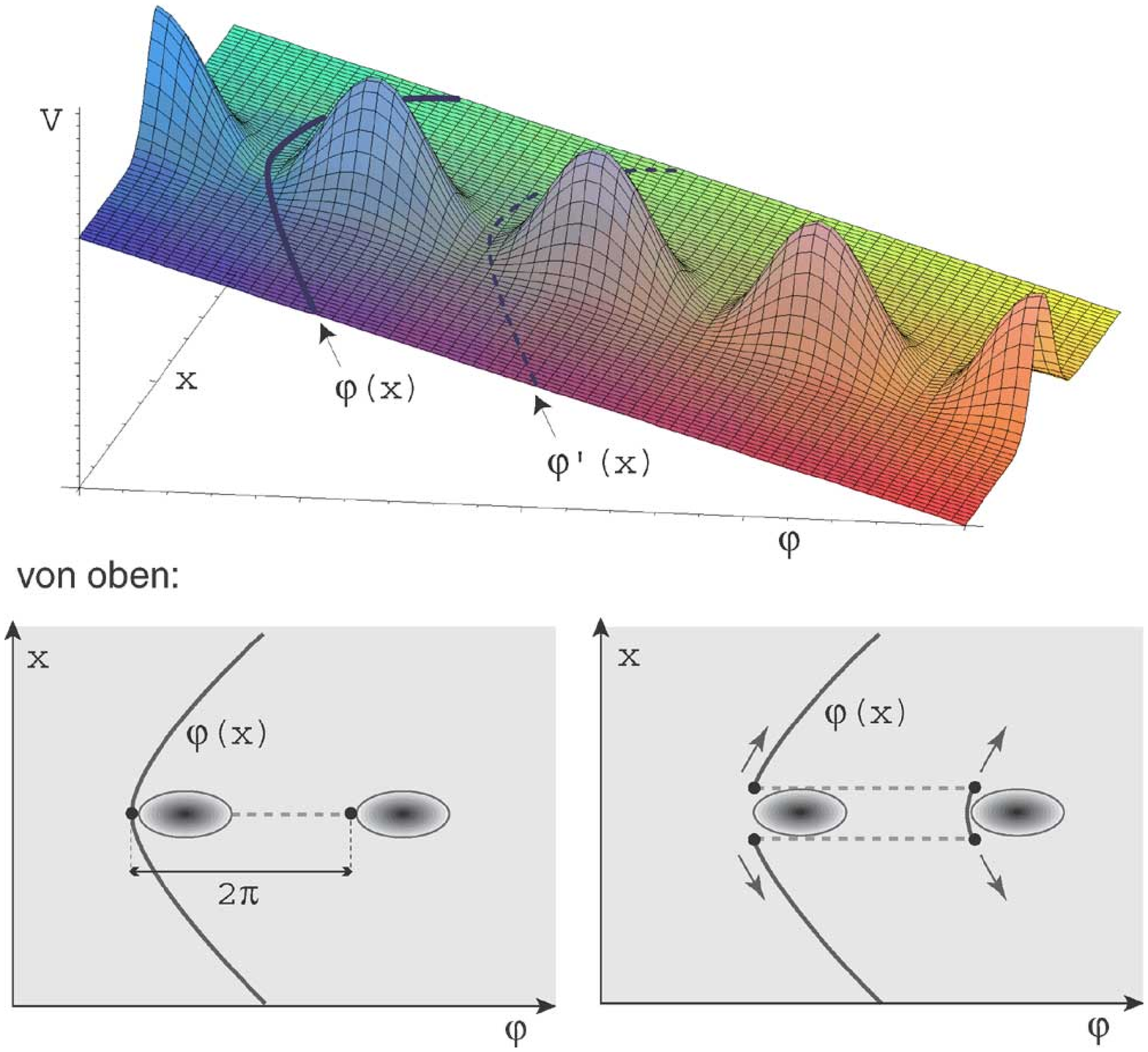}{0.75\linewidth}{\emph{Mechanismus der phase--slips.}
Im Bild oben ist eine (gepinnte) Phasenfeldkonfiguration im Potential eines Pinning-Zentrum und einem
externen Feld eingezeichnet. Links unten ist eine Aufsicht der Konfiguration mit Ausbildung eines
$\pm 2\pi$-Vortex-Paares gezeigt. Aufgrund des Energiegewinns kann das Vortex-Paar aus dem System
ged"angt werden (rechts unten).}{fig.phaseslip}

\emph{Wozu dient dieser ''Mechanismus''?} Dazu kann man sich folgendes Bild "uberlegen:
Man betrachtet ein System mit einer Verunreinigung, welche etwas um die Position $x_i$
(gau"sisch) ausgeschmiert ist, und einem an das CDW-System angelegten externen elektrischen Feld.
Die Phase ist zun"achst auf einen Wert $\varphi_i$ mit $\varphi_i|2\pi=Qx_i$ ''gepinnt''.
Nun kann die oben beschriebene Ausbildung einer Dislokation dazu f"uhren, dass das Phasenfeld
bei $x_i$ lokal ''entspannt'' wird und es sich weiterbewegen kann. In Abb \ref{fig.phaseslip}
ist dies etwas anschaulicher dargestellt.

Der Term kann auch auf eine andere Weise, wie folgt plausibel gemacht werden.
Man geht jetzt von einem $1+1$-dimensionalen (XY-)Modell
$\tilde {\cal S}_{XY}=K/2\int d\tilde\tau\,\int d\tilde x\,\left[(\partial_{\tilde \tau}\phi)^2+(\partial_{\tilde x}\phi)^2\right]$
aus und nimmt die hier nicht ber"ucksitigten \emph{Vortex}--Konfigurationen (in Raum und Zeit) mit
$\oint_{{\cal C}_j}d\phi=2\pi q_j$ hinzu ($q_j\in\mathbb{Z}$ ist die Windungszahl des Vortex und
${\cal C}_j$ ein beliebiger geschlossener Weg um den Vortex herum).
Folgt man der Rechnung von Kosterlitz \cite{CM:Kosterlitz73}, findet man, dass
diese Vortices einen zus"atzlichen Energiebeitrag, welcher einer Coulomb--WW in
zwei Dimensionen entspricht, liefert.
Man erh"alt somit ein $1+1$--dimensionales \emph{Coublomb--Gas--Modell}.
Jos{\'e}, Kadanoff, Kirkpatrick und Nelson \cite{STAT:Kadanoff77} haben gezeigt,
dass sich dieses Coulomb--Gas--Modell (mit $q=2$) f"ur tiefe Temperaturen auf ein
\emph{Sinus--Gordon--Modell} der Form (vgl. mit (4.5) in deren Arbeit)
\begin{equation}
 \int dx\,\int d\tilde \tau\,\frac{K}{2}\left[(\partial_{\tilde \tau}\tilde \phi(x,\tilde \tau))^2
+(\partial_{x}\tilde \phi(x,\tilde \tau))^2\right]+2y_0\cos(2\pi K\tilde\varphi)
\end{equation}
abbilden l"a"st. Dies kann mit $\tilde{\cal S}_{\theta}$ verglichen werden, indem man
$\tilde\theta=\frac{\hbar v}{\pi c}\theta$ und $\tilde \tau=v\tau$ setzt:
\begin{equation}
 \tilde {\cal S}_{\theta}=\int\limits_{0}^{\lambda_T}d\tilde\tau\,\int dx
 \left[\frac{1}{2\tilde g}\left((\partial_{\tilde \tau}\tilde \theta)^2
+(\partial_{x}\tilde \theta)^2\right)+
\frac{\str{\lambda}}{\tilde g}\cos(q\pi/\tilde g\tilde\theta)\right]
\end{equation}
mit $\tilde g=\hbar v/c$, $\str{\lambda}=\lambda/c$ (siehe auch n"achstes Kapitel).
Identifiziert man $K=\tilde g^{-1}$ und $\tilde\lambda/\tilde g=2y_0$ sieht man,
dass beide Modelle dieselbe Form haben (mit $q=2$).
Der Parameter $y_0$ kontrolliert die Ausbildung der Vortices und l"a"st diese f"ur kleine
Werte unwahrscheinlicher werden. D.h. f"ur unsere phase--slips,
wie schon eingangs erw"ahnt, dass diese erst
f"ur gro"se Werte von $\lambda$ wahrscheinlich sind.


\subsection{harmonisches Modell}

In diesem Abschnitt werde ich die Berechnung der Paarkorrelationsfunktion
$\tmw{ (\varphi(x,\tau)-\varphi(x^{\prime},\tau^{\prime}))^2}_0$ bzw.
$\tmw{ (\theta(x,\tau)-\theta(x^{\prime},\tau^{\prime}))^2}_0$
f"ur das harmonische Modell ($V=\lambda\equiv 0$)
\begin{equation}\label{eq.hqmodel}
 {\cal S}_0=\int_0^{\hbar\beta}d\tau\,\int_0^L dx\,
 \left[\frac{c}{2}\left((\partial_x\varphi)^2+\frac{1}{v^2}(\partial_{\tau}\varphi)^2\right)\right]
\end{equation}
(entsprechend in $\theta$) darstellen. Da die Rechnung "ahnlich zu der in \ref{app.cfelmod}
verl"auft, beschr"anke ich mich auf wenige Zwischenschritte.
Mit $\varphi(x,\tau)=\frac{1}{L(\hbar\beta)}\sum_{k,n}e^{\imath(kx+\omega_n\tau)}\varphi_{k,n}$
($\varphi_{k,n}=\int dx\,\int d\tau\,e^{-\imath(kx+\omega_n\tau)}\varphi(x,\tau)$, $k=2\pi m/L$ und
$\omega_n=2\pi n /(\hbar\beta)$ sind die \emph{Matsubara}-Frequenzen) erh"alt man
\begin{eqnarray}
  \tilde {\cal S}_0&=&\frac{1}{2T L (\hbar\beta)}\sum\limits_{k,n} J(k,n)|\varphi_{k,n}|^2 \quad\text{mit}\nn\\
  J(k,n)&=&\frac{c}{\hbar\beta}(k^2+v^{-2}\omega_n^2)\,.
\end{eqnarray}
Zus"atzlich f"uhrt man den Quellterm $\tilde {\cal S}_Q[j]\equiv -\int dx\,\int d\tau\, j(x,\tau)\varphi(x,\tau)$
ein, womit die Zustandssumme lautet:
\begin{eqnarray}
 {\cal Z}_0[j]&=&\int\pd\varphi\,e^{-\tilde {\cal S}_0-\tilde {\cal S}_Q [j]}\nn\\
   &=&\prod_{k,n}\iint da_{k,n}\,db_{k,n}\,e^{-\frac{J(k,n)}{2T L (\hbar\beta}(a_{k,n}^2+b_{k,n}^2)+
     \frac{1}{L(\hbar\beta)}(a_{k,n}\alpha_{k,n}+b_{k,n}\beta_{k,n})}
\end{eqnarray}
mit $\varphi_{k,n}=a_{k,n}+\imath b_{k,n}$ und $j_{k,n}=\alpha_{k,n}+\imath \beta_{k,n}$.
Die Korrelationsfunktion $\tmw{\varphi(0)\varphi(x)}_0$ l"a"st sich damit als Funktionalableitung
\begin{eqnarray}\label{eq.hkfphi}
 \tmw{\varphi(x,\tau)\varphi(\str{x},\str{\tau})}_0&=&\left.\frac{\delta^2}{\delta j(x,\tau)\delta j(\str{x},
    \str{\tau})}{\cal Z}_0[j]\right|_{j=0}
  =TJ^{-1}(\str{x}-x,\str{\tau}-\tau)\nn\\
 &=&\frac{T}{L(\hbar\beta)}\sum\limits_{k,n}e^{-\imath (k(\str{x}-x)+\omega_n(\str{\tau}-\tau))}J^{-1}(k,n)
\end{eqnarray}
schreiben und man erh"alt
\begin{equation}\label{eq.kfphi}
 \langle (\varphi(x,\tau)-\varphi(x^{\prime},\tau^{\prime}))^2\rangle_0=\frac{2v^2}{c L\beta}
  \sum_{n,k}\frac{1-e^{\imath\omega_n(\tau-\tau^{\prime})+\imath k(x-x^{\prime})}}{(vk)^2+\omega_n^2}
\end{equation}
oder analog f"ur $\theta$
\begin{equation}\label{eq.kfth}
 \langle (\theta(x,\tau)-\theta(x^{\prime},\tau^{\prime}))^2\rangle_0=\frac{2\pi^2c}{L(\hbar\beta)\hbar}
  \sum_{n,k}\frac{1-e^{\imath\omega_n(\tau-\tau^{\prime})+\imath k(x-x^{\prime})}}{(vk)^2+\omega_n^2}\,.
\end{equation}
F"ur weitere Zwecke formen wir (\ref{eq.kfth}) etwas weiter um:
\begin{equation}
\ldots=\frac{2\pi^2c}{L(\hbar\beta)\hbar}\frac{L}{2\pi}\int\limits_{-\Lambda}^{\Lambda}dk\,
         \left[\sum_n\frac{1}{(vk)^2+\omega_n^2}-
           e^{\imath k(x-x^{\prime})}\sum_n\frac{\cos(\omega_n(\tau-\tau^{\prime}))}{(vk)^2+\omega_n^2}\right]=\ldots
\end{equation}
mit (\ref{eq.kfsum}) und (\ref{eq.kfcsum}):
\begin{eqnarray}
\ldots&=&\frac{\pi c}{(\hbar\beta)\hbar}\int\limits_{-\Lambda}^{\Lambda}dk\,\left[\frac{\hbar\beta}{2v|k|}\left(
 \coth\big(\frac{\hbar\beta}{2}v|k|\big)-e^{\imath k(x-x^{\prime})}
 \frac{\cosh\big(\big(\frac{\hbar\beta}{2}-|\tau-\tau^{\prime}|\big)v|k|\big)}{\sinh\big(\frac{\hbar\beta}{2}v|k|\big)}\right)\right]\nn\\
&=&\frac{\pi c}{v\hbar}\int\limits_0^{\Lambda}\frac{dk}{k}\,\left(\coth\big(\frac{\lambda_T}{2}k\big)-\cos(k(x-x^{\prime}))
 \frac{\cosh\big((1-2|\tau-\tau^{\prime}|/(\hbar\beta))\frac{\lambda_T}{2}k\big)}
 {\sinh\big(\frac{\lambda_T}{2}k\big)}\right) \label{eq.thetakf}
\end{eqnarray}

%% file: RG.tex
\newpage

\section{Renormierungsgruppe}\label{sec.RG}
\setcounter{equation}{0}

\subsection{"Uberblick}
Das Ziel dieses Kapitels ist die Herleitung der Flussgleichungen f"ur die Parameter
des Modells (\ref{eq.qham}). Mit der Renormierungsgruppentransformation
(kurz: RG) werden Freiheitsgrade sukzessive aus dem System entfernt,
d.h. man bekommt effektive Parameter bzw. einen effektiven Hamiltonian
(oder Wirkung)
auf gr"o"seren L"angenskalen. Man l"a"st den Hamiltonian
durch den Hilbertraum der Hamiltonians ''flie"sen'', indem
sukzessive eine \emph{Partialspur} bzw. ein RG-Operator
auf den Hamiltonian angewendet und anschlie"send
das System wieder auf die urspr"ungliche Gr"o"se reskaliert wird.
Von besonderem Interesse sind Fixpunkte der RG, d.h.
Fixpunkte des RG-Operators, an welchen das entscheidende
Verhalten des Systems abgelesen werden kann.

Da f"ur die Renormierung unseres Modells eine anisotrope
\emph{Impulsschalen--RG} benutzt wurde, werde ich die
relevanten Schritte kurz beschreiben (siehe z.B. auch \cite{STAT:Cardy96} oder \cite{STAT:Natter00}):
\onefigure{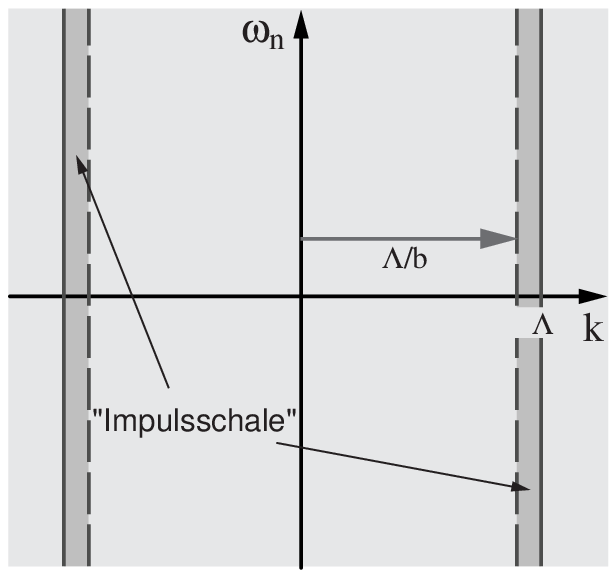}{6cm}{''Impulsschale'' in 1+1 Dimensionen}{fig.momshell}
\renewcommand{\labelenumi}{(\roman{enumi})}
\renewcommand{\labelenumii}{(\alph{enumii})}
\begin{enumerate}
\item $\tilde {\cal S}$ wird in $\str{\tilde {\cal S}}$ mittels der
   Transformation
   \begin{equation*}
     \tilde{\cal S}\rightarrow\str{\tilde {\cal S}}\equiv\hat {\cal R}[\tilde {\cal S}]
   \end{equation*}
   renormiert, wobei $\hat {\cal R}$ wie folgt als Partialspur definiert ist
   \begin{equation*}
    e^{-\str{\tilde{\cal S}}[\varphi^{<}]}\equiv\int\pd\varphi^{>}\,e^{-\tilde{\cal S}[\varphi]}\,,
    \end{equation*}
    mit der Zerlegung  $\varphi(x)=\varphi^{<}(x)+\varphi^{>}(x)$, wobei
    \begin{eqnarray}\label{eq.phisplit}
       \varphi^{<}(x,\tau)&=&((\hbar\beta)L)^{-1}\sum_n \sum\limits_{0\leq |k|\leq\Lambda/b}
           e^{\imath (kx+\omega_n\tau)}\varphi_{k,n}\nn\\
       \varphi^{>}(x,\tau)&=&((\hbar\beta)L)^{-1}\sum_n \sum\limits_{|k|>\Lambda/b}^{\Lambda}
          e^{\imath (kx+\omega_n\tau)}\varphi_{k,n}
     \end{eqnarray}
 (analog f"ur $\theta$).
    D.h. die Partialspur wird "uber Moden mit $\Lambda/b<|k|\leq\Lambda$
    gebildet; f"ur die \emph{Matsubara--Frequenzen} $\omega_n=2\pi n/(\hbar\beta)$ gibt es keinen Cutoff, d.h.
    dass das Feld $\varphi$ unabh"angig von $\omega_n$ in der Impulsschale ausintegriert wird.
     $b=e^{l}$ ist der Reskalierungsparameter mit $0<l\ll 1$.
    Der Cutoff f"ur $|k|$ ist durch $\Lambda\approx a^{-1}$
    gegeben. Die ''Impulsschale'' ist in Abb. \ref{fig.momshell} veranschaulicht.
\item Desweiteren m"ussen Reskalierungen von $x$, $\tau$ und $\varphi$
      vorgenommen werden, um die Dichte der Freiheitsgrade und der Fluktuationsst"arke
      zu erhalten. Diese sind gegeben durch
      \begin{eqnarray}
        x&\rightarrow&\str{x}=x/b\quad \text{entsprechend}\quad  k\rightarrow\str{k}=kb\nn\\
        \tau&\rightarrow&\str{\tau}=\tau/b^{z}\quad\qquad\qquad\quad\omega\rightarrow\str{\omega}=\omega b^{z}\nn\\
        \varphi&\rightarrow&\str{\varphi}(x/b)=\varphi b^{-\zeta}\,.
       \end{eqnarray}
\item Um die Flussgleichungen der Parameter in ${\cal S}$ zu erhalten, nehmen wir an,
  dass der Parameter $\gamma_j$ unter der RG (in erster Ordnung)
   zu $\str{\gamma_j}=\gamma_j(1-f(\{\gamma_i\})\ln b)$ "ubergeht, wobei $f(\{\gamma_i\})$ eine
Funktion von allen auftretenden Parametern sein kann. Ber"ucksichtigt man noch die
Reskalierung von $\gamma_j$: ${\str{\gamma_j}}_R=\gamma_j b^{\alpha}\approx\gamma_j (1+\alpha\ln b)$
erh"alt man mit $\str{\gamma_j}\equiv\gamma_j+d\gamma_j$ folgende Flussgleichung:
\begin{equation}
 \frac{d\gamma_j}{d\ln b}=(\alpha-f(\{\gamma_i\}))\gamma_j\,.
\end{equation}

\end{enumerate}

Bevor die RG f"ur unser Modell durchgef"uhrt wird, werde ich die
das System charakterisierenden dimensionslosen Parameter im n"achsten
Abschnitt einf"uhren.
Danach werden die Wirkungen ${\cal S}_{\varphi}$ und ${\cal S}_{\theta}$
renormiert.

\subsection{dimensionslose Parameter}\label{sec.param}

Da die relativistische \emph{de Broglie}--Wellenl"ange $\lambda_T$ eine wichtige L"angenskala ist,
auf der Quantenfluktuationen relevant werden, kann man diese L"ange mit der kleinsten
auftretenden L"ange $\Lambda^{-1}\approx a$ vergleichen:
\begin{equation*}
 \lambda_T\Lambda=\frac{\hbar\beta v}{c}\Lambda c=\underbrace{
\frac{c}{T\Lambda^{d-2}}}_{\tilde t^{-1}}\cdot\underbrace{\hbar\frac{v}{c}\Lambda^{d-1}}_{\tilde g}
\end{equation*}
Der so definierte Parameter  $\tilde t$ ist die reskalierte Temperatur und
$\tilde g$ kontrolliert die Quantenfluktuationen, da dieser Parameter im wesentlichen
den Koeffizienten des Impulsoperators in (\ref{eq.qham}) darstellt.

Die dimensionslosen Parameter, welche die Unordnungsst"arke und die
St"arke der Dislokationen beschreiben, sind im wesentlichen das
Verh"altnis der nicht reskalierten Parameter zur elastischen Konstante.

Zusammengefasst lauten die relevanten Parameter unseres Modells
\begin{eqnarray}
 \tilde t&=&\frac{T\Lambda^{d-2}}{c}\\
 \tilde g&=&\hbar\frac{v}{c}\Lambda^{d-1}\\
 \tilde V&=&\left(\frac{V^2}{c^2}\Lambda^{d-4}\right)^{1/2}\\
 \tilde \lambda&=&\frac{\lambda}{c \Lambda^2}
\end{eqnarray}

\subsection{Replica-Wirkung / Unordnungsmittel}\label{sec.replica}
Um den Unordnungsterm zu renormieren, geht man im allgemeinen
zum \emph{Replica--Hamiltonian} bzw. hier zur \emph{Replica--Wirkung}
"uber, d.h., dass der Unordnungsmittelwert ausgef"uhrt wird.
Der \emph{Replica--''Trick''} besteht in folgender Darstellung der freien
Energie
\begin{equation}
 \umw{{\cal F}}=-T\overline{\ln{\cal Z}}=-T\overline{\lim_{n\rightarrow 0}
\frac{1}{n}({\cal Z}^n-1)}\,.
\end{equation}
und der Vertauschung der zwei Grenzwertprozesse (Unordnungsmittelung
und Limes $n\rightarrow 0$)\footnote{which is ''believed to create no trouble''},
um die Mittelung des Logarithmus zu vermeiden:
\begin{equation}
 \overline{\ln{\cal Z}}=\lim_{n\rightarrow 0} \frac{1}{n}(\overline{{\cal Z}^n}-1)\,.
\end{equation}
Die \emph{Replica--Wirkung} ${\cal S}^n[\{\varphi^{\gamma}\}_{\gamma=1..n}]$
 wird definiert durch
\begin{eqnarray}
 \overline{{\cal Z}^n}=\overline{\prod\limits_{\gamma=1}^{n}\int{\cal D}[\varphi^{\gamma}]
e^{-\tilde{\cal S}[\varphi^{\gamma}]}}
 &=&\int\left(\prod\limits_{\gamma=1}^{n}{\cal D}[\varphi^{\gamma}]\right)
\overline{e^{-\sum\limits_{\alpha=1}^{n}\tilde{\cal S}[\varphi^{\alpha}]}}\nn\\
 &\equiv &\int\left(\prod\limits_{\gamma=1}^{n}{\cal D}[\varphi^{\gamma}]\right)
e^{-\tilde{\cal S}^n[\{\varphi^{\gamma}\}_{\gamma=1..n}]}\,.
\end{eqnarray}
${\cal S}^n$ ist also eine Wirkung in der jeder Freiheitsgrad $n$-fach repliziert ist
($\alpha$, $\gamma$ und $\nu$ sind die \emph{Replica--Indices}).
Bei einigen Systemen kann die Vertauschung von Unordnungsmittelung und $n\rightarrow 0$
zu Problemen f"uhren, in unserem Fall wird dies allerdings nicht angenommen.

F"ur die Wirkung (\ref{eq.actphi}) l"a"st sich die Replica--Version mit
$\umw{e^{g(\vec{r})}}=e^{\umw{g^2(\vec{r})}/2}$
f"ur gau"sverteiltes $\vec{r}$ leicht berechnen:
\begin{eqnarray}
 e^{-\tilde {\cal S}^n}&=&e^{-\sum_{\alpha} \tilde {\cal S}_0[\varphi^{\alpha}]}
e^{\frac{1}{2}\sum_{\alpha,\nu}\umw{ \tilde {\cal S}_{\textit{imp}}[\varphi^{\alpha},\{x,\tau\}]
\tilde {\cal S}_{\textit{imp}}[\varphi^{\nu},\{\str{x},\str{\tau}\}]}}\nn\\
 &=&e^{-\sum_{\alpha}( \tilde {\cal S}_0[\varphi^{\alpha}]-\frac{V^2}{2\hbar^2}\sum_{\nu}
 \int dx\,d\str{x}\,\int d\tau\,d\str{\tau}\, \umw{\cos(\varphi^{\alpha}+\alpha(x))
\cos(\str{{\varphi^{\nu}}}+\alpha(\str{x}))})}
\end{eqnarray}
wobei ${\cal S}_0$ der gau"ssche Anteil von ${\cal S}_{\varphi}$ (\ref{eq.hqmodel}) ist und
${\cal S}_{\textit{imp}}$ der Pinning-Anteil.

Mit (\ref{eq.phaseMW}) kann der Unordnungsmittelwert leicht berechnet werden
\begin{equation}
 \umw{\cos(\varphi^{\alpha}(x,\tau)+\alpha(x))\cos(\varphi^{\nu}(\str{x},\str{\tau})+\alpha(\str{x}))}=
\cos(\varphi^{\alpha}(x,\tau)-\varphi^{\nu}(\str{x},\str{\tau}))\delta(x-\str{x})\,.
\end{equation}
Die Replica--Wirkung zu (\ref{eq.actphi}) lautet also
\begin{eqnarray}\label{eq.ractphi}
 \tilde{\cal S}^n[\{\varphi^{\gamma}\}]=\sum\limits_{\alpha,\nu=1}^{n}\int_0^{\hbar\beta}d\tau\,\int_0^{L}dx\,
 \Bigg[ &\frac{c \delta_{\alpha,\nu}}{2\hbar}\left((\partial_x\varphi^{\alpha}(x,\tau))^2+
 v^{-2}(\partial_{\tau}\varphi^{\alpha}(x,\tau))^2\right)&\nonumber\\
 &+\frac{V^2}{2\hbar^2}\int_0^{\hbar \beta}d\str{\tau}\,
 \cos(p(\varphi^{\alpha}(x,\tau)-\varphi^{\nu}(x,\str{\tau})))\Bigg]&
\end{eqnarray}

\subsection{RG des Unordnungsterms}\label{sec.RGV}

In diesem Abschnitt werde ich die RG der Wirkung $\tilde {\cal S}_{\varphi}$
bzw. der Replica--Version (\ref{eq.ractphi})
etwas detailierter, insbesondere da dies bei endlichen Temperaturen
geschieht, und allgemeiner in $d+1$--Dimensionen ausf"uhren.

Zun"achst wird das Funktional
\begin{equation}
 R[f(\vec{r})]\equiv \frac{V^2}{2\hbar^2}\cos(p f(\vec{r}))
\end{equation}
definiert, womit der Unordnungsanteil der Replica--Wirkung (\ref{eq.ractphi})
als
\begin{equation}
 \tilde {\cal S}_{R}^{n}=-\iint\limits_0^{\hbar\beta} d\tau\,d\str{\tau}
\int d^d\vec{x}\sum_{\alpha,\nu} R[\varphi^{\alpha}(\vec{x},\tau)-\varphi^{\nu}(\vec{x},\str{\tau})]
\end{equation}
geschrieben werden kann.

Au"serdem wird folgende zeitlich nichtlokale Korrelationsfunktion ben"otigt (siehe (\ref{eq.hkfphi}))
\begin{equation}\label{eq.kfeqx}
 \tmw{\varphi(\vec{x},\tau)\varphi(\vec{x},\str{\tau})}=\frac{v^2}{L^d\beta c}\sum_{\vec{k},\omega_n}
 \frac{e^{\imath\omega_n(\tau-\str{\tau})}}{(v\vec{k})^2+\omega_n^2}\,.
\end{equation}
Die renormierte Wirkung wird durch
\begin{equation}
 e^{-\tilde{\cal S}^{\alpha,\nu}_{<}[\{\varphi_{<}^{\alpha}\}]}=\int\prod_{\gamma}\left(\pd\varphi^{\gamma}_{>}\right)\,
e^{-\tilde {\cal S}^{n}[\{\varphi_{<}^{\alpha},\varphi_{>}^{\alpha}\}]}
\end{equation}
definiert, wobei das Feld $\varphi^{\gamma}$ in einen langwelligen ($\varphi^{\gamma}_{<}$)
und kurzwelligen ($\varphi^{\gamma}_{>}$) Anteil, wie in (\ref{eq.phisplit}) angegeben,
aufgespalten wurde.

Mit
\begin{eqnarray}
  \tilde{\cal S}^n[\{\varphi^{\gamma}\}]&\equiv& \tilde{\cal S}^{n}_{0,<}[\{\varphi^{\gamma}_{<}\}]+
\tilde{\cal S}^{n}_{0,>}[\{\varphi^{\gamma}_{>}\}]+
\tilde{\cal S}^{n}_R[\{\varphi^{\gamma}_{\lessgtr}\}]\\
 \tilde{\cal S}^{n}_{0,\lessgtr}[\{\varphi^{\gamma}_{\lessgtr}\}]&=&
        \sum\limits_{\alpha=1}^{n}\int_0^{\hbar\beta}d\tau\,\int_{L^d}d^d\vec{x}\,
          \frac{c}{2\hbar}\left[(\nabla_{\vec{x}}\varphi^{\alpha}_{\lessgtr}(\vec{x},\tau))^2+
                                v^{-2}(\partial_{\tau}\varphi^{\alpha}_{\lessgtr}(\vec{x},\tau))^2\right]
\end{eqnarray}
erh"alt man so f"ur $\tilde{\cal S}^{n}_{<}[\{\varphi_{<}^{\gamma}\}]$
\begin{equation}
 \tilde{\cal S}^{n}_{<}[\{\varphi_{<}^{\gamma}\}]=
 \tilde{\cal S}^{n}_{0,<}[\{\varphi^{\gamma}_{<}\}]+\ln {\cal Z}^{n}_{0,<}+
 \kmw{e^{-\tilde {\cal S}_R^{n}[\{\varphi^{\gamma}_{\lessgtr}\}]}-1}_{0,>}\,,
\end{equation}
wobei $\kmw{O[\{\varphi^{\gamma}_{\lessgtr}\}]}_{0,>}=\ln\tmw{O[\{\varphi^{\gamma}_{\lessgtr}\}]}_{0,>}$
der Kumulantenmittelwert des Funktionals $O$ bzgl. $\tilde{\cal S}_{0,>}$ ist.
Diese Kumulante wird in eine St"orungsreihe entwickelt,
d.h. die Unordnung wird als kleine St"orung betrachtet,
und deshalb f"ur kleine $\epsilon=\frac{V^2}{2\hbar^2}$ entwickelt:
\begin{equation}\label{eq.Vexpansion}
 \kmw{e^{-\tilde {\cal S}_R^{n}[\{\varphi^{\gamma}_{\lessgtr}\}]}-1}_{0,>}\equiv
\sum\limits_{k=1}^{\infty} A_k\epsilon^k
\end{equation}
mit $A_1=\kmw{\sum\limits_{\alpha,\nu=1}^n\int d\tau\,d\str{\tau}\,\int d^d\vec{x}\,\cos(p(\varphi^{\alpha}-\str{{\varphi^{\nu}}}))}_{0,>}$.
Da wir die Flussgleichungen f"ur kleine Unordnungsst"arken herleiten wollen,
beschr"anken wir uns auf die niedrigste Ordnung der St"orungsreihe.
Die renormierte Wirkung ist in 1. Ordnung in $\epsilon$ gegeben durch:
\begin{equation}
 \tilde{\cal S}^{n}_{<}[\{\varphi_{<}^{\gamma}\}]\doteq\tilde{\cal S}^n_{0,<}-
 \underbrace{\int d\tau\,d\str{\tau}\,\int d^d\vec{x}\,
  \kmw{R[\varphi^{\alpha}-\str{{\varphi^{\nu}}}]}_{0,>}}_{-\tilde {\cal S}_{1,<}}\,.
\end{equation}
Um fortzufahren, definiere ich als Abk"urzung: $\varphi^{\alpha,\nu}\equiv\varphi^{\alpha,\nu}_{<}+\varphi^{\alpha,\nu}_{>}$ mit
\begin{equation}
 \varphi^{\alpha,\nu}_{\lessgtr}=\varphi^{\alpha}_{\lessgtr}(\vec{x},\tau)-{\varphi^{\nu}_{\lessgtr}}(\vec{x},\str{\tau})
\end{equation}
Damit wird $R[\varphi^{\alpha,\nu}]$ um $\varphi^{\alpha,\nu}_{<}$ in kleinen
$\varphi^{\alpha,\nu}_{>}$ entwickelt:
\begin{equation}
 R[\varphi^{\alpha,\nu}]=R[\varphi^{\alpha,\nu}_{<}]+R^{\prime}[\varphi^{\alpha,\nu}_{<}]\varphi^{\alpha,\nu}_{>}+
 \frac{1}{2}R^{\prime\prime}[\varphi^{\alpha,\nu}_{<}](\varphi^{\alpha,\nu}_{>})^2+{\cal O}((\varphi^{\alpha,\nu}_{>})^3)
\end{equation}
Aus der Definition von $R$ folgt sofort: $R^{(m)}[-f]=(-1)^mR^{(m)}[f]$
und nach einer Umbennenung der Replicaindices, kann man schreiben:
\begin{equation}
  R[\varphi^{\alpha,\nu}]=R[\varphi^{\alpha,\nu}_{<}]+2R^{\prime}[\varphi^{\alpha,\nu}_{<}]\varphi^{\alpha}_{>}
  +R^{\prime\prime}[\varphi^{\alpha,\nu}_{<}]((\varphi^{\alpha}_{>})^2-\varphi^{\alpha}_{>}\varphi^{\nu}_{>})+{\cal O}((\varphi^{\alpha,\nu}_{>})^3)
\end{equation}

Und f"ur $\kmw{R[\varphi^{\alpha,\nu}]}_{0,>}$ folgt
\begin{eqnarray}
  \kmw{R[\varphi^{\alpha,\nu}]}_{0,>}&=&R[\varphi^{\alpha,\nu}_{<}]+2R^{\prime}[\varphi^{\alpha,\nu}_{<}]
 \underbrace{\tmw{\varphi^{\alpha}_{>}}_{0,>}}_{=0}
  +R^{\prime\prime}[\varphi^{\alpha,\nu}_{<}]\tmw{(\varphi^{\alpha}_{>})^2-\varphi^{\alpha}_{>}\varphi^{\nu}_{>}}_{0,>}\nn\\
  &&+{\cal O}\left(\kmw{(\varphi^{\alpha,\nu}_{>})^3}_{0,>}\right)
\end{eqnarray}
(Zur Berechnung der Kumulante im quadratischen Teil wurde das Wick-Theorem benutzt.)
Obiger Ausdruck bedeutet, dass $\tilde{\cal S}^n_{1,<}$ au"ser $\tilde{\cal S}^n_{R,<}$
noch Korrekturen zu $\tilde V$ und, wie wir gleich sehen werden, auch zu
$\tilde g$ liefert.
Im folgenden wird nur noch der interessante Anteil
\begin{equation}\label{eq.intpart}
 R^{\prime\prime}[\varphi^{\alpha,\nu}_{<}]\tmw{(\varphi^{\alpha}_{>})^2-\varphi^{\alpha}_{>}\varphi^{\nu}_{>}}_{0,>}
\end{equation}
untersucht.
Betrachten wir zun"achst den Teilausdruck $R^{\prime\prime}[\varphi^{\alpha,\nu}_{<}]\tmw{(\varphi^{\alpha}_{>})^2}_{0,>}$.
Es ist $R^{\prime\prime}[\varphi^{\alpha,\nu}_{<}]=-p^2R[\varphi^{\alpha,\nu}_{<}]$ und
\begin{equation}
 \tmw{{\varphi^{\alpha}_{>}}^2}_{0,>}=\frac{v^2}{L^d\beta c}{\sum\limits_{\vec{k},n}}^{>}\frac{1}{(v\vec{k})^2+\omega_n^2}
 =\frac{v^2}{\beta c}\int^{>}\frac{d^d\vec{k}}{(2\pi)^d}\sum\limits_n\frac{1}{(v\vec{k})^2+\omega_n^2}
\end{equation}
wobei das ${}^{>}$ andeuten soll, dass die Summe bzw. das Integral
nur "uber Impulse mit $\Lambda/b<|\vec{k}|\leq \Lambda$ l"auft.
Die Frequenzsumme l"a"st sich mit (\ref{eq.kfsum}) ausrechnen
\footnote{da wir bei endlichen Temperaturen sind, darf hier nicht zum Integral $\int d\omega$ "ubergegangen werden!}
\begin{eqnarray}
 \ldots&=&\frac{v^2}{\beta c}\left(\frac{\hbar\beta}{2\pi}\right)^2\int^{>}\frac{d^d\vec{k}}{(2\pi)^d}
 \frac{\pi}{|a|}\coth(\pi|a|)\quad ,\quad a=v|k|\frac{\hbar\beta}{2\pi}\nn\\
 &=&\frac{v\hbar}{c}\frac{S_d}{(2\pi)^d}\int\limits_{\lambda/b}^{\Lambda}\frac{k^{d-1}}{k}
 \coth\left(\frac{\lambda_T}{2}k\right)dk\nn\\
 &=&\frac{v\hbar}{c}K_d\Lambda^{d-1}\coth(\lambda_T\Lambda/2)\ln(b)\,.
\end{eqnarray}
Wir erhalten folgende Korrektur des Unordnungsparameters $\epsilon$:
\begin{equation}
 \boxed{\epsilon\rightarrow\str{\epsilon}=\epsilon\left(1-p^2\tilde g K_d
\coth(\tilde g/(2\tilde t))\ln b\right)}\,.
\end{equation}
Nun kommen wir zum zweiten Term von (\ref{eq.intpart}), welcher eine Korrektur zum Parameter $\tilde g$ liefert.
Dazu wird eine Gradientenentwicklung von
$R^{\prime\prime}[\varphi^{\alpha,\nu}_{<}]$ durchgef"uhrt, wozu man zu neuen
Koordinaten (''Relativ- und Schwerpunktskoordinaten in $\tau$-Richtung'')
\begin{eqnarray}
 \tau+\str{\tau}&\equiv& 2u\quad , \quad u\in[0,\hbar\beta]\nn\\
 \tau-\str{\tau}&\equiv& 2w\quad , \quad w\in[-\hbar\beta/2,\hbar\beta/2]\nn
\end{eqnarray}
"ubergeht und $\varphi^{\alpha,\alpha}_{<}$ um $u$ in $w$ entwickelt:
\begin{equation}\label{eq.Entphi}
 \varphi^{\alpha,\alpha}_{<}=\varphi^{\alpha}_{<}(\vec{x},u+w)-\varphi^{\alpha}_{<}(\vec{x},u-w)=
   0+2\frac{\partial\varphi^{\alpha}_{<}}{\partial u}(\vec{x},u)w+{\cal O}(w^2)
\end{equation}
Die Replicaindices sind hier gleich, da nach (\ref{eq.kfeqx}) ein
Kronecker--Delta in der Korrelationsfunktion $\tmw{\varphi^{\alpha}_{>}\varphi^{\nu}_{>}}_{0,>}$
auftritt.
Mit (\ref{eq.Entphi}) wird $R^{\prime\prime}$ um $0$ entwickelt:
\begin{equation}\label{eq.EntR}
 R^{\prime\prime}[\varphi^{\alpha\alpha}_{<}]=
  R^{\prime\prime}[0]+\frac{1}{2}R^{\prime\prime\prime\prime}[0]
  \left(2\frac{\partial\varphi^{\alpha}_{<}}{\partial u}(\vec{x},u)w\right)^2
  +{\cal O}\left[w^3,\left(2\frac{\partial\varphi^{\alpha}_{<}}{\partial u}(\vec{x},u)w\right)^3\right]
\end{equation}
{\bf Bemerkung: } W"are $\varphi^{\alpha\alpha}_{<}$ eine zeitlich lokale Funktion,
w"urde $R^{\prime\prime}[\varphi^{\alpha\alpha}_{<}]=R^{\prime\prime}[0]=\textit{const.}$
von den langwelligen Anteilen des Feldes unabh"angig sein, d.h. man w"urde nur eine
unbedeutende Konstante als Korrektur bekommen, die auch in (\ref{eq.EntR}) als
Term 0. Ordnung auftaucht, welche ich im weiteren nicht mehr beachte.$\diamond$

Der volle Korrekturterm lautet (bei der Integral-Transformation zu $u$ und $w$
wurde die Jacobi-Determinante von $1/4$ ber"ucksichtigt):
\begin{equation}
 \frac{1}{2}\int\limits_{0}^{\hbar\beta}du\,\int\limits_{-\hbar\beta/2}^{\hbar\beta/2}dw\,\int d^d\vec{x}\,
  R^{\prime\prime\prime\prime}[0]\left(\frac{\partial\varphi^{\alpha}_{<}}{\partial u}(\vec{x},u)w\right)^2
  \frac{v^2}{L^d\beta c}{\sum\limits_{\vec{k},n}}^{>}\frac{e^{\imath\omega_n 2w}}{(v\vec{k})^2+\omega_n^2}
\end{equation}

Die Frequenzsumme kann nun mit (\ref{eq.kfcsum}) und anschlie"send das $w$-Integral
mit (\ref{eq.x2cosh}) berechnet werden, wobei insbesondere die Bedingung in (\ref{eq.x2cosh})
beachtet werden mu"s. Zu bermerken ist, dass der Wert der Frequenzsumme mit
$w$ auf dem Integrationsintervall exponetiell abf"allt und somit die Entwicklung
(\ref{eq.Entphi}) bzw. (\ref{eq.EntR}) rechtfertigt.
Letzendlich erh"alt man mit $R^{\prime\prime\prime\prime}[0]=\frac{V^2}{2\hbar^2}p^4$ ($u\rightarrow\tau$)

\begin{equation}
 \int\limits_{0}^{\hbar\beta}d\tau\,\int d^d\vec{x}\,(\partial_{\tau}\varphi^{\alpha}_{<})^2
 \left(\frac{p^4}{2}K_d \frac{V^2}{\hbar v^2 c}\int\limits_{\Lambda/b}^{\Lambda} dk\,\frac{k^{d-1}}{k^4}
 \left[1-\frac{\lambda_T k/2}{\sinh(\lambda_T k/2}\right]\right)\,.
\end{equation}

Der (Massen--)Koeffizient des kinetischen Terms wird in erster Ordnung wie folgt durch den
Unordnungsterm renormiert:
\begin{equation}
 \boxed{\left(\frac{c}{2\hbar v^2}\right)\rightarrow\left(\frac{c}{2\hbar v^2}\right)^{\prime}=
 \left(\frac{c}{2\hbar v^2}\right)\left(1+p^4K_d\tilde V^2\left[1-\frac{\tilde g/(2\tilde t)}{\sinh(\tilde g/(2\tilde t)}\right]\ln b\right)}
\end{equation}

Zu bemerken ist, dass es zur elastischen Konstante in dieser Ordnung keine
RG-Korrekturen durch die Unordnung gibt. Diese w"urde man erst mit $A_2$ bekommen.
Die entg"ultigen Flussgleichungen werden in Abschnitt \ref{sec.flussgl} bestimmt.

\subsection{RG der ''phase--slips''}\label{sec.RGl}

In diesem Abschnitt werden die Renormierungskorrekturen zu $\tilde g$, $\tilde t$ und $\tilde \lambda$
durch den phase--slip Term hergeleitet. Gleichzeitig wird auch die RG eines
$1+1$--dimensionalen \emph{Sinus--Gordon}--Modells bei endlichen Temperaturen behandelt,
welches auch durch $\tilde{\cal S}_{\theta}$ beschrieben wird. In diesem Abschnitt
wird der Index ${}_{\theta}$ weggelassen.

Analog zu (\ref{eq.Vexpansion}) wird auch in diesem Fall der Kumulantenmittelwert
des phase--slip--Terms in eine St"orreihe in $\lambda/\hbar$ entwickelt:
\begin{equation}
 \kmw{e^{-\tilde{\cal S}_{\lambda}[\{\theta_{\lessgtr}\}]}-1}_{0,>}\equiv\sum\limits_{k=1}^{\infty}B_k\left(\frac{\lambda}{\hbar}\right)^k
\end{equation}
wobei
\begin{equation}
 \tilde{\cal S}_{\lambda}[\{\theta_{\lessgtr}\}]=\frac{\lambda}{\hbar}\int d\tau\,\int dx\,\cos(q\theta)\,.
\end{equation}
In erster Ordnung ($B_1$) erh"alt man eine einfache Korrektur zu $\tilde \lambda$ und die zweite
Ordnung ($B_2$) liefert Korrekturen zu $\tilde g$ und $\tilde t$.

Beginnen wir zun"achst mit
\begin{equation}
 B_1=\kmw{\int d\tau\,\int dx\,\cos(q\theta)}_{0,>}\,,
\end{equation}
und setzen wieder $\theta=\theta_{<}+\theta_{>}$. Mit (\ref{eq.sincosMW})
erh"alt man sofort:
\begin{eqnarray}
 \kmw{\cos(q\theta)}_{0,>}&=&\cos(q\theta_{<})\kmw{\cos(q\theta_{>})}_{0,>}+\sin(q\theta_{<})\kmw{\sin(q\theta_{>})}_{0,>}\nn\\
 &=&\cos(q\theta_{<})e^{-\frac{q^2}{2}\tmw{\theta_{>}}_{0,>}}\,,
\end{eqnarray}
und mit (\ref{eq.thetakf}):
\begin{equation}
 \tmw{\theta_{>}}_{0,>}=\frac{\pi c}{2\hbar v}\int_{\Lambda/b}^{\Lambda}\frac{dk}{k}\,\coth(\lambda_T k/2)
= \frac{\pi c}{2\hbar v}\coth(\lambda_T\Lambda/2)\ln b\,.
\end{equation}
Der Parameter $\lambda$ wird in erster Ordnung durch
\begin{equation}
 \boxed{\lambda\rightarrow\str{\lambda}=\lambda\left(1-\frac{q^2 \pi}{4\tilde g}\coth(\tilde g/(2\tilde t))\ln b\right)}
\end{equation}
renormiert.

F"ur die Betr"age aus der zweiten Ordnung ist die Rechnung etwas aufwendiger,
da hier die einfache Gradientenentwickung, wie sie im vorigen Abschnitt
benutzt wurde, nur divergente Beitr"age liefern w"urde.
Der Koeffizient $B_2$ ist gegeben durch
\begin{equation}
 B_2=\frac{1}{2}\int d\tau\,d\str{\tau}\,\int dx\,d\str{x}\,\kmw{\cos(q\theta(x,\tau))\cos(q\theta(\str{x},\str{\tau}))}_{0,>}\,.
\end{equation}
Mit Hilfe des Wick'schen Theorems und der Entwicklung aller $\cos$--Terme in $\theta_{>}$ vereinfacht sich
dieser Ausdruck zu
\begin{eqnarray}\label{eq.kumb2}
 \kmw{\cos(q\theta)\cos(q\str{\theta})}_{0,>}&=&\tmw{\cos(q\theta)\cos(q\str{\theta})}_{0,>}
 -\tmw{\cos(q\theta)}_{0,>}\tmw{\cos(q\str{\theta})}_{0,>}\nn \\
 &\approx&q^2\sin(q\theta_{<})\sin(q\str{\theta}_{<})\tmw{\theta_{>}\str{\theta}_{>}}_{0,>}\nn\\
 &=&\frac{q^2}{2}(\cos(q(\theta_{<}-\str{\theta}_{<}))-\cos(q(\theta_{<}+\str{\theta}_{<})))\tmw{\theta_{>}\str{\theta}_{>}}_{0,>}\,,
\end{eqnarray}
mit $\str{\theta}\equiv\theta(\str{x},\str{\tau})$.
Der Term  $\cos(q(\theta_{<}+\str{\theta}_{<}))$ wird im weiteren
nicht mehr ber"ucksichtigt, da dieser im Fall $x=\str{x}$ und
$\tau=\str{\tau}$, die zweite Harmonische zur ''Grundschwingung''
wird und daher unter der RG weniger relevante Korrekturen liefert
als der erste Term in (\ref{eq.kumb2}) \cite{STAT:Kadanoff77,CDW:Knops80}.

Die Korrelationsfunktion in (\ref{eq.kumb2}) ist mit (\ref{eq.thetakf}) wiederum einfach zu erhalten:
\begin{equation}
 \tmw{\theta_{>}\str{\theta}_{>}}_{0,>}=\frac{\pi c}{2v\hbar}\cos(\Lambda(x-\str{x}))
 \frac{\cosh\left((1-2|\tau-\str{\tau}|/(\hbar\beta))\frac{\lambda_T\Lambda}{2}\right)}{\sinh(\lambda_T\Lambda/2)}\ln b\,.
\end{equation}

\trenner

Nun betrachten wir den Term $\cos(q(\theta_{<}-\str{\theta}_{<}))$.
Wie schon erw"ahnt liefert die einfache, naive Entwicklung dieses Cosinus--Terms
bis zur zweiten Ordnung in $\Delta \theta_{<}\equiv \theta_{<}-\str{\theta}_{<}$
divergente Korrekturterme (siehe z.B. Diskussion in \cite[Kap. 4]{CM:Godreche92}
von P. Nozi{\`e}res).

Um diesen Term besser zu erfassen, m"ussen wir eine (einfache) \emph{Operatorproduktentwicklung}
durchf"uhren, welche alle Ordnungen der Cosinus--Reihe mit ber"ucksichtigt.
Dazu wird aus jedem Term der Form $(\Delta \theta_{<})^{2n}$ f"ur $n>1$ der
relevante marginale Anteil $(\Delta \theta_{<})^2$ extrahiert.
Dies erreicht man, indem man zun"achst folgenden Mittelwert betrachtet:
\begin{equation}
 \tmw{\cos(q\Delta \theta_{<})}_{0,<}=1+\tmw{(q\Delta \theta_{<})^2\sum\limits_{n=1}^{\infty}\frac{(-1)^{n}}{(2n)!}(q\Delta \theta_{<})^{2n-2}}_{0,<}
\end{equation}
Jeden Summanden in dieser Darstellung kann man nun teilweise auswerten:
\begin{equation}
 \tmw{(q\Delta \theta_{<})^2\frac{(-1)^{n}}{(2n)!}(q\Delta \theta_{<})^{2n-2}}_{0,<}=
\binom{2n}{2}\frac{(-1)^{n}}{(2n)!}\tmw{(q\Delta \theta_{<})^2}_{0,<}\tmw{(q\Delta \theta_{<})^{2n-2}}_{0,<}\,.
\end{equation}
Der zu untersuchende Term kann mit dieser "Uberlegung geschrieben werden als
\begin{equation}
 \cos(q\Delta \theta_{<})\approx 1+(q\Delta \theta_{<})^2\left(\sum\limits_{n=1}^{\infty}
 \binom{2n}{2}\frac{(-1)^{n}}{(2n)!}\tmw{(q\Delta \theta_{<})^{2n-2}}_{0,<}\right)\,,
\end{equation}
d.h. f"ur Terme h"oherer Ordnung wird jeweils ein Teilmittelwert ausgef"uhrt,
um den marginalen Anteil zu erhalten. Mit Hilfe des Wick--Theorems kann man
\begin{equation}
 \tmw{(q\Delta \theta_{<})^{2n-2}}_{0,<}=\frac{(2n-2)!}{2^{n-1}(n-1)!}\tmw{(q\Delta \theta_{<})^2}_{0,<}^{n-1}
\end{equation}
ausrechnen \cite{STAT:Kadanoff79,CDW:Knops80} und bekommt nun als Entwicklung des Cosinus--Terms:
\begin{equation}
 \cos(q\Delta \theta_{<})=1-\frac{1}{2}(q\Delta \theta_{<})^2e^{-\frac{q^2}{2}\tmw{(\Delta \theta_{<})^2}_{0,<}}\,.
\end{equation}

\trenner

Da Korrekturen zu $\tilde t$ und $\tilde g$ berechnet werden sollen, wird,
"ahnlich dem Vorgehen im vorigen Abschnitt, $(q\Delta \theta_{<})^2$
in neue Koordinanten
\begin{eqnarray}
 2s&\equiv &x+\str{x}\quad ,\quad 2r \equiv x-\str{x}\nn\\
 2u&\equiv &\tau+\str{\tau}\quad ,\quad 2w \equiv \tau-\str{\tau}\nn
\end{eqnarray}
umgeschrieben (Jacobideterminante $(1/4)^2$) und in den ''Relativ''--Koordinaten
$r$ und $w$ entwickelt. Als relevante Terme bleiben "ubrig
\begin{equation}
 (q\Delta \theta_{<})^2=q^2\left[\left(2\frac{\partial \theta_{<}(s,u)}{\partial s}r\right)^2+
 \left(2\frac{\partial \theta_{<}(s,u)}{\partial u}w\right)^2+\ldots\right]\,.
\end{equation}

Insgesamt erh"alt man nach einigen Umformungen und den Umbennungen $u\rightarrow\tau$ und $s\rightarrow x$
den Korrekturterm der RG zu $\tilde{\cal S}_0[\theta_{<}]$:
\begin{eqnarray}\label{eq.rg_l2}
 &&-\frac{\pi c}{2 v\hbar}\left(\frac{\lambda}{\hbar}\right)^2\frac{q^4}{\sinh(\lambda_T\Lambda/2)}
 \int_0^{\hbar\beta}d\tau\,\int_0^L dx\,\int_0^{\hbar\beta/4}dw\,\int_0^{\infty}dr\,\nn\\
 &&\times \left[\left(r\frac{\partial\theta_{<}(x,\tau)}{\partial x}\right)^2+
       \left(w\frac{\partial\theta_{<}(x,\tau)}{\partial \tau}\right)^2\right]
       e^{-\frac{q^2}{2}\tmw{(\theta_{<}(x+r,\tau+w)-\theta_{<}(x-r,\tau-w))^2}_{0,<}}\nn\\
 &&\times\cos(2\Lambda r)\cosh\left(\left(1-\frac{4w}{\hbar\beta}\right)\lambda_T\Lambda/2\right)\ln b
\end{eqnarray}
Die Korrelationsfunktion wurde schon in (\ref{eq.thetakf}) berechnet und lautet f"ur $T>0$ mit
der Substitution $\tilde k=\lambda_T k/2$ und $\tilde r=4 r/\lambda_T$, $\tilde w=4w/(\hbar\beta)$:
\begin{eqnarray}
 &&\tmw{(\theta_{<}(x+r,\tau+w)-\theta_{<}(x-r,\tau-w))^2}_{0,<}\nn\\
 &&=\frac{\pi}{\tilde g}\int\limits_0^{\tilde g/(2\tilde t)}\frac{d\tilde k}{\tilde k}
 \left(\coth(\tilde k)-\cos(\tilde r\tilde k)\frac{\cosh((1-\tilde w)\tilde k)}{\sinh(\tilde k)}\right)
 \equiv \frac{\pi}{\tilde g}f_c(\tilde g/(2\tilde t);\tilde r,\tilde w)
\end{eqnarray}
Da die Integrale "uber $r$ und $w$ nicht weiter berechnet werden k"onnen, definiere ich
folgende Funktionen:
\begin{eqnarray}
 f_x(y,z)&\equiv&\frac{1}{64}\int_0^{\infty}dr\int_0^1dw r^2 e^{-\frac{q^2\pi}{2 y}f_c(z;r,w)}\cos(zr)\cosh((1-|w|)z)\nonumber\\
 f_{\tau(y,z)}&\equiv&\frac{1}{64}\int_0^{\infty}dr\int_0^1dw w^2 e^{-\frac{q^2\pi}{2y}f_c(z;r,w)}\cos(zr)\cosh((1-|w|)z)\nonumber
\end{eqnarray}
Damit lassen sich die renormierte elastische Konstante und Masse in $\tilde{\cal S}$
in kompakter Form schreiben als:
\begin{equation}
 \boxed{\left(\frac{\hbar v^2}{2\pi^2 c}\right)\rightarrow\str{\left(\frac{\hbar v^2}{2\pi^2 c}\right)}=
  \left(\frac{\hbar v^2}{2\pi^2 c}\right)\left(1-\frac{\pi^3}{4}\frac{\tilde \lambda^2}{\tilde t^4}
  \frac{q^4}{\sinh(\tilde g/(2\tilde t))}f_x(\tilde g,\tilde g/(2\tilde t))\ln b\right)}
\end{equation}
und
\begin{equation}
 \boxed{\left(\frac{\hbar}{2\pi^2 c}\right)\rightarrow\str{\left(\frac{\hbar}{2\pi^2 c}\right)}=
  \left(\frac{\hbar}{2\pi^2 c}\right)\left(1-\frac{\pi^3}{4}\frac{\tilde \lambda^2}{\tilde t^4}
  \frac{q^4}{\sinh(\tilde g/(2\tilde t))}f_{\tau}(\tilde g,\tilde g/(2\tilde t))\ln b\right)}\,,
\end{equation}
woraus Betr"age zu den im n"achsten Abschnitt dargestellten Flussgleichungen
f"ur $\tilde g$ und $\tilde t$ folgen.

Vorher schauen wir uns die Korrekturen bei $T=0$ an.
Dazu gehen wir zu (\ref{eq.rg_l2}) zur"uck und bilden in diesem
Ausdruck den Limes $T\rightarrow 0$ und finden:
\begin{equation}\label{eq.limitt0}
 \frac{\cosh\left(\left(1-\frac{4w}{\hbar\beta}\right)\lambda_T\Lambda/2\right)}{\sinh(\lambda_T\Lambda/2)}
  \underset{\beta\rightarrow \infty}{\rightarrow}e^{-2vw\Lambda}\,.
\end{equation}
Mit den Substitutionen $\tilde k=k/\Lambda$, $\tilde r=2\Lambda r$, $\tilde w=2\Lambda vw$
und (\ref{eq.limitt0}) f"ur $k$ statt $\Lambda$,
l"a"st sich die Korrelationsfunktion (\ref{eq.thetakf}) wie folgt schreiben:
\begin{eqnarray}
 &&\tmw{(\theta_{<}(x+r,\tau+w)-\theta_{<}(x-r,\tau-w))^2}_{0,<}\nn\\
 &&=\frac{\pi}{\tilde g}\int_0^1\frac{d\tilde k}{\tilde k}\,\left(1-\cos(\tilde k\tilde r)e^{-\tilde k\tilde w}\right)\\
 &&\underset{*}{=}\frac{\pi}{\tilde g}\left(\ln(\tilde w^2+\tilde r^2)/2+\gamma+\Omega(\tilde r,\tilde w)\right)\quad\text{mit}\\
 \Omega(\tilde r,\tilde w)&\equiv&e^{-\tilde w}\int_{\tilde r}^0\frac{x\cos x+\tilde r\sin x}{x^2+\tilde r^2}dx+
  \int_{\tilde r}^{\infty}\frac{e^{-x}}{x}dx\nn
\end{eqnarray}
mit $\gamma=0.577216...$, der \emph{Eulerschen Konstanten}\footnote{Zur Herleitung der Identit"at $*$ wurde Mathematica benutzt.}.
Die Funktion $\Omega(\tilde r,\tilde w)$ ist so gew"ahlt, dass sie f"ur gro"se $\tilde r$ und $\tilde w$ verschwindet,
d.h. das asymptotische Verhalten der Korrelationsfunktion ist, wie in zwei (r"aumlichen) Dimensionen, logarithmisch
mit $|\vec{r}|$.

Mit den Definitionen
\begin{eqnarray}
 f^0_{x}(\tilde g)&\equiv&\frac{1}{16}\int_0^{\infty} d\tilde r\,\int_0^{\infty} d\tilde w\,\tilde r^2 e^{-\frac{q^2\pi}{2\tilde g}
 f^0_c(\tilde w,\tilde r)}\cos(\tilde r)e^{-\tilde w}\label{eq.fx0}\\
 f^0_{\tau}(\tilde g)&\equiv&\frac{1}{16}\int_0^{\infty} d\tilde r\,\int_0^{\infty} d\tilde w\,\tilde w^2 e^{-\frac{q^2\pi}{2\tilde g}
 f^0_c(\tilde w,\tilde r)}\cos(\tilde r)e^{-\tilde w}\quad\text{mit}\label{eq.ft0}\\
 f^0_c(\tilde w,\tilde r)&\equiv&\left(\ln(\tilde w^2+\tilde r^2)/2+\gamma+\Omega(\tilde r,\tilde w)\right)\label{eq.fc0}
\end{eqnarray}
erh"alt man bei $T=0$ folgende RG--Korrekturen:
\begin{eqnarray}
 \left(\frac{\hbar v^2}{2\pi^2 c}\right)&\rightarrow&\str{\left(\frac{\hbar v^2}{2\pi^2 c}\right)}=
  \left(\frac{\hbar v^2}{2\pi^2 c}\right)\left(1-\frac{q^4\pi^3}{2}\frac{\tilde \lambda^2}{\tilde g^4}f^0_{x}(\tilde g)\right)\\
\left(\frac{\hbar }{2\pi^2 c}\right)&\rightarrow&\str{\left(\frac{\hbar }{2\pi^2 c}\right)}=
  \left(\frac{\hbar }{2\pi^2 c}\right)\left(1-\frac{q^4\pi^3}{2}\frac{\tilde \lambda^2}{\tilde g^4}f^0_{\tau}(\tilde g)\right)
\end{eqnarray}

\subsection{Reskalierung und Flussgleichungen}\label{sec.flussgl}

Wir k"onnen nun zun"achst einmal die Flussgleichungen f"ur $\lambda=0$
in $d+1$ Dimensionen aufschreiben.
Zuvor mu"s aber der Beitrag der Reskalierung (Punkt (ii) in dem
eingangs beschriebenen Schema) bestimmt werden.
Die Reskalierung von $\tilde {\cal S}^{n}_{\varphi}$ liefert:
\begin{eqnarray}
 \str{c}&=&cb^{d+z-2+2\zeta}\quad,\quad \str{T}=Tb^z\nn\\
 \str{\left(\frac{c}{v^2}\right)}&=&\left(\frac{c}{v^2}\right)b^{d+z-2z+2\zeta}\quad \text{und}\nn\\
 {\str{V}}^2&=&V^2b^{d+2z}\quad\Rightarrow\quad \str{V}=Vb^{d/2+z}\,.\nn
\end{eqnarray}
F"ur die dimensionslosen Parameter folgt:
\begin{eqnarray}
 \str{\tilde t}&=&\frac{\str{T}}{\str{c}}\Lambda^{d-2}=\tilde t b^{2-d-2\zeta}\nn\\
 \str{\tilde g}&=&\hbar\left(\str{c}\str{\left(\frac{c}{v^2}\right)}\right)^{-1/2}\Lambda^{d-1}
   =\tilde g b^{1-d-2\zeta}\nn\\
 \str{\tilde V}&=&\str{V}(\str{c})^{-1}\Lambda^{d/2-2}=\tilde V b^{2-d/2-2\zeta}\nn
\end{eqnarray}
Die Flussgleichungen lauten damit und den in Abschnitt \ref{sec.RGV} hergeleiteten
RG Korrekturen (s. Punkt (iii) im RG--Schema):

\bigskip
\framebox[\linewidth]{
\begin{Beqnarray}
 \frac{d\tilde t}{d\ln b}&=&\left[2-d-2\zeta\right]\tilde t\\
 \frac{d\tilde g}{d\ln b}&=&\left[1-d-2\zeta-\frac{p^4}{2}K_d\tilde V^2
      \left(1-\frac{\tilde g/(2\tilde t)}{\sinh(\tilde g/(2\tilde t))}\right)\right]\tilde g\\
 \frac{d\tilde V}{d\ln b}&=&\left[2-d/2-2\zeta-\frac{p^2}{2}\tilde g K_d\coth(\tilde g/(2\tilde t))\right]\tilde V
\end{Beqnarray}
}
\bigskip

Mit $\lim_{x\rightarrow\infty}\coth(x)=1$ kann man diese Gleichungen f"ur $T=0$
aufschreiben:
\begin{eqnarray}
 \frac{d\tilde g}{d\ln b}&=&\left[1-d-2\zeta-\frac{p^4}{2}K_d\tilde V^2\right]\tilde g\\
 \frac{d\tilde V}{d\ln b}&=&\left[2-d/2-2\zeta-\frac{p^2}{2}\tilde g K_d\right]\tilde V
\end{eqnarray}
\trenner

Durch die Reskalierung von $\tilde {\cal S}_{\theta}$ erh"alt man folgende Beziehungen
($\str{\theta}=\theta b^{\str{\zeta}}$):
\begin{eqnarray}
 \str{\left(\frac{1}{c}\right)}&=&\left(\frac{1}{c}\right)b^{1-z+2\str{\zeta}}\nn\\
 \str{\left(\frac{v^2}{c}\right)}&=&\left(\frac{v^2}{c}\right)b^{z-1+2\str{\zeta}}\nn\\
 \str{\lambda}&=&\lambda b^{1+z}\nn
\end{eqnarray}
Aus der Kommutatorrelation f"ur $\theta$ und $\varphi$, welche auf allen L"angenskalen die gleiche Form haben soll,
folgt sofort $\str{\zeta}=-\zeta$ und damit f"ur die dimensionslosen Parameter folgt:
\begin{equation}
 \str{\tilde t}=\tilde t b^{1-2\zeta}\quad ,\quad
 \str{\tilde g}=\tilde g b^{-2\zeta}\quad ,\quad
 \str{\tilde \lambda}=\tilde \lambda b^{2-2\zeta}\,,
\end{equation}
d.h. in $d=1$ f"ur $\tilde t$ und $\tilde g$ dieselben wie f"ur $\tilde {\cal S}^{n}_{\varphi}$.
Mit den im vorigen Abschnitt hergeleiteten RG--Korrekturen erh"alt man folgende
Flussgleichungen f"ur $\tilde t$, $\tilde g$ und $\tilde \lambda$:

\bigskip
\framebox[\linewidth]{
\begin{Beqnarray}
 \frac{d\tilde t}{d\ln b}&=&\left(1-2\zeta-\frac{\pi^3}{4}\frac{\tilde \lambda^2}{\tilde t^4}
  \frac{q^4}{\sinh(\tilde g/(2\tilde t))}f_{\tau}(\tilde g,\tilde g/(2\tilde t))\right)\tilde t\\
 \frac{d\tilde g}{d\ln b}&=&\left(-2\zeta-\frac{\pi^3}{8}\frac{\tilde \lambda^2}{\tilde t^4}
  \frac{q^4}{\sinh(\tilde g/(2\tilde t))}(f_{x}(\tilde g,\tilde g/(2\tilde t))+f_{\tau}(\tilde g,\tilde g/(2\tilde t)))\right)\tilde g\\
 \frac{d\tilde \lambda}{d\ln b}&=&\left(2-2\zeta-\frac{q^2\pi}{4\tilde g}\coth(\tilde g/(2\tilde t))\right)\tilde\lambda
\end{Beqnarray}
}
\bigskip

oder f"ur $T=0$:
\begin{eqnarray}
 \frac{d\tilde g}{d\ln b}&=&\left(-2\zeta-\frac{q^4\pi^3}{4}\frac{\tilde \lambda^2}{\tilde g^4}
  (f^0_x(\tilde g)+f^0_{\tau}(\tilde g))\right)\tilde g\label{eq.flgl_glt0}\\
 \frac{d\tilde \lambda}{d\ln b}&=&\left(2-2\zeta-\frac{q^2\pi}{4\tilde g}\right)\tilde \lambda
\end{eqnarray}
 In den Flussgleichungen f"ur $\tilde\lambda$ wurden jeweils nur Korrekturen bis zur ersten Ordnung
in $\tilde\lambda$ ber"ucksichtigt.

\trenner

In $d=1$ k"onnen die Flussgleichungen aller vier Parameter f"ur $T>0$ kombiniert werden
und man erh"alt zusammenfassend folgenden Satz von partiellen Differentialgleichungen:

\bigskip
\framebox[\linewidth]{
\begin{Beqnarray}
\frac{d\tilde t}{d\ln b}&=&\left[1-2\zeta - F_{\tau}\right]\tilde t \label{eq.flgl_t}\\
 \frac{d\tilde g}{d\ln b}&=&\left[-2\zeta - \frac{p^4}{2\pi}\tilde V^2\left(1-\frac{\tilde g/(2\tilde t)}{\sinh(\tilde g/(2\tilde t))}\right)-(F_x+F_{\tau})/2\right]\tilde g\label{eq.flgl_g} \\
 \frac{d\tilde V}{d\ln b}&=&\left[\frac{3}{2}-2\zeta-\frac{p^2}{2\pi}\tilde g\coth(\tilde g/(2\tilde t))\right]\tilde V\label{eq.flgl_V} \\
 \frac{d\tilde \lambda}{d\ln b} &=&\left[2-2\zeta-\frac{q^2\pi}{4}\tilde g^{-1}\coth(\tilde g/(2\tilde t))
  \right]\tilde \lambda\label{eq.flgl_l}
\end{Beqnarray}
}
\bigskip

wobei $F_x$ und $F_\tau$ wie folgt definiert wurden
\begin{eqnarray}
 F_{x/\tau}&=&\frac{\pi^3}{4}\frac{\tilde \lambda^2}{\tilde t^4}
  \frac{q^4}{\sinh(\tilde g/(2\tilde t))}f_{x/\tau}(\tilde g,\tilde g/(2\tilde t))\quad\text{mit}\\
 f_x(y,z)&=&\frac{1}{64}\int_0^{\infty}dr\int_0^1dw r^2 e^{-\frac{q^2\pi}{2 y}f_c(z;r,w)}h(z;r,w)\label{eq.fx}\\
 f_{\tau}(y,z)&=&\frac{1}{64}\int_0^{\infty}dr\int_0^1dw w^2 e^{-\frac{q^2\pi}{2 y}f_c(z;r,w)}h(z;r,w)\quad\text{mit}\label{eq.ft}\\
 f_c(z;r,w)&=&\int_0^{z}\frac{dt}{t\sinh(t)}\left[\cosh(t)-\cos(t r)\cosh((1-|w|)t)\right]\label{eq.fc}\\
 h(z;r,w)&=&\cos(zr)\frac{\cosh((1-|w|)z)}{\sinh(z)}\nonumber\,.
\end{eqnarray}

Korrekturen zu $\tilde V$ bzgl. $\tilde \lambda$ und umgekehrt wurden nicht ber"ucksichtigt,
da dies Korrekturen min. 2. Ordnung w"aren und man daher davon ausgehen kann, dass diese
nur kleine Beitr"age liefern, andererseits sind diese Korrekturen im Hinblick
auf den in Abschnitt \ref{sec.phaseslip} beschriebenen (m"oglichen) phase--slip--Mechanismus
interessant.
Allerdings ist aufgrund der unterschiedlichen Beschreibung der Unordnung und der phase--slips
eine Berechnung (wahrscheinlich) sehr aufwendig.

\emph{Zur Wahl von $\zeta$}: Die beiden Wirkungen ${\cal S}_{\varphi}$
und ${\cal S}_{\theta}$ sind invariant unter dem Shift $\varphi,\theta\rightarrow (\varphi+2n\pi),(\theta+2n\pi)$.
Andererseits liefert die Reskalierung
$\varphi=\str{\varphi}b^{\zeta}$ im nichtlinearen Cosinus--Term
einen zus"atzlichen Term $\propto \ln (b) \zeta\varphi\sin(p\varphi+\alpha(x))$
(analog f"ur $\theta$, jeweils in erster Ordnung in $\ln b$), sodass
die reskalierten Wirkungen nicht mehr diese Symmetrie aufweisen.
Daher sollte $\zeta=0$ sein.

\subsection{numerische Auswertung und Flussdiagramme}\label{sec.flussdia}

In diesem Abschnitt sollen die Flussgleichungen und Phasendiagramm ausf"uhrlich
untersucht werden. Dazu beginnen wir mit dem Fall $\lambda=0$.

Bisher war der Parameter $p$ beliebig, um m"oglichst allgemein zu bleiben.
Um die Flussgleichungen aber weiter untersuchen zu k"onnen, muss $p$ fixiert
werden. Im Fall der CDW ist $p=1$, da das Unordnungspotential an die
Ladungsdichte $\rho(x)$ koppelt und in $\rho$ $p=1$ ist. Au"serdem wird
$\zeta=0$ gesetzt.

Betrachten wird zun"achst den einfachsten
Fall $T=0$ in $d=1$.
\fourfigureso{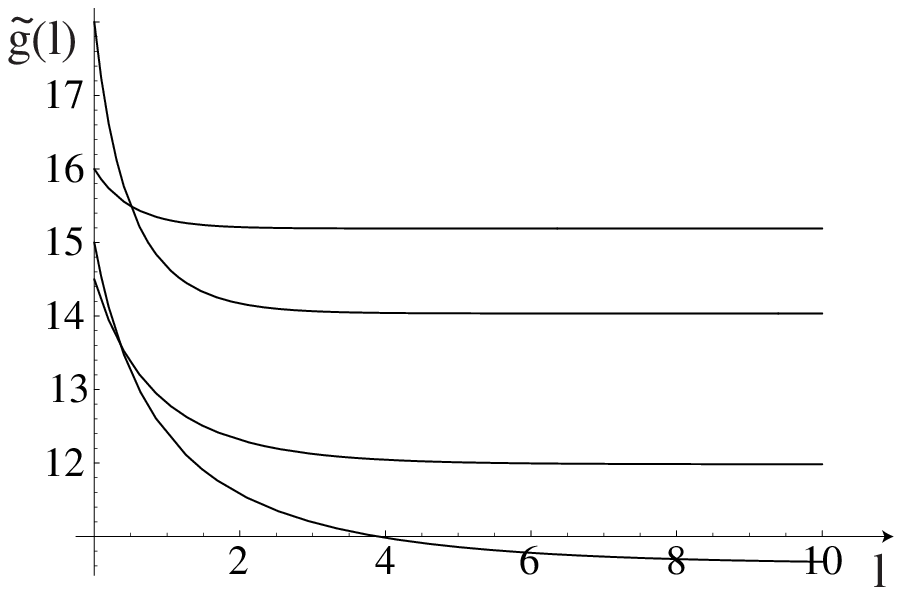}{0.45\linewidth}{Fluss von $\tilde g$ f"ur
verschiedene Startwerte in der delokalisierten Phase.}{fig.f_t0l0_1}
{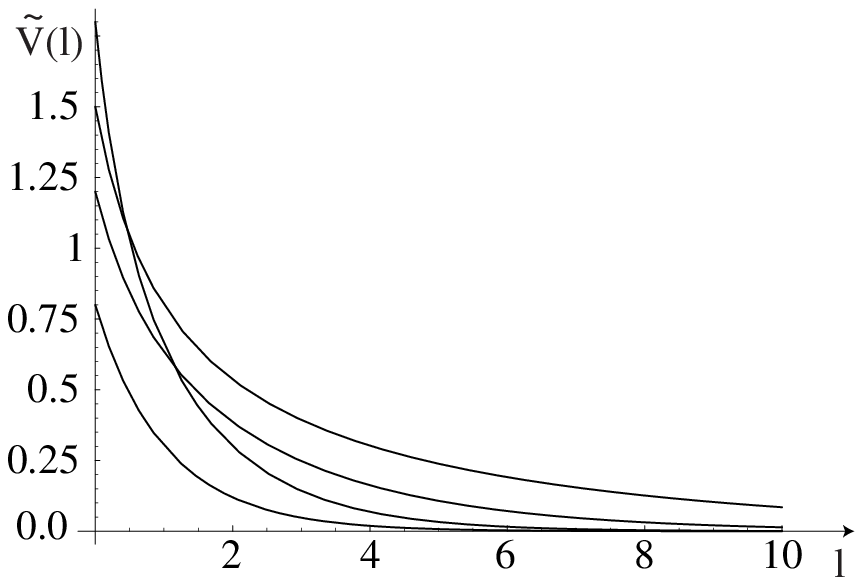}{0.45\linewidth}{Fluss von $\tilde V$ in der delokalisierten Phase.}{fig.f_t0l0_2}
\fourfiguresu{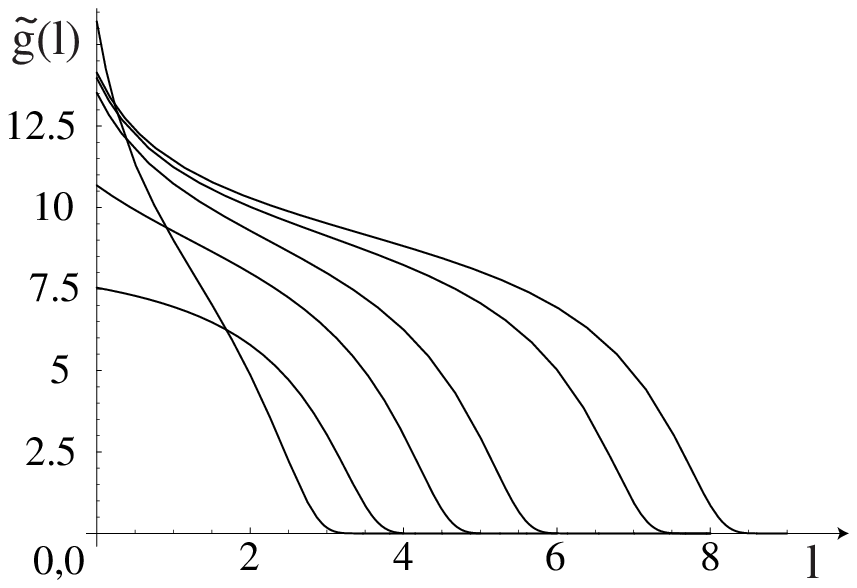}{0.45\linewidth}{Fluss von $\tilde g$ in der lokalisierten Phase.}{fig.f_t0l0_3}
{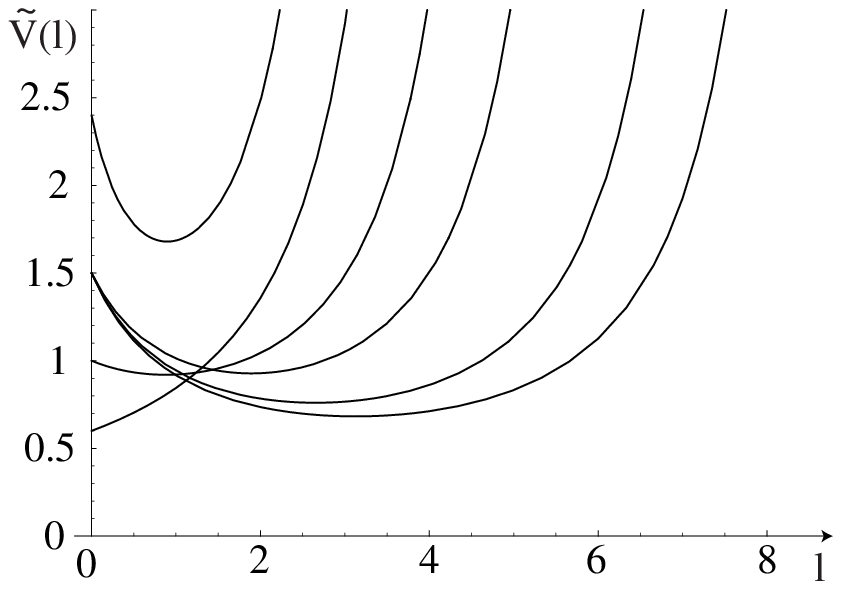}{0.45\linewidth}{Fluss von $\tilde V$ in der lokalisierten Phase.}{fig.f_t0l0_4}

Die Flussgleichungen werden mit einem \emph{Runge--Kutta}--Verfahren numerisch integriert.
Typische Fl"usse der Parameter $\tilde g$ und $\tilde V$ sind in den Abbildungen \ref{fig.f_t0l0_1}--\ref{fig.f_t0l0_4}
aufgezeichnet.

\onefigure{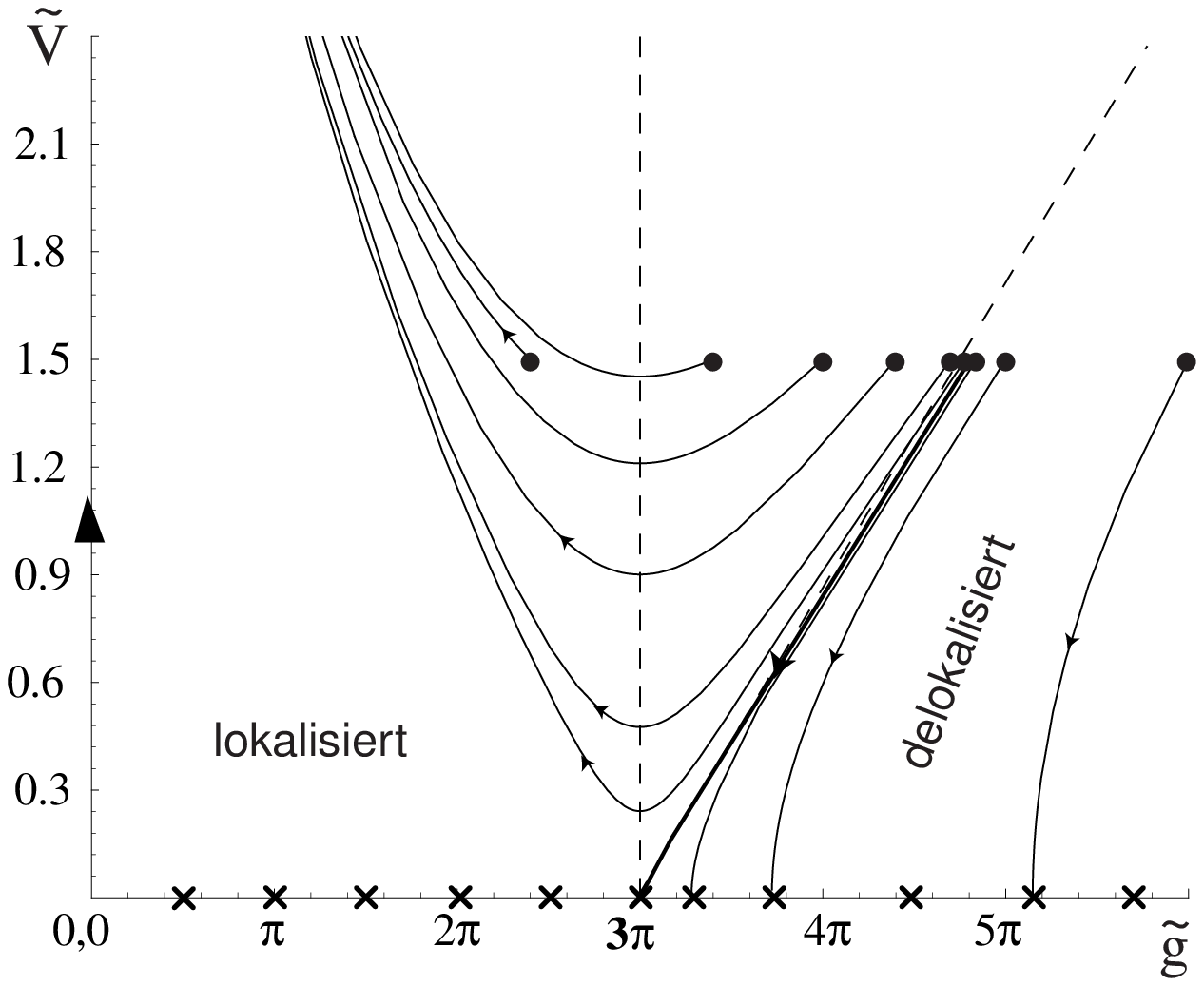}{0.8\linewidth}{Typische Fl"usse bei $T=0$ in der $\tilde V$--$\tilde g$--Ebene.
Anfangswerte sind mit einem Kreis gekennzeichnet, die Flussrichtung mit einem Pfeil. Kreuze stehen f"ur Fixpunkte
der Parameter. Die Separatix ist fett gezeichnet.}{fig.fluss_t0l0}

Zeichnet man die Fl"usse in der $\tilde g$--$\tilde V$--Ebene erh"alt man das in Abb. \ref{fig.fluss_t0l0}
numerisch bestimmte Phasendiagramm. Wie man sieht, gibt es zwei unterschiedliche Bereiche (Phasen) der
Startwerte: eine {\bf lokalisierte Phase}, in der die Unordnung relevant wird und $\tilde V$ f"ur jeden
Startwert aus der Phase gegen unendlich flie"st (und $\tilde g$ gegen $0$), und eine
{\bf delokalisierte Phase}, in der die Unordnung irrelevant wird. Beide Phasen werden durch
eine Separatrix voneinander getrennt, welche im Fixpunkt ($\tilde g_{c_1}=3\pi$, $\tilde V=0$) endet.

Diese einfachen Fl"usse in der $\tilde g$--$\tilde V$--Ebene, d.h. die Funktion $\tilde V(\tilde g)$
kann auch analytisch bestimmt werden. Dazu werden die beiden partiellen Differentialgleichungen (PDGLn)
durcheinander dividiert, womit man die einfache gew"ohnliche DGL
\begin{equation}
 \frac{d\tilde V}{d\tilde g}=\frac{1}{\tilde V}\left(1-\frac{3\pi}{\tilde g}\right)
\end{equation}
erh"alt. Diese DGL hat f"ur $\tilde V\geq 0$ folgende L"osung \footnote{einfach mit Separationsansatz integrierbar.}
\begin{equation}
 \tilde V(\tilde g)=\sqrt{2(\tilde g-3\pi\ln(\tilde g)+c_g)}\,,
\end{equation}
wobei $c_g$ eine Konstante ist, welche von den Anfangswerten abh"angt.

F"ur die Separatrix mit $\tilde V(3\pi)=0$ erh"alt man folgende Gleichung:
\begin{equation}
 \tilde V_{\textit{sep}}(\tilde g)=\left[2\tilde g-6\pi(1+\ln(\tilde g/(3\pi)))\right]^{1/2}\quad,\quad \tilde g\geq 3\pi\,.
\end{equation}
Entwickelt man diesen Ausdruck um den Fixpunkt, erh"alt man einen kritischen Exponenten der Separatrix an diesem Fixpunkt
von $\kappa=1$:
\begin{equation}
 \tilde V_{\textit{sep}}(\tilde g)\propto (\tilde g-3\pi)^{\kappa}\quad,\quad\kappa=1\,,
\end{equation}
was auch gut in Abb. \ref{fig.fluss_t0l0} (gestrichelte Linie) zu erkennen ist.

\onefigure{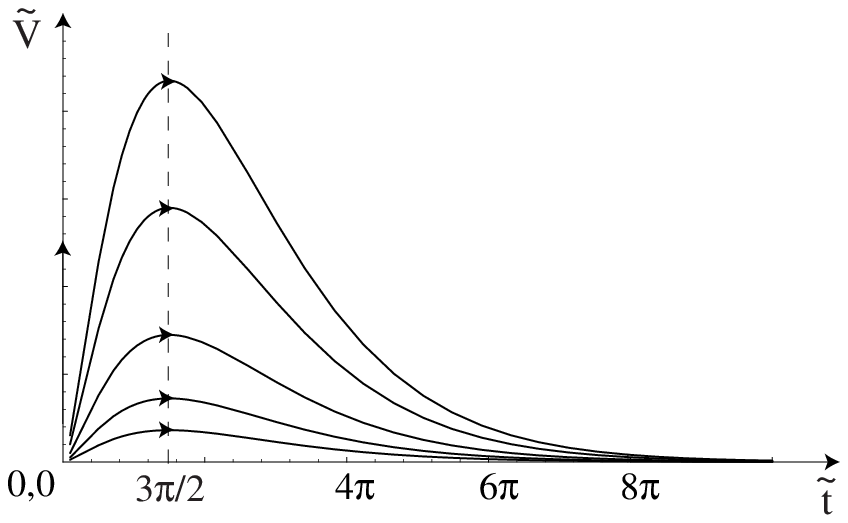}{0.6\linewidth}{Typische Fl"usse in der $\tilde t$--$\tilde V$--Ebene.}{fig.fluss_g0l0}
Nun gehen wir zu $T>0$ "uber und betrachten zun"achst den Fluss in der $\tilde t$--$\tilde V$--Ebene mit $\tilde g=0$:
In diesem Fall haben die DGLn wieder eine einfache Form und k"onnen analytisch gel"ost werden:
\begin{eqnarray}
 \tilde t(l)&=&c_1e^{l}\\
 \tilde V(l)&=&c_2e^{-c_1e^l/\pi+3/2 l}
\end{eqnarray}
oder
\begin{equation}
 \tilde V(\tilde t)=c_t\tilde t^{3/2} e^{-\tilde t/\pi}\,.
\end{equation}
Der Wert der Konstanten $c_t$ h"angt wieder von den Anfangswerten ab.
Typische Fl"usse sind in Abb. \ref{fig.fluss_g0l0} dargestellt. Man sieht, dass bei beliebig kleiner, aber
endlicher Temperatur die Unordnung irrelevant wird.

\twofigures{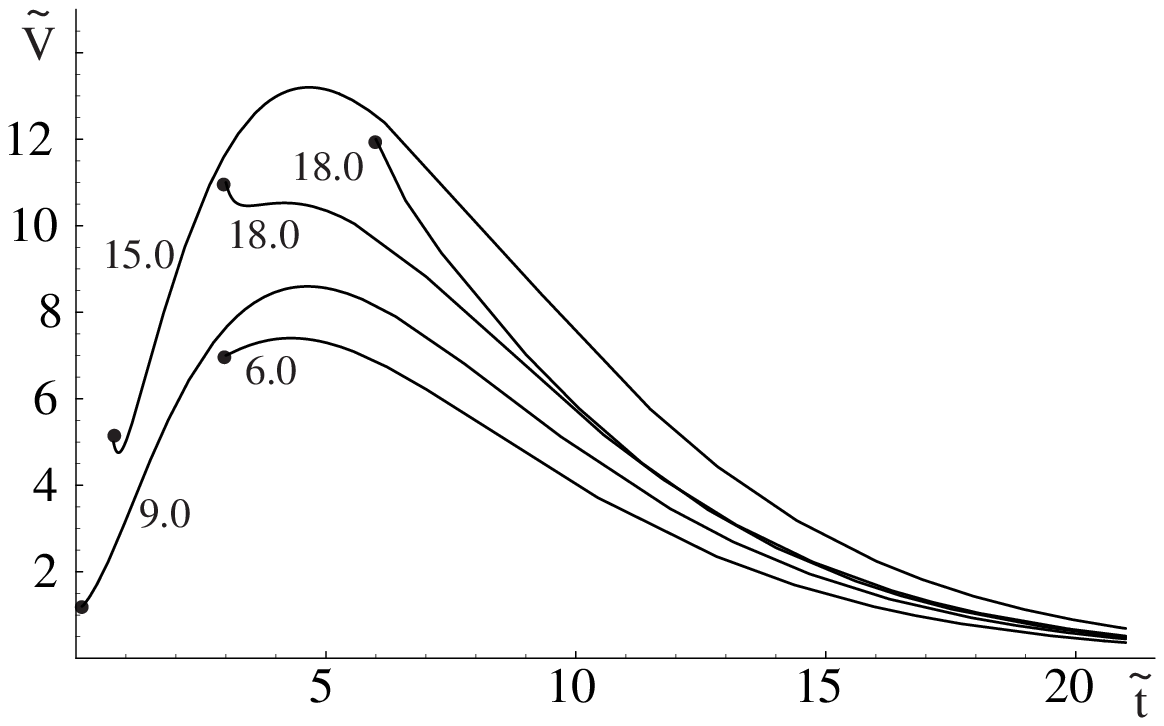}{0.45\linewidth}{Fl"usse in $\tilde t$--$\tilde V$--Ebenen
f"ur verschiedene Werte von $\tilde g>0$. An den Graphen stehen die Werte von
$\tilde g(0)$. Startpunkte sind mit einem Kreis gekennzeichnet.
(Da hier der Fluss zu kleinen $\tilde g$ nicht
dargestellt wird, sieht es so aus, als ob $\tilde V(\tilde t)$ immer bei
$\tilde t\approx 3\pi/2$ ein Maximum hat, was allerdings nicht der Fall ist - siehe Text.)
}{fig.fluss_l0_1}{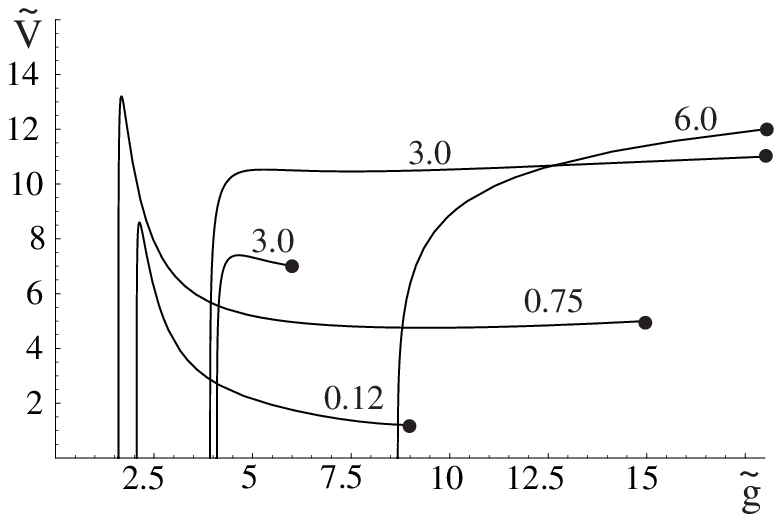}{0.45\linewidth}{Fl"usse in $\tilde g$--$\tilde V$--Ebenen
f"ur verschiedene Werte von $\tilde t>0$. An den Graphen stehen die Werte von
$\tilde t(0)$. }{fig.fluss_l0_2}

Dass dies auch bei endlichem $\tilde g$ der Fall ist, wird in Abb. \ref{fig.fluss_l0_1} und \ref{fig.fluss_l0_2}
verdeutlicht, in denen der Fluss einmal in Ebenen mit konstantem $\tilde g>0$ projiziert ist,
und in Abb. \ref{fig.fluss_l0_2} in Ebenen mit konstantem $\tilde t$.

In Abb. \ref{fig.fluss_g0l0} erkennt man au"serdem, dass es eine ausgezeignete Temperatur
$\tilde t_c=3\pi/2$ bei $\tilde g=0$  gibt, bei der sich der Fluss von $\tilde V$ mit $\tilde t$
umkehrt.
\onefigure{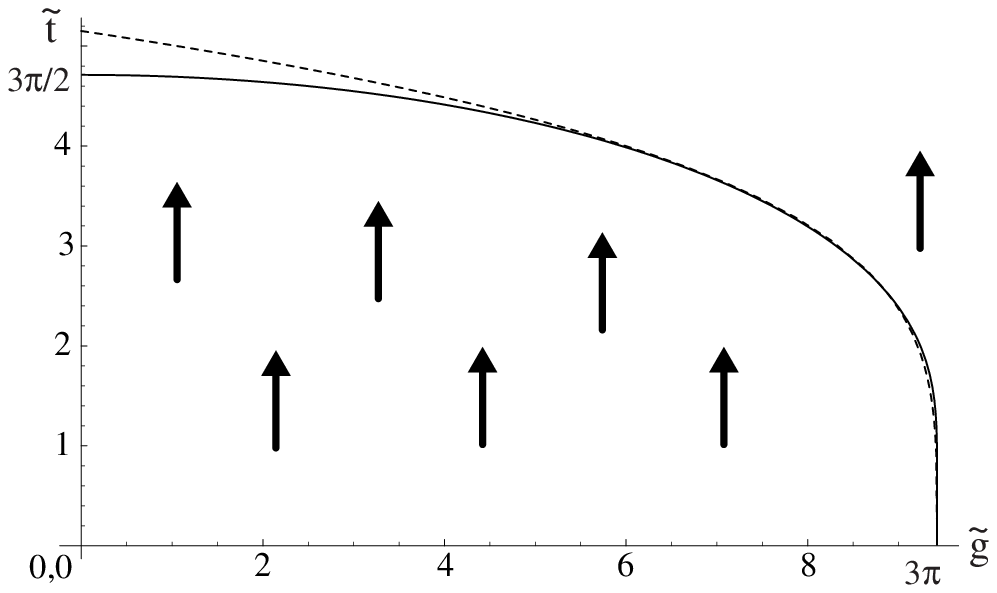}{0.7\linewidth}{kritische Linie in der $\tilde V=0$ Ebene. Gestrichelt ist die Kurve
$\tilde t_{c}(\tilde g)\propto (3\pi-\tilde g)^{\sigma}$ mit $\sigma=0.250\pm 0.02$ am
Fixpunkt $(\tilde g=3\pi,\tilde V=0)$ angefittet. Die Pfeile sollen den trivialen Fluss
in dieser Ebene verdeutlichen.}{fig.critline_V0}

F"ur endliche $\tilde g$ wird $\tilde t_c(\tilde g)$ (bzw. $\tilde g_{c_1}(\tilde t)$) durch die Gleichung
$3/2-\tilde g/(2\pi)\coth(\tilde g/(2\tilde t_c(\tilde g)))=0$ definiert.
Die L"osungskurve $\tilde t_c(\tilde g)$ ist in Abb. \ref{fig.critline_V0}
dargestellt. Im Bereich des Fixpunktes ($\tilde g_{c_1}=3\pi,\tilde t=0$) kann der
Verlauf durch $\tilde g_{c_1}(\tilde t)\propto (3\pi-\tilde t)^{\sigma}$
mit dem kritischen Exponenten $\sigma=0.250\pm 0.002$ angen"ahert werden.

\onefigure{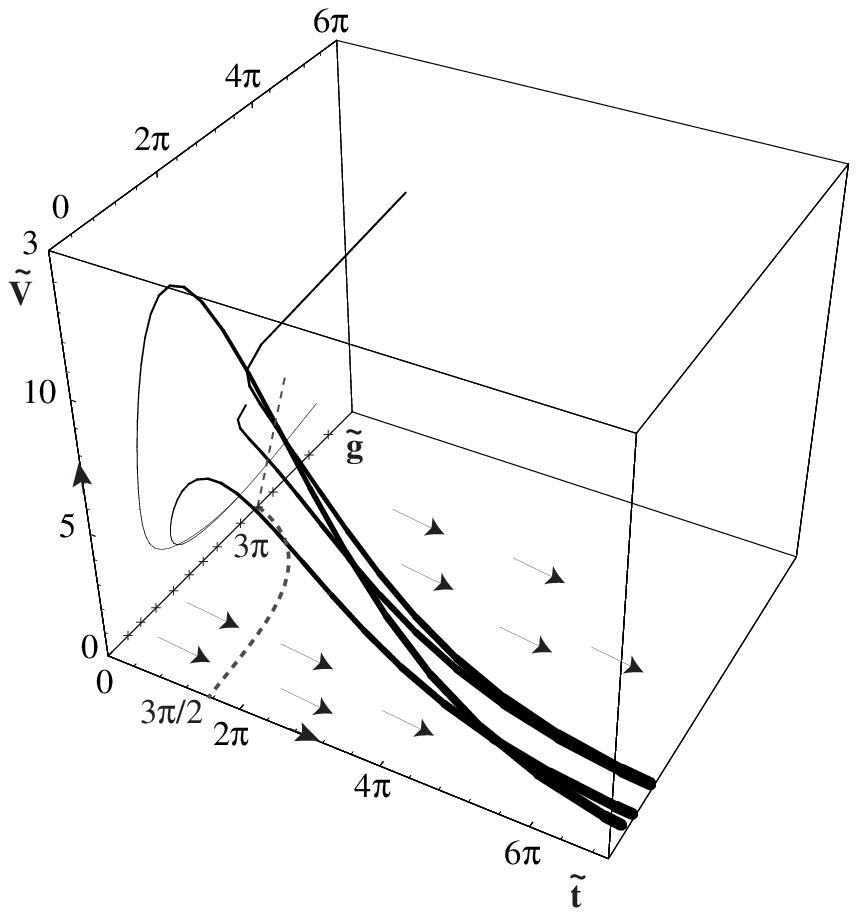}{0.6\linewidth}{Dreidimensionale Darstellung des Flusses mit
eingezeichneter Separatrix und $\tilde t_c(\tilde g)$ (gestrichelt).}{fig.fluss_l0}

In Abb. \ref{fig.fluss_l0} ist der Fluss aller drei Parameter
(also $\vec{\Phi}(l)\equiv (\tilde t(l),\tilde g(l),\tilde V(l))$)
dreidimensional dargestellt.

\trenner

\fourfigureso{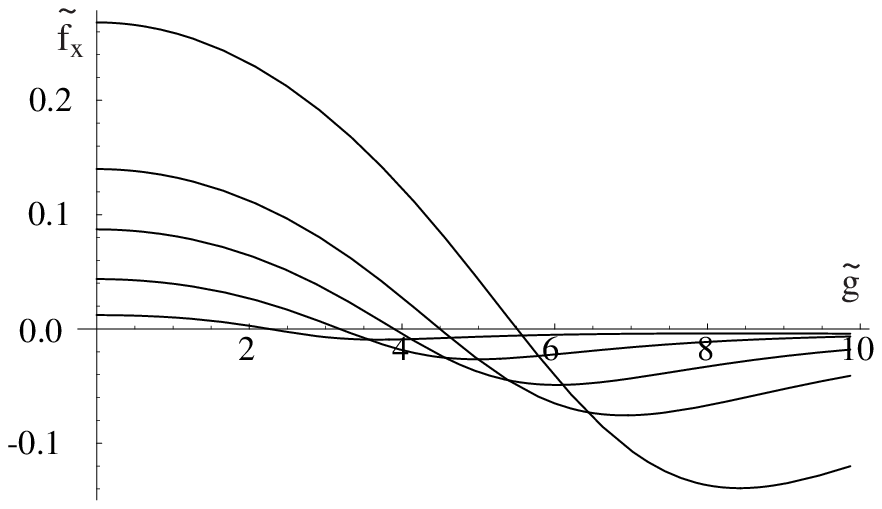}{0.5\linewidth}{Funktionsverlauf von $f_x$ mit $\tilde g$
bei verschiedenen Temperaturen $\tilde t$. Der Wert der Nullstelle dieser Funktion
geht f"ur $\tilde t\rightarrow 0$ gegen den Wert $\approx 0.92$ (NS von $f_x^0$) und
wird f"ur gr"o"sere $\tilde t$ gr"o"ser.}{fig.fx1}{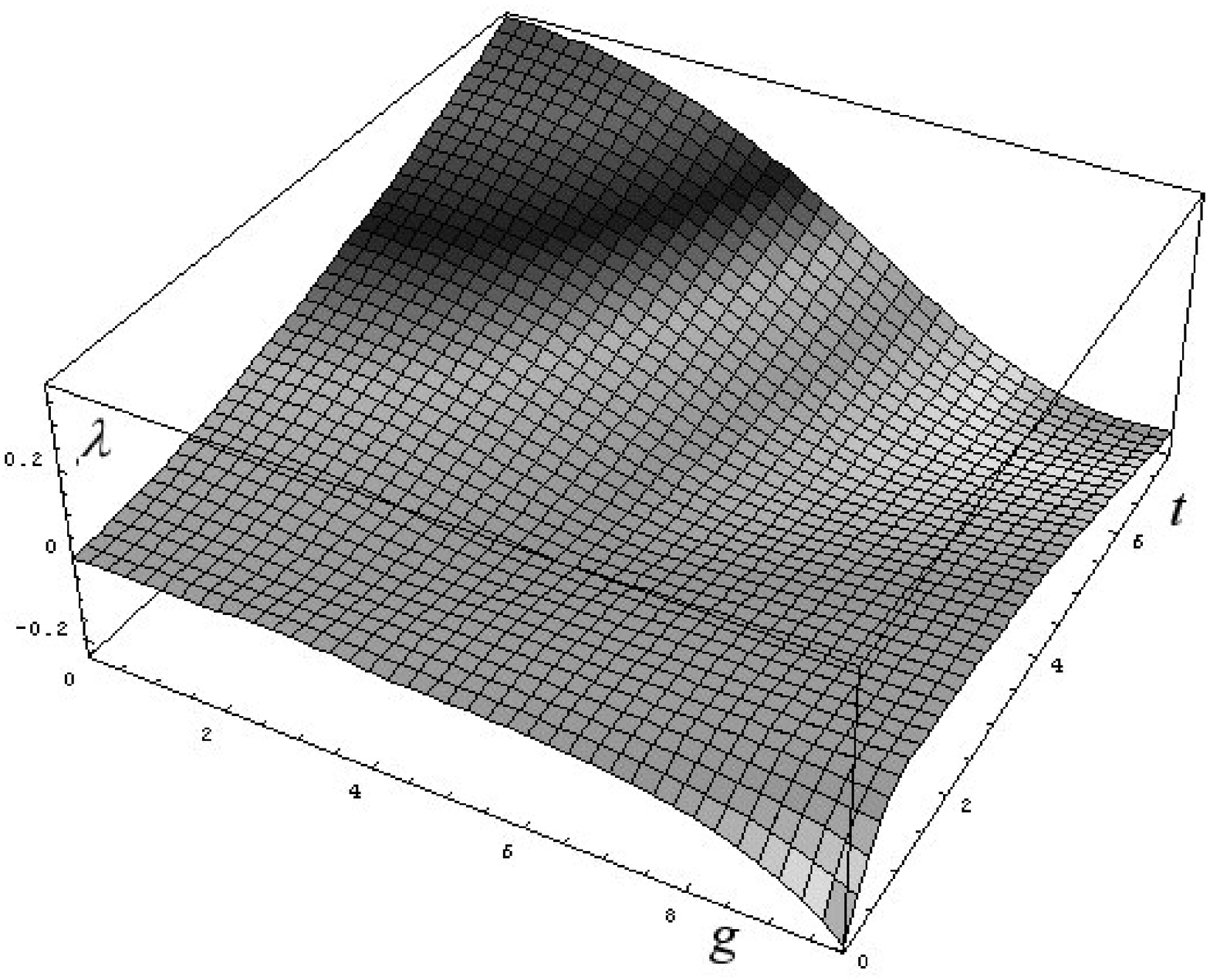}
{0.4\linewidth}{Funktionsverlauf von $f_x$ mit $\tilde g$ und $\tilde t$.}{fig.fx2}
\fourfiguresu{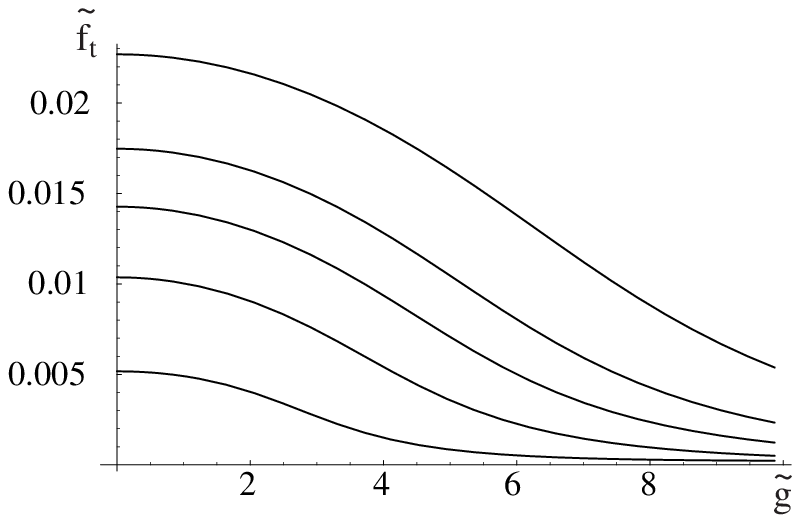}{0.5\linewidth}{Funktionsverlauf von $f_{\tau}$ mit $\tilde g$
bei verschiedenen Temperaturen $\tilde t$.
$f_{\tau}(\tilde g=0,\tilde t)$ w"achst mit $\tilde t$.}{fig.ft1}{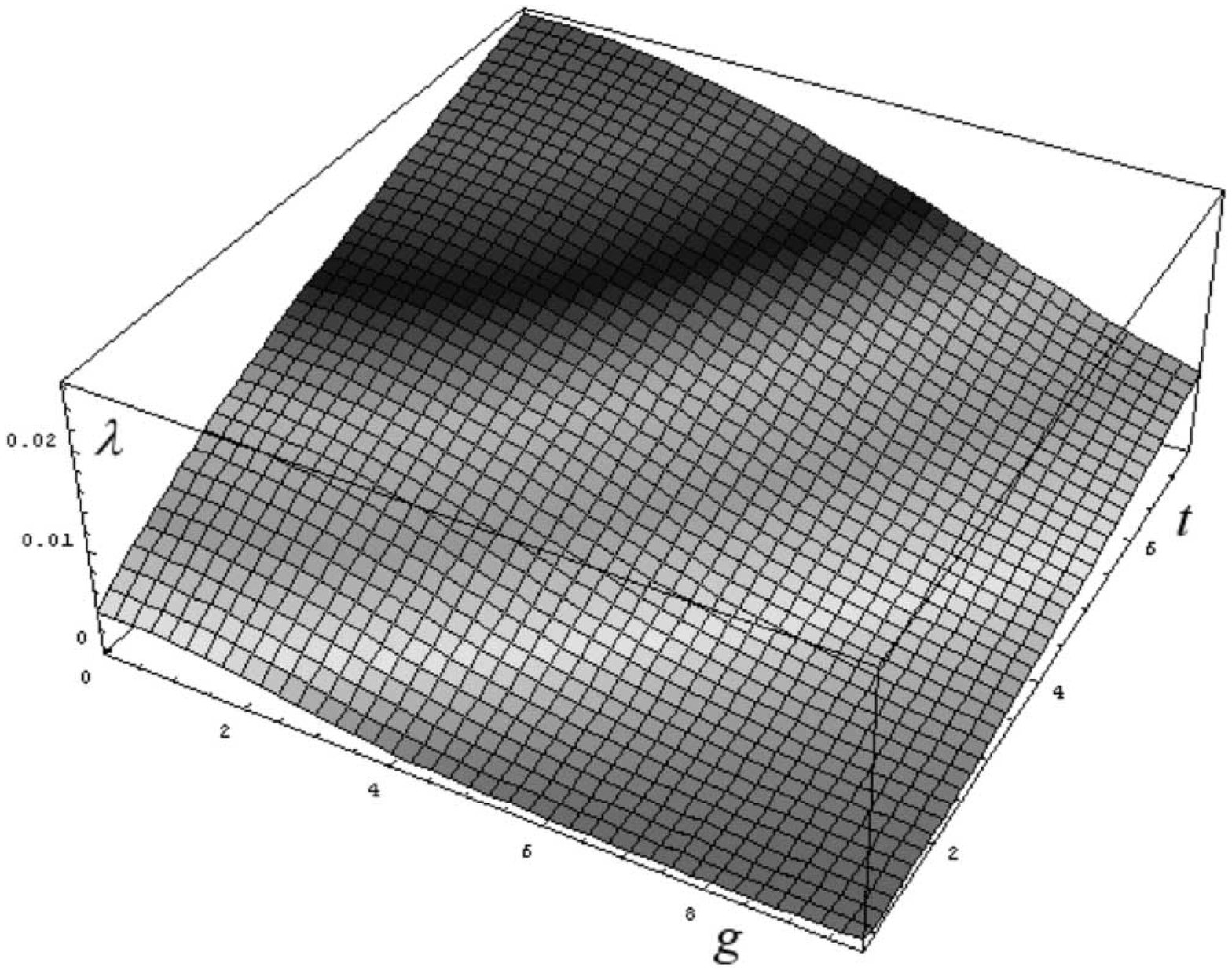}
{0.4\linewidth}{Funktionsverlauf von $f_{\tau}$ mit $\tilde g$ und $\tilde t$.}{fig.ft2}

Bevor wir die Flussgleichungen f"ur $\tilde V=0$, $\tilde\lambda>0$ untersuchen k"onnen,
mu"s der Verlauf der Funktionen $f_x$ (\ref{eq.fx}) und $f_{\tau}$ (\ref{eq.ft}) bestimmt werden.
Dieser wurde numerisch f"ur verschiedene
Werte von $\tilde t$ und $\tilde g$ bestimmt. In den Abbildungen \ref{fig.fx1}--\ref{fig.ft2}
ist der Graph dieser Funktionen dargestellt. Zwischen zwei benachbarten Werten, wurden die Funktionen
kubisch interpoliert. Mit diesen Funktionen werden im folgenden die Flussgleichungen numerisch
untersucht.
Da die numerische Berechnung dieser Funktionen allerdings sehr zeitaufwendig ist
(insbesondere f"ur gro"se $\tilde g$ und $\tilde g/(2\tilde t)$), ist
hier eine numerische Untersuchung der Flussgleichungen nur begrenzt m"oglich.

{\bf Bemerkung 1:} Da die Funktion $f_c$ bei endlichen Temperaturen f"ur gro"se $\tilde r$ linear mit
$\tilde r$ ansteigt - d.h., dass die Paarkorrelationsfunktion sich aufgrund der beschr"ankten
$\tau$--Ausdehnung des Systems f"ur gro"se $\tilde r$ quasi eindimensional verh"alt -
konvergiert das $\tilde r$--Integral in $f_x$ und $f_{\tau}$ und rechtfertigt wiederum die Entwicklung
f"ur kleine $\tilde r$. Die Konvergenz der $\tilde w$--Integrale ist in jedem Fall,
durch die exponentiell abfallenden Funktionen, garantiert.
Bei $T=0$ ist die Konvergenz des $\tilde r$--Integrals in $f_x^0$ (\ref{eq.fx0}) wegen des
logarithmischen Verhaltens der Paarkorrelationsfunktion (s. (\ref{eq.fc0})) f"ur bestimmte
$\tilde g$--Werte nicht mehr gegeben, da das Integral
\begin{equation*}
 I_a(\alpha)\equiv\int\limits_{a}^{\infty}\tilde r^{-\alpha}\cos(\tilde r) d\tilde r\quad,\quad a>0
\end{equation*}
f"ur $\alpha\leq 0$ nicht mehr existiert (mit $\lim_{\alpha\rightarrow 0}I_a(\alpha)=-\sin(a)$).
Diese Bedingung definiert mit (\ref{eq.fx0}) einen
kritischen Wert f"ur $\tilde g$, denn f"ur gro"se $\tilde r$ gilt $\alpha\approx 2-q^2\pi/4$,
womit $f_x^0$ nur f"ur $\tilde g<q^2\pi/8$ existiert. $\diamond$

{\bf Bemerkung 2:} Wie schon in Abschnitt \ref{sec.phaseslip} und Abb. \ref{fig.phaseslip}
angedeutet, setzen wir $q=2$. Diese Wahl ist plausibel, da zum einen die Werte von
$q\in2\mathbb{Z}\backslash\{0\}$ sein m"ussen (siehe Abb. \ref{fig.phaseslip}) und au"serdem
$q$ den minimalen Wert annimmt, da die Wahrscheinlichkeit f"ur phase--slips
$\varphi\rightarrow\varphi+q\pi$ f"ur gr"o"sere $q=4,6,\ldots$ bei tiefen Temperaturen
sehr klein sein wird. $\diamond$

\onefigure{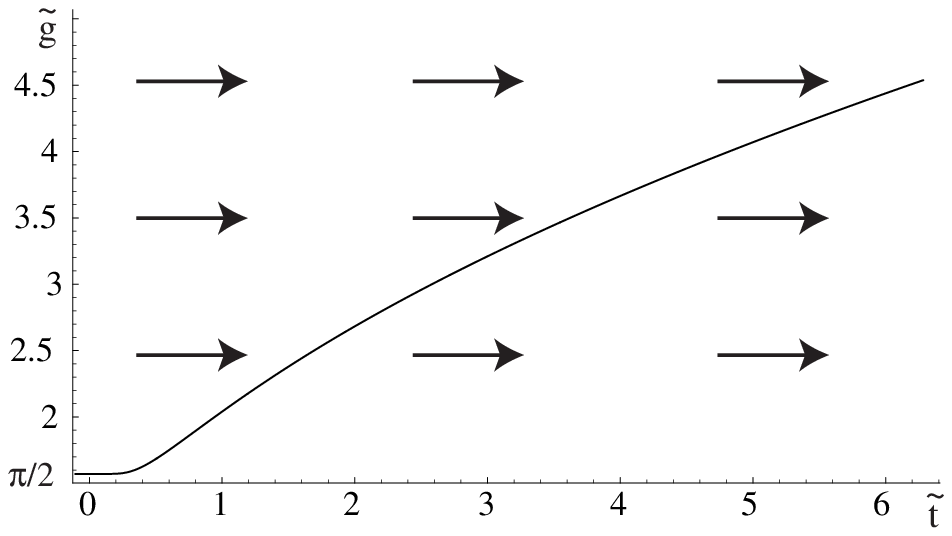}{0.6\linewidth}{kritische Linie in der $\tilde \lambda=0$ Ebene.
Der Fluss in dieser Ebene ist einfach und geht zu gro"sen $\tilde t$ hin (Pfeile)}{fig.critlineL0}

Kommen wir jetzt zur Untersuchung der Fl"usse bei $T=0$. An den Flussgleichungen
f"ur $\tilde \lambda$ (\ref{eq.flgl_l}) sieht man, dass es wieder eine Linie
$\tilde g_{c_2}(\tilde t)$ gibt, bei der sich diesmal der Fluss von $\tilde\lambda$
umkehrt. Diese ist durch die Gleichung $2-\pi/\tilde g_{c_2}(\tilde t)\coth(\tilde g_{c_2}(\tilde t)/(2\tilde t))=0$
definiert. F"ur $\tilde t=0$ erh"alt man einen kritischen Wert von $\tilde g_{c_2}=\pi/2$.
Der Verlauf ist in Abb. \ref{fig.critlineL0} dargestellt. Diese Werte von $\tilde g$
trennen also, "ahnlich wie die Kurve $\tilde t_c(\tilde g)$ f"ur $\tilde V$, zwei Regionen
in denen $\tilde \lambda$ gegen null ($\tilde g<\tilde g_{c_2}(\tilde t)$) bzw. gegen unendlich
($\tilde g>\tilde g_{c_2}(\tilde t)$) flie"st.
Der Wert von $\tilde g_{c_2}$ stimmt mit dem Wert von $\tilde g$ "uberein, bei dem $f_x^0$
(unbestimmt) divergiert, was bedeutet, dass die Flussgleichung f"ur $\tilde g$ (\ref{eq.flgl_glt0})
bei $T=0$ nur bis $\tilde g=\pi/2$ g"ultig ist. F"ur $\tilde g>\pi/2$ ist die gezeigte RG f"ur $\tilde g$
nicht mehr ausreichend. Um dennoch Aussagen f"ur $\tilde g>\pi/2$  bei $T=0$ machen zu k"onnen,
vernachl"assigen wir in diesem Bereich die RG--Korrekturen bzgl. $\tilde \lambda$ zu $\tilde g$,
was bedeutet, dass $\tilde g(l)$ nicht mehr ''flie"st'' ($d\tilde g/dl = 0 $) und somit $\tilde \lambda$
f"ur alle $\tilde g>\pi/2$ einfach gegen unendlich flie"st (sofern $\tilde\lambda(0)>0$ war).

\onefigure{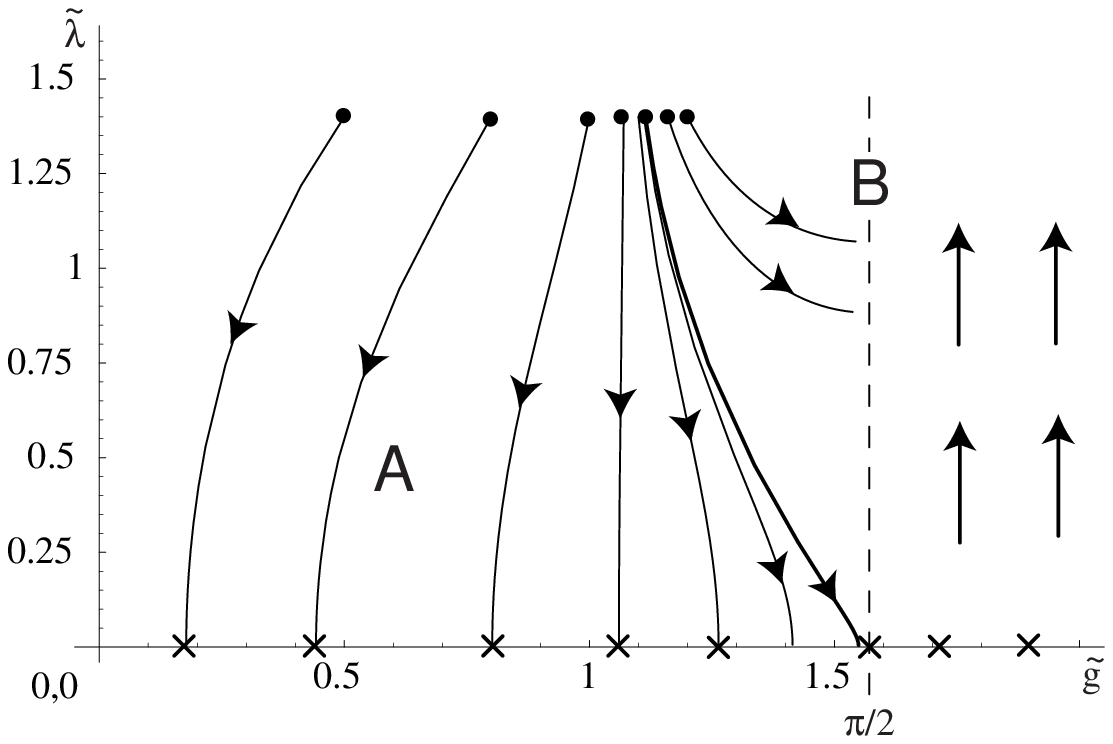}{0.8\linewidth}{Fluss bei $\tilde t=0$ und $\tilde V=0$. In der Phase
A werden die phase--slips irrelevant und in B relevant. Die (n"aherungsweise) Separatrix ist
fett dargestellt.}{fig.fluss_t0v0}

Der komplette Fluss bei $T=0$ ist in Abb. \ref{fig.fluss_t0v0} dargestellt. Man findet wiederum zwei
Phasen (A und B), in denen einmal $\tilde\lambda$ gegen $0$ flie"st (A), d.h. die phase--slips irrelevant
werden, und in der Phase B die phase--slips relevant werden. Da $f_x^0+f_{\tau}^0$ zwischen $0$ und $\pi/2$
bei $\approx 1.07$ eine Nullstelle besitzt (s. Abb. \ref{fig.fx1}), ist die Steigung der Separatrix der
beiden Phasen negativ. Aufgrund der Divergenz von $f_x^0(\tilde g)$ am Fixpunkt ($\tilde g=\tilde g_{c_2}$, $\tilde \lambda=0$),
ist ein kritischer Exponent der Separatrix am Fixpunkt nicht bestimmbar. Untersucht man
allerdings die in Abb. \ref{fig.fluss_t0v0} fett dargestellte Flusslinie, welche n"aherungsweise die
Separatrix ist, findet man, dass die Separatrix n"aherungsweise linear in den Fixpunkt verl"auft.

\twovfigures{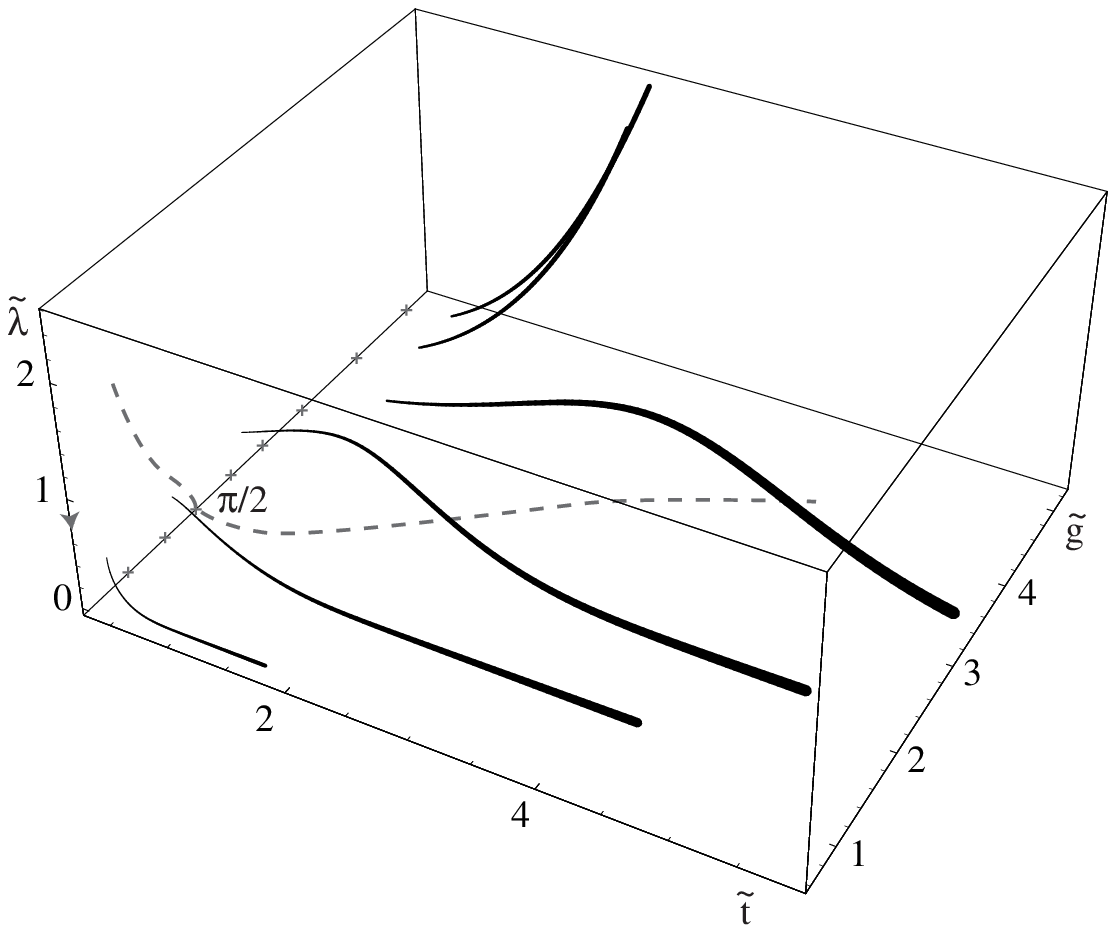}{0.8\linewidth}{Fl"usse f"ur $\tilde V=0$. Gestrichelt ist die Separatrix der beiden Phasen
A und B und die kritische Linie $\tilde g_{c_2}(\tilde t)$ eingezeichnet.}{fig.fluss_V0}
{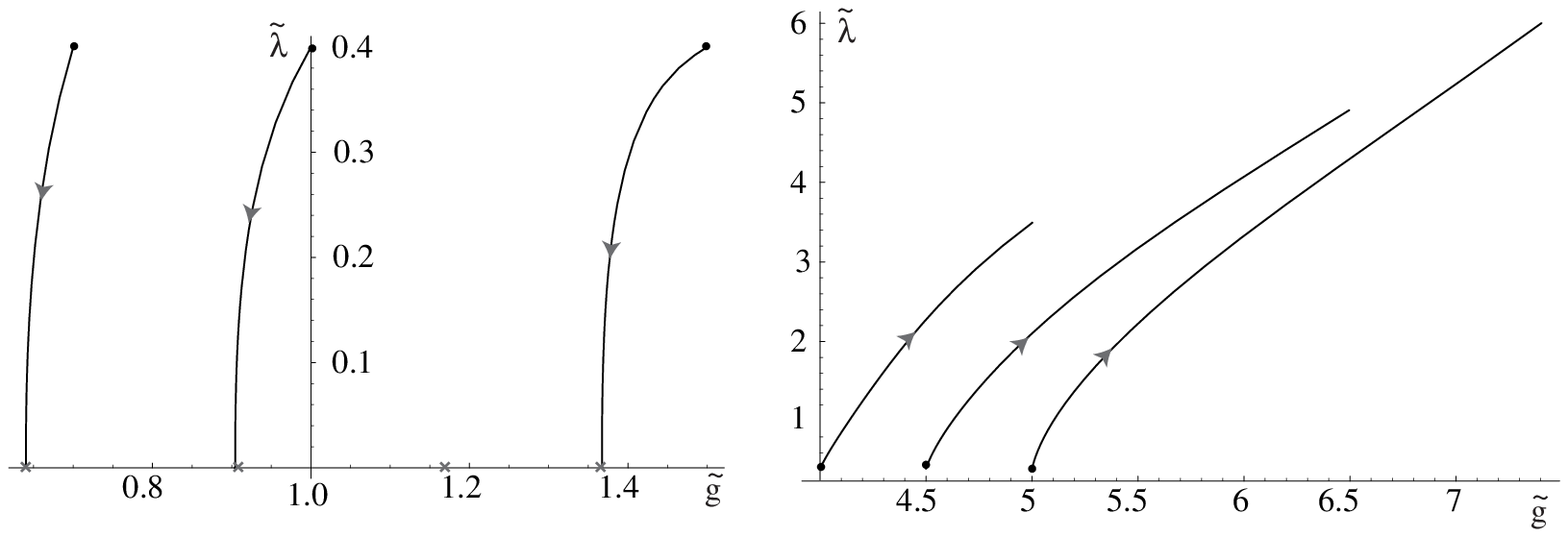}{\linewidth}{Fluss f"ur $\tilde V=0$ und $T>0$ in Ebenen mit konstantem $\tilde t$
projiziert f"ur $\tilde g(0)<\tilde g_{c_2}$ (links) und $\tilde g(0)>\tilde g_{c_2}$ (rechts). }{fig.fluss_V0_1}

Der Fluss f"ur endliche $\tilde t$ ist in Abb. \ref{fig.fluss_V0} oder auch in Abb. \ref{fig.fluss_V0_1}, in
Ebenen mit konstantem $\tilde t$ projiziert, dargestellt.
In Abb. \ref{fig.fluss_V0_1} (rechts) und f"ur die zwei Flusslinien in Abb. \ref{fig.fluss_V0} bei gro"sen $\tilde g$
ist zu beachten, dass die Differentialgleichungen, aus den anf"anglich genannten Gr"unden, nur bis zu einem
bestimmten Wert von $\tilde g$ integriert werden konnten und nicht klar ist
(aber wahrscheinlich, siehe n"achstes Kapitel), ob der Fluss, wie die zwei ''mittleren'' Flusslinien in Abb. \ref{fig.fluss_V0},
die kritische Linie $\tilde g_{c_2}(\tilde t)$ kreuzen und somit $\tilde \lambda$ letztendlich auch gegen null flie"st,
wie dies bei kleineren Startwerten von $\tilde g(l)$ der Fall ist.
Dies bedeutet also, dass auch die phase--slips, zumindest f"ur kleine Ausgangswerte von $\tilde g$,
bei endlichen Temperaturen irrelevant werden.

\trenner

\onefigure{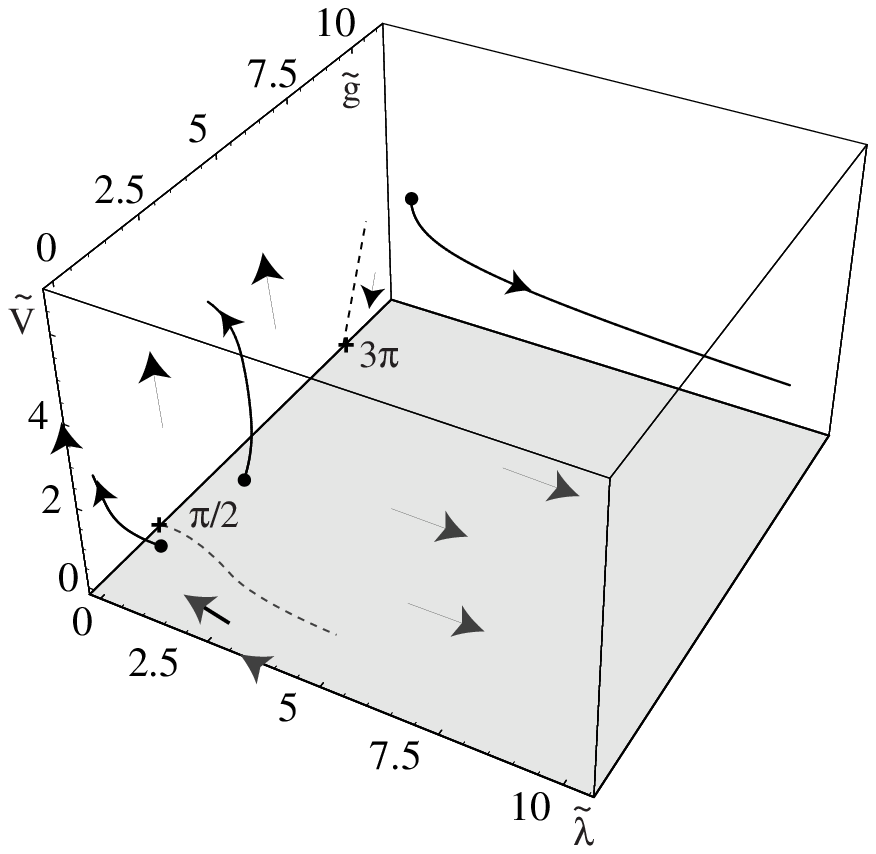}{0.8\linewidth}{Fluss bei $\tilde t=0$.}{fig.fluss_t0}

Zum Schlu"s schauen wir uns noch den kompletten Fluss bei $T=0$ f"ur $\tilde V$, $\tilde\lambda$
und $\tilde g$ an,
welcher in Abb. \ref{fig.fluss_t0} visualisiert ist. Man kann drei Bereiche identifizieren:
F"ur $\tilde g\lesssim\tilde g_{c_2}$ ist die Unordnung relevant und die phase--slips irrelevant, im
Zwischenbereich $\tilde g_{c_2}\lesssim\tilde g\lesssim\tilde g_{c_1}$ sind beide Fluktuationen
relevant (wenn $\tilde\lambda=0$ oder $\tilde V=0$) und den Bereich $\tilde g\gtrsim\tilde g_{c_1}$
in dem die phase--slips relevant werden, aber die Unordnung irrelevant wird.
Zur Berechnung der Flusslinien wurde f"ur $\tilde g\geq\tilde g_{c_2}$ $f_x^0=0$ gesetzt.

Da in den Flussgleichungen f"ur $\tilde\lambda$ und $\tilde V$ keine RG--Korrekturen der
jeweils andere Parameter ber"ucksichtigt wurden, sind die Aussagen f"ur die Fl"usse bei
endlichem $\tilde\lambda$ und  $\tilde V$ mit Vorsicht zu geniessen.
In den zwei Bereichen  $\tilde g\lesssim\tilde g_{c_2}$ und $\tilde g\gtrsim\tilde g_{c_1}$
sollte der Verlauf der Flusslinien durch die beiden ''"au"seren'' Flusslinien in
Abb. \ref{fig.fluss_t0} richtig erfasst werden, da diese eine sinnvolle Erweiterung
der bisher gefundenen Fl"usse sind.
Im ''Zwischenbereich'' wird $\tilde g$ zu kleineren Werten flie"sen und $\tilde V$ damit
relevant bleiben, $\tilde \lambda$ kann dann aber wieder in den ''irrelevanten Bereich''
flie"sen, was durch den Verlauf der berechneten Flusslinie unterst"utzt wird (''mittlere''
Flusslinie in Abb. \ref{fig.fluss_t0}).
Allerdings scheint dieser Fluss physikalisch nicht sinnvoll zu sein, was
im n"achsten Kapitel mit einem einfachen Argument dargestellt wird.

%% file: result.tex
\newpage
\section{Resultate und Relevanz f"ur verwandte Systeme}\label{sec.results}
\setcounter{equation}{0}

In diesem Kapitel m"ochte ich die physikalischen Konsequenzen der
Resultate aus dem vorigen Kapitel etwas ausf"uhrlicher besprechen und
teilweise auch mit denen schon aus
h"oheren Dimensionen oder anderen Modellen bekannten vergleichen.

Bei $T=0$ im Fall ohne Dislokationen haben wir 
f"ur zunehmende Quantenfluktuationen einen Phasen"ubergang von einer
lokalisierten, in der die Unordnung unter der RG w"achst, zu einer
delokalisierten Phase, in der die Unordnung zu null renormiert wird,
gefunden.
Dieser Phasen"ubergang findet bei einem Wert von $\tilde g_{c_1}=3\pi/p^2$
in einer Dimension statt, wobei die Grenze der Phasen sowohl von der anf"anglichen
Unordnungsst"arke, als auch von der St"arke der Quantenfluktuationen abh"angt
(siehe Abb. \ref{fig.fluss_t0l0}).
Dieses Phasendiagramm wurde schon von Giamarchi und Schulz \cite{CDW:GiaSchulz88}
f"ur ein eindimensionales wechselwirkendes Elektronengas im Zufallspotential gefunden.
In diesem Paper entspricht der Parameter $K$, welcher
die Wechselwirkungsst"arke kontrolliert (siehe auch Abschnitt \ref{sec.TLM}),
dem Parameter $\tilde g^{-1}$ bei uns (bis auf einen numerischen Vorfaktor).

Nun k"onnen wir uns die Relevanz dieses Phasen"ubergangs f"ur Ladungsdichtewellen anschauen.
Der Hamiltonian ohne Unordnung hatte die Form
\begin{equation*}
 \hat {\cal H}=\int dx\, \frac{c}{2}\left[\left(\frac{v}{c}\right)^2\hat P^2(x)
+(\partial_x\hat\varphi(x))^2\right]\quad\,,
\quad \komm{\hat P(x)}{\hat \varphi(\str{x})}=\frac{\hbar}{\imath}\delta(x-\str{x})\,.
\end{equation*}
F"ur CDWs ist $c\approx \frac{\hbar v_F}{2\pi}$ (siehe Kapitel \ref{sec.klassmod})
und $v\approx v_F\left(\frac{m_e}{m^{\ast}}\right)^{1/2}$ \cite{CDW:Gruener94},
wobei $m_e$ die Elektronenmasse und $m^{\ast}$ die effektive Masse ist.
F"ur bekannte Ladungsdichtewellen--Systeme ist $m^{\ast}/m_e$ von der
Gr"o"senordnung $10^3$ \cite{CDW:Gruener88}. Vergleicht man nun die Gr"o"senordnungen
von $\left(\frac{v}{c}\right)^2\hat P^2=\tilde g^2\left(\frac{\hat P}{\hbar}\right)^2$
und $(\partial_x\hat\varphi(x))^2$ unter Beachtung der
Kommutatorrelation ($\hat P/\hbar$ und $\partial_x\hat\varphi$
haben dieselbe Gr"o"senordnung) findet man, dass der kinetische Term um einen Faktor $\sim 10^3$
kleiner ist als der elastische Term, was die klassische Behandlung des Modellls
in Kapitel
\ref{sec.klassmod} rechtfertigt. F"ur den Wert des Parameters $\tilde g$ hei"st das daher, 
dass dieser in realisierbaren CDW--Systemen im Bereich
von $10^{-2}-10^{-1}$ liegt, also wesentlich kleiner als $\tilde g_{c_1}$ ist.
Dies stimmt mit der bisherigen Beobachtung "uberein, dass die Unordung
in Ladungsdichtewellen immer relevant, d.h. die delokalisierte Phase nicht realisiert ist.

Betrachtet man den phase--slip Term bzw. das entsprechende Phasendiagramm,
findet man ebenfalls einen Phasen"ubergang bei $T=0$ bei einem Wert von $\tilde g_{c_2}=q^2\pi/8$
von einer Phase, in der die phase--slips irrelevant sind (A), zu einer
Phase, in der diese relevant sind (B), siehe Abb. \ref{fig.fluss_t0v0}.
Die Phasengrenze h"angt wiederum von den Startwerten der Quantenfluktuationen
und der phase--slip Wahrscheinlichkeiten ab.

Aufgrund der Struktur von ${\cal S}_{\theta}$, ist ein Vergleich mit dem Phasendiagramm
eines zweidimensionalen Sinus--Gordon Modells bei $T=0$ angebracht (siehe z.B. \cite{CM:Sachdev99}).
Dieses Modell hat zwei Phasen, in denen der nichtlineare Term relevant bzw.
irrelevant wird. In letzterem Fall flie"st der Koeffizient
auf die Fixpunktlinie $\tilde\lambda=0$, $\tilde g<\tilde g_{c_2}$,
d.h. auf gro"sen L"angenskalen kann das System durch eine freie Gau"ssche Theorie
beschrieben werden. Das Phasendiagramm stimmt demnach (qualitativ) mit
unserem "uberein.
Da, wie in Abschnitt \ref{sec.phaseslip} angedeutet, ein XY--Modell
mit Vortices auf ein Sinus--Gordon--Modell abgebildet werden kann, l"a"st sich dieses
Phasendiagramm auch mit einem Kosterlitz--Thouless "Ubergang \cite{CM:Kosterlitz74} vergleichen.

In unserem Fall hat dieser "Ubergang entscheidene Konsequenzen auf den
Lokalisierung--Delokalisierungs--"Ubergang, da
\begin{equation*}
 \pi/2=\tilde g_{c_2}<\tilde g_{c_1}=3\pi\,.
\end{equation*}
Im Bereich $\tilde g\lesssim \tilde g_{c_2}$ wird in jedem
Fall die Unordnung relevant und die Dislokationen irrelevant sein,
und f"ur $\tilde g\gtrsim \tilde g_{c_1}$ dominieren die phase--slips.
Allerdings ist das Verhalten des Systems auf gro"sen L"angenskalen
 im Zwischenbereich $\tilde g_{c_2}<\tilde g<\tilde g_{c_1}$
unklar. Die numerischen Ergebnisse sind
in diesem Bereich, aufgrund der im vorigen Kapitel geschilderten
Schwierigkeiten, nicht sehr verl"asslich. Die Untersuchung
lieferte, zumindest im Gebiet in dem die Flussgleichungen gelten,
dass die Dislokationen in diesem Zwischenbereich irrelevant werden,
was allerdings physikalisch nicht sehr plausibel ist, wie man sich
folgenderma"sen "uberlegen kann: Angenommen wir haben ein System in
welchem phase--slips relevant sind, d.h. die Korrelationen in
der Phase durch die Spr"unge zerst"ort werden. ''Schaltet'' man
nun schwache Unordnung ein, wird dies keinen Einfluss
auf die Phasenspr"unge haben, und diese auch auf gro"sen L"angenskalen
nicht unterdr"ucken.
Daher ist es wahrscheinlich, dass der Lokalisierungs--
Delokalisierungs--"Ubergang bei $\tilde g_{c_1}$
 verschwindet und der "Ubergangswert
von $\tilde g$ um immerhin einen Faktor $6$ zu $\tilde g_{c_2}$
 vermindert wird.
F"ur reale CDW--Systeme wird dies wiederum keine Konsequenzen haben,
da wie schon erw"ahnt, realistische Werte f"ur $\tilde g$ kleiner
als $\tilde g_{c_2}$ sind.

Bei endlichen Temperaturen werden sowohl die Unordnungsfluktuationen
(siehe Abb. \ref{fig.fluss_l0}) als auch die phase--slips irrelevant,
sodass nur thermische Fluktuationen auf gro"sen L"angenskalen eine Rolle spielen.
Im Fall der Unordnung ist dies klar, da der $\coth$--Term in der
Flussgleichung f"ur gro"se $\tilde t$ linear mit $\tilde t$ verl"auft
und somit, aufgrund des negativen Vorzeichens, $\tilde V$ gegen null geht.
Dieses Verhalten haben wir schon in Abschnitt \ref{sec.klassham}
anhand der Temperaturabh"angigkeit der Korrelationsfunktion gesehen.

F"ur die phase--slips kann dies folgenderma"sen erkl"art werden:
da der Fluss von $\tilde g$ f"ur gro"se $\tilde t$ sehr stark
abnimmt (siehe Def. von $F_x$, $F_{\tau}$ und den Verlauf von $f_x$ und
$f_{\tau}$ mit $\tilde t$), wird $\tilde\lambda(\tilde t)$
irgendwann auf jeden Fall die kritische Linie $\tilde g_{c_2}(\tilde t)$
(bei entsprechend gro"sen Startwerten von $\tilde g$) kreuzen und somit gegen null flie"sen.
Hierbei muss aber ber"ucksichtigt werden, dass die Flussgleichungen
nur f"ur kleine $\tilde\lambda$ g"ultig sind.

\subsection{Vergleich mit Fl"ussen der Unordnung in $d>1$}

\onefigure{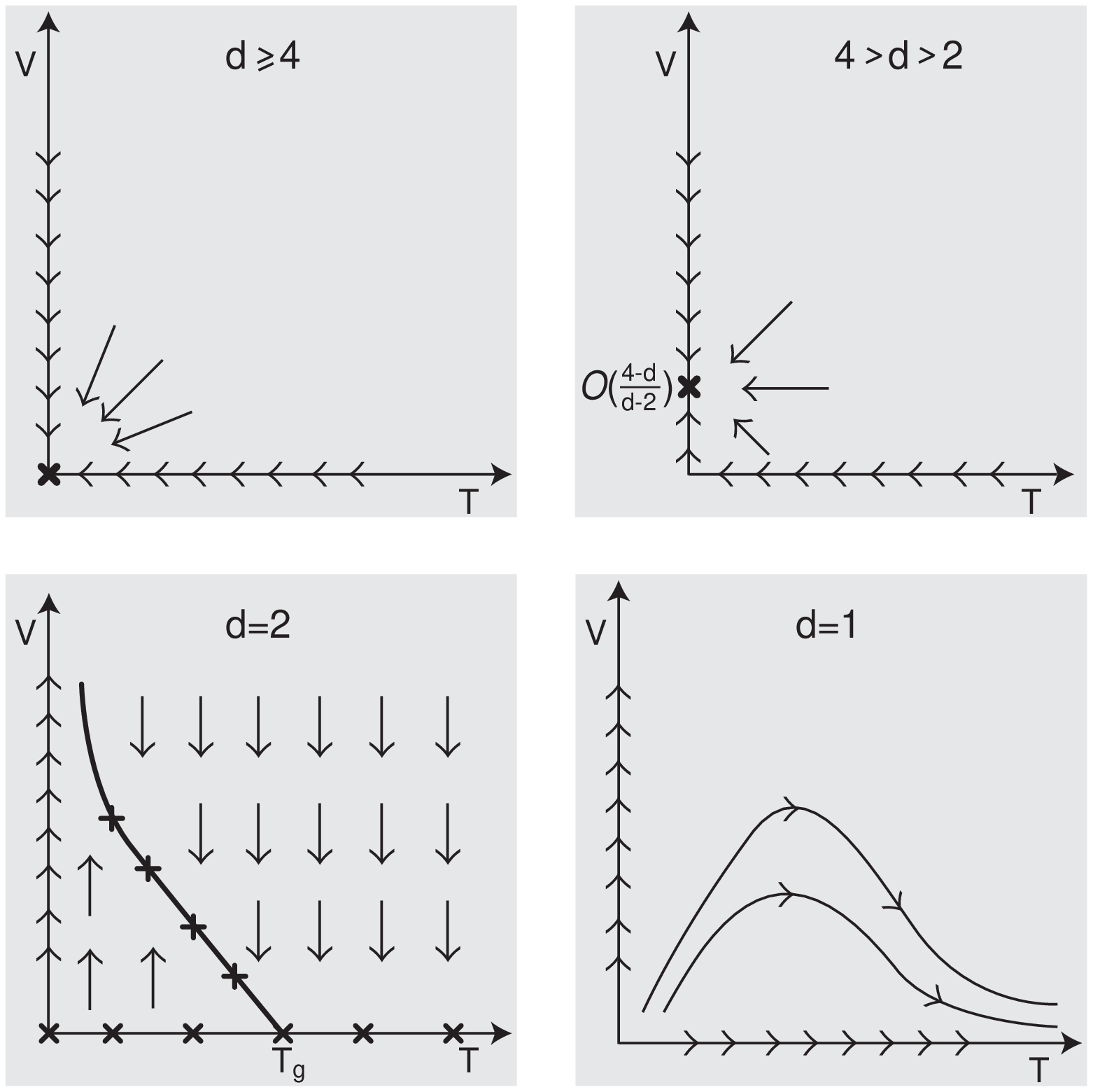}{0.8\linewidth}{Karikatur der Fl"usse in $d$ Dimensionen in der $T$--$V$--Ebene.}{fig.fluessedim}

Den Fluss in der $\tilde t$--$\tilde V$--Ebene ohne Dislokationen ($\tilde \lambda=0$)
und ohne Quantenfluktuationen ($\tilde g=0$), d.h. die Relevanz der Unordnung in
eindimensionalen elastischen Medien bei endlichen Temperaturen,
k"onnen wir mit dem Fluss in $d\geq 2$ vergleichen.

In vier oder mehr Dimensionen werden sowohl Unordnungs- als auch thermische Fluktuationen
irrelevant und man hat demnach einen Fixpunkt bei  ($T=0$, $V=0$). Die Paarkorrelationsfunktion
geht gegen einen konstanten Wert, d.h. es gibt keine Rauhigkeit.

In Dimensionen $2<d<4$ bleiben thermische Fluktuationen irrelevant,
der Fixpunkt ''wandert'' zu einem endlichen Wert von $V$
(wobei zu beachten ist, dass hier funktionale RG benutzt werden muss),
welcher in geeigneten Einheiten von der Ordnung ${\frac{4-d}{2-d}}$ ist.
Die Rauhigkeit ist hier logarithmisch.

In $d=2$ gibt es eine endliche "Ubergangstemperatur $T_g$ (Glas"ubergangstemperatur)
oberhalb welcher die Unordnung irrelevant wird, und unterhalb welcher sie
auf eine Fixpunktlinie
$V^{\ast}\propto (T_g-T)$ flie"st. Die Glasphase ist durch ein superrauhes
Verhalten gekennzeichnet (die Paarkorrelationsfunktion ist $\propto (\ln r)^2$).
Siehe dazu auch \cite{CDW:NatSchei}.

Die Fl"usse in $d\geq 4$, $2<d<4$, $d=2$ und in einer Dimension, in welcher,
wie im vorigen Kapitel gesehen, die thermischen Fluktuationen relevant
und die Unordnungsfluktuationen irrelevant werden, sind in Abb. \ref{fig.fluessedim}
zusammengestellt.

\subsection{Tomonaga--Luttinger Modell}\label{sec.TLM}

Aufgrund der Formulierung der Unordnung und der phase--slips in zwei verschiedenen
Feldern, bietet sich ein Vergleich mit dem Tomonaga--Luttinger (TL) Modell an,
mit welchem man (ohne Unordnung und Vortices) eindimensionale Elektronensysteme bei
tiefen Temperaturen ($k_B T\ll\epsilon_F$) beschreiben kann,
in welchen die Dispersionsrelation an der Fermikante linearisiert werden kann.
Der zun"achst fermionische Hamiltonian kann in diesem Fall
mit Hilfe der Bosonisierung (auf die ich hier nicht weiter eingehe) auf ein
TL Modell abgebildet werden. Die Wirkung wird meist in einem von zwei
zueinander ''dualen'' Feldern beschrieben, wobei das jeweils andere ausintegriert
wurde - je nachdem f"ur welche Korrelationen man sich interessiert.
Die autretenden Wirkungen sind au"serdem die gleichen, wie f"ur ein masseloses, skales Feld.

Identifiziert man
$\tilde g/\pi$ mit dem Luttingerparameter $K$, welcher ein Ma"s f"ur die
Dichte--Dichte--Wechselwirkung ist ($K=1$: keine WW, $K>1$: attraktive WW,
$K<1$: repulsive WW), und $v$ mit der renormalisierten Fermi--Geschwindigkeit,
kann man $\tilde {\cal S}_{\varphi,0}$ und $\tilde {\cal S}_{\theta,0}$
schreiben als
\begin{eqnarray}
 \tilde {\cal S}_{\varphi,0}&=&\frac{1}{2\pi K v}\int dx\,\int d\tau\,
   \left[(\partial_{\tau}\varphi)^2+v^2(\partial_x\varphi)^2\right]\quad\text{bzw.}\nn\\
 \tilde {\cal S}_{\theta,0}&=&\frac{K}{2\pi v}\int dx\,\int d\tau\,
   \left[(\partial_{\tau}\theta)^2+v^2(\partial_x\theta)^2\right]\nn
\end{eqnarray}
d.h. als dieselbe Wirkung, wenn man $K$ gegen $1/K$ austauscht.
Das Feld $\varphi$ hat im TL Modell die Bedeutung eines
''phononischen'' Verschiebungsfeldes und $\theta$ ist das zu
$\varphi$ ''duale'' Feld, hat aber keine direkte physikalische
Bedeutung. Der Gradient von $\theta$, also der zu $\varphi$ konjugierte
Impuls, misst die Differenz in der Dichte von links- und
rechts- laufenden Teilchen \cite{CM:Sachdev99}, also den Strom
(vgl. $\partial_x\varphi\propto \rho$).

Im TL Modell wird der Einfluss der Unordnung (insbes. \emph{backscattering})
ebenfalls durch einen Cosinus--Term mit zuf"alliger Phase beschrieben
(s. z.B. \cite{CDW:GiaSchulz88}). Vortices im Feld $\varphi$
k"onnen auch durch einen Cosinus--Term im Feld $\theta$
(oder umgekehrt) beschrieben werden \cite{CM:Sachdev99}.

%% file: summary.tex
\newpage
\section{Zusammenfassung und Ausblicke}

Ladungsdichtewellen--Systeme werden sp"atestens seit den 1970er
Jahren intensiv theoretisch untersucht. Das auch heute noch neue interessante
Resultate und Ergebnisse gefunden werden, spricht daf"ur, dass diese Systeme physikalisch
von hohem Interesse sind.
Den Beitrag dieser Arbeit dazu will ich hier nocheinmal zusammenfassen:

\bi
\item In Kapitel \ref{sec.klassmod} haben wir den Einfluss der Unordung auf die
Temperaturabh"angigkeit der Paarkorrelationsfunktion
$C(x)=\umw{\tmw{(\varphi(x)-\varphi(0))^2}}$ im Fall von schwacher ($V_i\rho_1\ll\hbar v_F c_{\textit{imp}}$)
und starker ($V_i\rho_1\gg\hbar v_F c_{\textit{imp}}$) Unordung untersucht:

\item Im Fall von schwacher Unordnung haben wir mit einem einfachen RG-Argument gefunden, dass
die Temperaturabh"angigkeit dieser Korrelationsfunktion bei kleinen Temperaturen konstant ist, d.h.
die Unordnung relevant ist, wohingegen bei hohen Temperaturen die Korrelationsfunktion linear mit $T$
verl"auft, d.h. die Unordnung irrelevant wird.
Eine quantitativ bessere Erfassung der Temperaturabh"angigkeit wurde mit Hilfe einer selbstkonsistenten
Rechnung im Rahmen einer Burgers--Gleichung erreicht.
Im Gegensatz zu den Resultaten von Villain und Fernandez~\cite{CDW:ViFer84} und Feigel'man~\cite{CDW:Feigel80}
haben wir somit ein Verhalten gefunden, welches bei endlichen Temperaturen gilt.
Zus"atzlich wurde das Ergebnis numerisch mit einer Monte--Carlo--Rechnung best"atigt.

\item Im Fall von starker Unordnung konnte die Korrelationsfunktion bei $T=0$,
unter Ber"ucksichtigung der ''geordneten'' Statistik der Positionen der
Verunreinigungen, exakt berechnet werden.

\item Der Hauptpunkt der Behandlung des klassischen Modells war die Untersuchung
der Dynamik. Insbesondere wurde die bisher nicht studierte Kriechdynamik in einer
Dimension untersucht und folgender Ausdruck f"ur die Kriechgeschwindigkeit
gefunden
\begin{equation*}
   v(E,T)\propto Te^{-c_B(T^{\ast}/T)}\sinh\left({\frac{c_EEL_c}{T}}\right)\,,
\end{equation*}
also im Gegensatz zu h"oheren Dimensionen ein analytisches Verhalten.
Diese Formel konnte mit Hilfe von numerischen Simulationen gut best"atigt werden
und l"a"st sich auf experimentelle Ergebnisse von S. Zaitsev--Zotov \cite{CDW:ZZ93}
anwenden.

\item Ab Kapitel \ref{sec.qmodel} wurde das klassische Modell zu einem
quantenmechanischen mit kinetischem Term, welcher in der klassischen
Beschreibung nicht ber"ucksichtigt wurde, und einen Term, der
Dislokationen bzw. Raum--Zeit--Vortices im Phasenfeld beschreibt, verallgemeinert.

\item In den Kapiteln \ref{sec.RG} und \ref{sec.results} wurden die Phasendiagramme dieses ph"anomenologischen
Modells bei tiefen Temperaturen im Fall von eindimensionalen Ladungsdichtewellen
ausf"uhrlich besprochen - mit folgenden Ergebnissen:

 \bi

  \item Bei $T=0$ gibt es f"ur die Unordnungsfluktuationen einen Lokalisierungs--Delokalisierungs--"Ubergang bei einem
  kritischen Wert der Quantenfluktuationen $\tilde g_{c_1}$.

  \item Au"serdem findet man f"ur die Dislokationen zwei Phasen in denen phase--slips irrelevant bzw. relevant
  werden. Der Phasen"ubergang findet bei einem kritischen Wert $\tilde g_{c_2}$ statt.
  Ist $\tilde g_{c_1}/\tilde g_{c_2}=24/(pq)^2$ gr"o"ser als 1 hat man drei Phasen:
  f"ur kleine $\tilde g$ eine lokalisierte Phase ($\tilde g<\tilde g_{c_1}$),
  eine Phase in der sowohl Dislokationen als auch Unordnungsfluktuationen
  irrelevant sind und eine Phase in der phase--slips relevant sind ($\tilde g>\tilde g_{c_2}$).

  \item \emph{Im Fall der Ladungsdichtewelle ist allerdings $\tilde g_{c_2}<\tilde g_{c_1}$,
  was zur Folge hat, dass der Lokalisierungs--Delokalisierungs"ubergang verschwindet
  und man direkt von der lokalisierten Phase in die phase--slip--bestimmte Phase
  "ubergeht (bei $\tilde g_{c_2}$).}

 \ei

\ei

Die Ergebnisse aus Kapitel \ref{sec.klassmod} wurden teilweise
zur Ver"offentlichung in \emph{Physical Review B} submittiert.
Ein Vordruck dieses Papers ist im Anhang zu finden.

\trenner

Abschlie"send will ich kurz darstellen, was ausgehend von den erzielten Resultaten u.a.
untersucht werden kann:
\bi
\item Bisher wurde der Einfluss der phase--slips, wie dies schon in \ref{sec.phaseslip}
angedeutet wurde, nicht im Fall eines extern angelegten Feldes untersucht.
Wenn phase--slips in dem zu betrachtenden System relevant sind,
wird dies Konsequenzen f"ur die Dynamik (z.B. die Leitf"ahigkeit) haben.

\item Man kann zus"atzlich noch dissipative Effekte ber"ucksichtigen und deren
Relevanz f"ur das System untersuchen.

\item Desweiteren k"onnen noch Leitf"ahigkeiten oder Effekte von Tunnelprozessen
untersucht werden.

\ei

%% file: appendix.tex
\newpage
\appendix

\renewcommand{\theequation}{\thesubsection.\arabic{equation}}

\section{Anhang}

\subsection{Symbole, Einheiten und Notationen}\label{app.sym}
\setcounter{equation}{0}

\subsubsection{verwendete Symbole}

\begin{tabular}{|p{0.2\textwidth}|p{0.5\linewidth}|p{0.15\textwidth}|}
  \hline
  {\bf Symbol} & {\bf Bedeutung} & {\bf Einheit} \\ \hline\hline
  $\lambda_T=v\hbar\beta$ & relativistische de Broglie Wellenl"ange & $L$ \\ \hline
  $L_c$ & Fukuyama--Lee--L"ange & $L$ \\ \hline
  $c$ & elastische Konstante & $E L^{2-d}$ \\ \hline
  $c_{\textit{imp}}^{-1}$ & mittlerer Abstand der Verunreinigungen & $L$ \\ \hline
  $V$ & Pinning--Potential--''Dichte'' & $EL^{-d/2}$ \\ \hline
  $\lambda$ & Wahrscheinlichkeit zur Ausbildung von Dislokationen & $EL^{-d}$ \\ \hline
  $k_F=Q/2$ & Fermi Wellenvektor & $L^{-1}$ \\ \hline
  $\Lambda$ & (UV-) Cutoff f"ur $k$ & $L^{-1}$ \\ \hline
  $(v)$, $v_F$ & (renormalisierte) Fermi Geschwindigkeit & $LT^{-1}$ \\ \hline
  $\rho$ & Teilchendichte & $ L^{-d}$\\ \hline
  $K$ & Luttinger Parameter & $1$ \\ \hline
  $\varphi, \theta$ & Phasenfeld und ''duales'' Feld & $1$ \\ \hline
  $\hat P$ & zu $\hat\varphi$ konjugierter Impulsoperator & $ML^{2-d}T^{-1}$ \\ \hline
  $\alpha(x)$ & Zufallsphase (gleichwahrscheinlich in $[0,2\pi[$ verteilt) & $1$ \\ \hline
  \hline
   $\tilde{\cal S}=\hbar^{-1}{\cal S}$ & ''reduzierte'' Wirkung & $1$ \\ \hline
   $\tilde{\cal H}=T^{-1}{\cal H}$ & ''reduziertet'' Hamiltonian & $1$ \\ \hline
   $\tilde t,\,\tilde g,\,\tilde \lambda,\,\tilde V$ & dimensionslose Parameter (Bedeutung s. \ref{sec.param}) & $1$ \\ \hline
  \hline
  $S_d=(2\pi)^d K_d=2\pi^{d/2}/\Gamma(d/2)$ & Oberfl"ache der d-dimensionalen Einheitssph"are  \\ \cline{1-2}
  ${\cal M}_n(\mathbb{K})$ ${\cal M}_{n\times m}(\mathbb{K})$ & Menge der $n\times n$
   bzw. $n\times m$ Matrizen "uber einem K"orper $\mathbb{K}$\\ \cline{1-2}
  $\Theta(x)$ & Stufenfunktion ($0$ f"ur $x<0$, sonst $1$) \\ \cline{1-2}
\end{tabular}

\subsubsection{Notation}

\begin{itemize}
 \item $\overline{\cal A}$: Unordnungsmittelwert
 \item $\left\langle {\cal A}\right\rangle$: thermischer Mittelwert
 \item $\langle\langle{\cal A}\rangle\rangle$: Kumulanten-Mittelwert
 \item $\gb{x}$: Gau"sklammern (Kapitel \ref{sec.strongpin})
 \item $\hat {\cal A}$: Operator / Fouriertransformierte
 \item $\mat{A}$: Matrix
 \item $\vec{v}$: Vektor ($\vec{v}\cdot\vec{w}$ ist das Skalarprodukt oder Matrixmultiplikation)
 \item $a|b$: $a$ modulo $b$
 \item $\imath$: imagin"are Einheit (um Verwechselungen mit $i$ zu vermeiden)
\end{itemize}

\trenner

{\bf Fouriertransformierte und Fourierreihen}
\bi
 \item Die \emph{Fouriertransformierte} ist in dieser Arbeit
  wie folgt definiert:

  Sei $f(\vec{r})\in \textit{L}^2(\mathbb{R}^d)$ und $\vec{r},\vec{k}\in\mathbb{R}^d$,
  dann ist die Fouriertransformierte von $f(\vec{r})$ gegeben durch
  \begin{equation}
   \hat f(\vec{k})\equiv\int_{\mathbb{R}^d} d^d\vec{r}\, f(\vec{r})e^{-\imath\vec{r}\cdot\vec{k}}
  \end{equation}
   mit der Umkehrtransformation
  \begin{equation}
   f(\vec{r})=\frac{1}{(2\pi)^d}\int_{\mathbb{R}^d} d^d\vec{k}\, \hat f(\vec{k})e^{\imath\vec{r}\cdot\vec{k}}\quad\text{kurz:}\quad\int_{\vec{k}} \hat f(\vec{k})e^{\imath\vec{r}\cdot\vec{k}}
\end{equation}
\item \emph{Fourierreihen}:

  Sei $f(\vec{r})\in \textit{L}^2(L^d)$ eine periodische $\textit{L}^2$--integrierbare Funktion, mit Periodizit"atsvolumen $L^d$. Sei $\vec{k}=\frac{2\pi}{L}(n_1,\ldots,n_d)$
mit $n_i\in\mathbb{Z}$, dann ist die Fourierreihe zu $f$ wie folgt definiert:
\begin{equation}
 f(\vec{r})=L^{-d}\sum\limits_{\vec{k}}e^{\imath \vec{k}\cdot\vec{r}} f_{\vec{k}}
\end{equation}
mit $\sum_{\vec{k}}\equiv\sum_{n_1,\ldots,n_d=-\infty}^{\infty}$ und den
Fourierkoeffizienten
\begin{equation}
 f_{\vec{k}}=\int_{L^d}d^d{\vec{r}}\,f(\vec{r})e^{\imath\vec{k}\cdot\vec{r}}\,.
\end{equation}
F"ur $L\rightarrow\infty$:
\begin{equation}\label{eq.FR_Linf}
  \sum_{\vec{k}}\rightarrow \int\frac{d^d\vec{k}}{(\Delta k)^d}\quad\text{mit}\quad \Delta k= 2\pi L^{-1}\,,
\end{equation}
d.h. man gelangt wieder zur Fouriertransformierten.
\ei

\subsection{n"utzliche Formeln}\label{app.formeln}
\setcounter{equation}{0}

\bi
\item Zur Berechnung der n-ten Momente der Abstandsverteilung in Kapitel
\ref{sec.strongpin} wird
\begin{equation}
\left.\frac{\partial^n}{\partial x^n}\right|_{x=1}\frac{e^{x-1}}{x}=
(-1)^n n!\left(\sum\limits_{k=1}^{n}\frac{(-1)^k}{k!}+1\right)
\label{eq.expdiff}
\end{equation}
benutzt.

\item Laurent-Reihe von $\csch(x)$ bei $x=0$ bis zur 6. Ordnung:
\begin{equation}\label{eq.csch}
 \csch(x)=\frac{1}{\sinh(x)}=\frac{1}{x}-\frac{x}{6}+\frac{7x^3}{360}-\frac{31x^5}{15120}+{\cal O}(x^7)
\end{equation}

\item Zur Berechnung der Korrelationsfunktionen und der RG  sind folgende Formeln hilfreich:
\begin{equation}\label{eq.kfsum}
 \sum\limits_{n=-\infty}^{\infty}\frac{1}{n^2+a^2}=\frac{\pi}{|a|}\coth(\pi |a|)
\end{equation}
\begin{equation}\label{eq.kfcint}
  \int_{\mathbb{R}}dx\, \frac{e^{\imath k x}}{x^2 +a^2}=\frac{\pi}{|a|}e^{-|a| |k|}
\end{equation}
\begin{equation}\label{eq.kfcsum}
  \sum\limits_{n=-\infty}^{\infty}\frac{\cos(n x)}{n^2+a^2}=
   \frac{\pi}{|a|}\frac{\cosh((\pi-x)a)}{\sinh(\pi |a|)}\quad 0\le x\le 2\pi
\end{equation}
\item au"serdem f"ur die RG des Unordnungsterms
\begin{equation}\label{eq.x2cosh}
 \int_0^1dx\,x^2\cosh(ax+b)=\frac{1}{a^3}\left[(2+a^2)\sinh(a+b)-
2(a\cosh(a+b)+\sinh(b))\right]
\end{equation}

\item Poisson'sche Summenformel
\begin{equation}\label{eq.PSF}
 \sum\limits_{m=-\infty}^{\infty} f(m)=\sum_{n=-\infty}^{\infty}\int_{\mathbb{R}}dx\, f(x)e^{-\imath 2\pi n x}
\end{equation}

\item n-dimensionales Gau"sintegral:
\begin{equation}\label{eq.nGI}
 \int_{\mathbb{R}^n} d^n\vec{r}\, e^{-\frac{1}{2}\vec{r}^T\cdot\mat{A}\cdot\vec{r}+\vec{v}^T\vec{r}}=(2\pi)^n(\det\mat{A})^{-1/2}e^{\frac{1}{2}\vec{v}^T\cdot\mat{A}^{-1}\cdot\vec{v}}\,,
\end{equation}
wobei $\mat{A}\in {\cal M}_n(\mathbb{R})$ positiv definit und symmetrisch ist und $\vec{v}\in\mathbb{R}^n$.

\ei

\trenner

F"ur ein gau"sverteiltes Feld $u$ mit den Mittelwerten $\langle u\rangle =0$
und $\langle u^2\rangle\ne 0$ folgt mit dem Wick-Theorem:
\begin{eqnarray}
 \tmw{u^{2n}} &=&\frac{(2n)!}{2^n n!}\langle u^2\rangle^n\nonumber\\
 \tmw{u^{2n+1}} &=&0\nonumber
\end{eqnarray}
und daher
\begin{equation}
 \tmw{e^{g(u)} }=e^{\tmw{ g^2(u)}/2}\,.
\end{equation}
Insbesondere gilt:
\begin{eqnarray}\label{eq.sincosMW}
 \tmw{ \cos(p u)} &=&e^{-p^2\langle u^2\rangle/2}\\
 \tmw{\sin(p u)}  &=&0\nonumber
\end{eqnarray}

\subsection{Tilt Symmetrie}\label{app.tiltsym}
\setcounter{equation}{0}
In diesem Anhang wird gezeigt, dass aufgrund der sogenannten \emph{statistischen Tilt--Symmetrie}
$\varphi\rightarrow\varphi+ax$ des Modells (\ref{eq.xyrmodel}) die elastische Konstante $\hbar v_F$
nicht renormiert wird.
Dazu schaut man sich den \emph{Replica--Hamiltonian} zu (\ref{eq.xyrmodel}) an
(siehe dazu Abschnitt \ref{sec.replica}):
\begin{eqnarray}
   {\cal H}_n&=&\sum\limits_{\alpha,\beta=1}^n\int\limits_0^Ldx\,\left\{
   \frac{1}{2}\hbar v_F\left(\frac{\partial}{\partial x}\varphi_{\alpha}
   \right)^2\delta_{\alpha\beta}\right.
   - \frac{(\hbar v_F)^2\sigma}{2T}\left(\frac{\partial}{\partial x}\varphi_{\alpha}\right)
   \left(\frac{\partial}{\partial x}\varphi_{\beta}\right)\\
   &&-\left.\frac{V^2}{4T}\cos{(\varphi_{\alpha}-\varphi_{\beta})}\right\}\,.\nonumber
   \label{eq.replham}
\end{eqnarray}
In dieser Darstellung sieht man, dass unter o.g. Symmetrietransformation nur ein Gradiententerm
generiert wird, der durch eine Umdefinition von $\varphi$ wieder ''absorbiert'' werden kann.
Nun zeigen wir dies explizit und addieren einen Term $-g_0\int_0^L(\partial\varphi/\partial x)dx$ zu ${\cal H}$,
wie er unter einem Renormierungsgruppenschritt erzeugt werden w"urde,
und schreiben den Replica--Hamiltonian wie folgt auf:
\begin{equation}
 {\cal H}_n=\int\limits_0^Ldx\,\left(\frac{1}{2}\vec{\phi}\cdot\mat{A}\cdot\vec{\phi}^T-
 \vec{\phi}\cdot\vec{g_0}-\frac{V^2}{4T}\cos(\varphi_{\alpha}-\varphi_{\beta})\right)
\end{equation}
wobei folgende Vektoren und Matrizen verwendet wurden:
\begin{eqnarray}
 \boxed{1}&\equiv&\begin{pmatrix} 1 & \ldots & 1\\ \vdots & \ddots & \vdots\\ 1&\ldots&1\end{pmatrix}\in {\cal M}_n(\mathbb{R})\nn\\
 E_n\;&:&\quad\text{n-dimensionale Einheitsmatrix}\nn\\
 \mat{A}&\equiv&\hbar v_F E_n-\frac{\sigma (\hbar v_F)^2}{T}\boxed{1}\nn\\
 \vec{\phi}&\equiv&\left(\partial_x\varphi_1,\ldots,\partial_x\varphi_n\right)\nn\\
 \vec{g_0}&\equiv&(g_0,\ldots,g_0)^T\nn
\end{eqnarray}
Mit Hilfe der Identit"aten
\begin{eqnarray}
 (aE_n+b\boxed{1})^{-1}&=&a^{-1}E_n-\frac{b}{a(a+nb)}\boxed{1} \;,\; a\neq -nb\quad\text{und}\nn\\
 -\frac{1}{2}\vec{\phi}\cdot\mat{A}\cdot\vec{\phi}^T&=&-\frac{1}{2}\left((\vec{\phi}+\vec{w}^T)\cdot\mat{A}
\cdot(\vec{\phi}^T+\vec{w})-\vec{g_0}\cdot\mat{A}^{-1}\cdot\vec{g_0}^T\right)\; ,\;\vec{w}=-\mat{A}^{-1}\cdot\vec{g_0}^T\nn
\end{eqnarray}
erh"alt man
\begin{equation}
 {\cal H}_n=-\frac{1}{2}g_0^2 L C(n)+\int dx\, \left(\vec{\tilde\phi}\cdot\mat{A}\cdot\vec{\tilde\phi}^T-
 \frac{V^2}{4T}\cos(\varphi_{\alpha}-\varphi_{\beta})\right)\,,
\end{equation}
wobei $C(n)=n/(\hbar v_F)+\frac{\sigma n^2}{T -n\sigma\hbar v_F}$ und
$\vec{\tilde\phi}\equiv 2^{-1/2}(\vec{\phi}+\vec{w}^T)$.

Die ''renormierte'' elastische Konstante $\hbar\tilde v_F$ folgt dann aus der freien Energie:
\begin{equation}
   (\hbar\tilde v_F)^{-1}=-\frac{1}{L}
   \frac{\partial^2\bar F}{\partial g^2_0}\Big|_{g_0=0}\,.
   \label{eq.renelkon}
\end{equation}
Mit $-\frac{1}{L}\frac{\partial^2}{\partial g^2_0}\Big|_{g_0=0}(-T{\cal Z}_n)=C(n)$
und $\bar F=-T\lim_{n\rightarrow 0}\frac{1}{n}({\cal Z}_n-1)$ bekommt man schlie"slich
\begin{equation}
 (\hbar\tilde v_F)^{-1}=\lim_{n\rightarrow 0}\frac{1}{n} C(n)=(\hbar v_F)^{-1}
\end{equation}
d.h. die elastische Konstante wird tats"achlich nicht renormiert.

\subsection{Korrelationsfunktion f"ur gau"ssches Modell}\label{app.cfelmod}
\setcounter{equation}{0}
In diesem Abschnitt wird die Berechnung der Paarkorrelationsfunktion
$\tmw{(\phi(0)-\phi(\vec{r}))^2}=2\tmw{\phi^2(0)}-2\tmw{\phi(0)\phi(\vec{r})}$
f"ur ein einfaches gau"ssches bzw. elastisches Modell
\begin{equation}
 H_0=\int_{L^d} d^d\vec{r}\,\frac{c}{2}\big(\nabla_{\vec{r}}\phi(\vec{r})\big)^2
\end{equation}
kurz beschrieben.
Dazu wird zun"achst $H_0/T$ in der Zustandssumme
um einen linearen Quellterm $\tilde H_Q=-\int d^d\vec{r}\,j(\vec{r})\phi(\vec{r})$
erweitert. Geht man zur Fourierdarstellung $\phi(\vec{r})=L^{-d}\sum_{\vec{k}} e^{\imath\vec{k}\cdot\vec{r}}\phi_{\vec{k}}$
(entsprechend f"ur $j$) "uber, l"a"st sich die erweiterte Hamiltonfunktion schreiben als ($J(\vec{k})=c\vec{k}^2$)
\begin{equation}
 \tilde H=H_0/T+\tilde H_Q=L^{-d}\sum_{\vec{k}}\left(\frac{J(\vec{k})}{2T}|\phi_{\vec{k}}|^2-\phi_{\vec{k}}j_{-\vec{k}}\right)\,.
\end{equation}
Mit $\phi_{\vec{k}}=a_{\vec{k}}+\imath b_{\vec{k}}$ und
$j_{\vec{k}}=\alpha_{\vec{k}}+\imath\beta_{\vec{k}}$ l"a"st sich die Zustandssumme
als Integral "uber die reelen Koeffizienten $a_{\vec{k}}$ und $b_{\vec{k}}$ darstellen:
\begin{eqnarray}
Z[j]&=&\int\pd[\phi]\,e^{-\tilde H}\nn\\
    &=&\prod\limits_{\vec{k}}\int_{\mathbb{R}}da_{\vec{k}}\int_{\mathbb{R}}db_{\vec{k}}\,
    e^{-L^{-d}\left(\frac{J(\vec{k})}{2T}(a_{\vec{k}}^2+b_{\vec{k}}^2)-a_{\vec{k}}\alpha_{\vec{k}}-b_{\vec{k}}\beta_{\vec{k}}\right)}\,.
\end{eqnarray}
Nach einer quadratischen Erg"anzung des Exponeten:
$a_{\vec{k}}^2-\frac{2T}{J(\vec{k})}a_{\vec{k}}\alpha_{\vec{k}}=
  \left(a_{\vec{k}}-\frac{T\alpha_{\vec{k}}}{J(\vec{k})}\right)^2-\left(\frac{T\alpha_{\vec{k}}}{J(\vec{k})}\right)^2$
(analog f"ur $b_{\vec{k}}$)
und der Substitution $\tilde a_{\vec{k}}=a_{\vec{k}}-\frac{T\alpha_{\vec{k}}}{J(\vec{k})}$ bzw.
$\tilde b_{\vec{k}}=b_{\vec{k}}-\frac{T\beta_{\vec{k}}}{J(\vec{k})}$ sind die Integrale "uber
$\tilde a_{\vec{k}}$ bzw. $\tilde b_{\vec{k}}$ nur noch
einfache Gau"sintegrale, und man erh"alt ($Z_0\equiv Z[j\equiv0]$):
\begin{equation}
Z[j]=Z_0\underbrace{e^{\frac{T}{2L^d}\sum_{\vec{k}}j_{\vec{k}}j_{-\vec{k}}J^{-1}(\vec{k})}}_{\tilde Z[j]}\,.
\end{equation}
Schreibt man den Exponenten in $\tilde Z[j]$ wieder im realen Raum auf, l"a"st sich die Korrelationsfunktion
$\tmw{\phi(\str{\vec{r}})\phi(\vec{r})}$ als Funktionalableitung von $\tilde Z[j]$ schreiben:
\begin{eqnarray}
\tmw{\phi(\vec{r})\phi(\str{\vec{r}})}&=&\frac{\delta^2}{\delta j_{\vec{r}}\delta j_{\str{\vec{r}}}}\tilde Z[j]=TJ^{-1}(\vec{r}-\str{\vec{r}})\nn\\
&=&TL^{-d}\sum_{\vec{k}}e^{\imath \vec{k}\cdot(\vec{r}-\str{\vec{r}})}J^{-1}(\vec{k})\,.
\end{eqnarray}
Damit folgt f"ur $\tmw{(\phi(\str{\vec{r}})-\phi(\vec{r}))^2}$ sofort:
\begin{equation}
\tmw{(\phi(\str{\vec{r}})-\phi(\vec{r}))^2}=2TL^{-d}\sum_{\vec{k}}\frac{1-e^{\imath \vec{k}\cdot(\vec{r}-\str{\vec{r}})}}{c\vec{k}^2}\,.
\end{equation}
Insbesondere f"ur $d=1$ und $c=\hbar v_F$ folgt:
\begin{eqnarray}
 \tmw{(\phi(0)-\phi(x))^2}&=&\frac{2T}{L\hbar v_F}\sum_{k}\frac{1-e^{\imath k\cdot x}}{k^2}\nn\\
 &=&\frac{2T}{\hbar v_F}\int\limits_{-\Lambda}^{\Lambda}\frac{dk}{2\pi}\,\frac{1-\cos(k x)}{k^2}\nn\\
 &=&\frac{4T}{\pi\hbar v_F}\int_0^{\Lambda}\sin^2(kx/2)\underset{\Lambda\rightarrow\infty}{=}\frac{T}{\hbar v_F}|x|\,.
\end{eqnarray}
Beim "Ubergang zum Integral wurde (\ref{eq.FR_Linf}) benutzt.

\subsection{Berechnung des Integrals aus \ref{sec.strongpin}}\label{app.cfint}
\setcounter{equation}{0}

Berechnung des Integrals:
\begin{equation}
 \textit{Cf}=\int_{0}^{\infty}dx\,e^{-x}\left(\frac{x}{2\alpha}-\gb{\frac{x}{2\alpha}}\right)^2\,.
\end{equation}
Der quadratische Term in $\textit{Cf}$ wird zun"achst erweitert, was auf die folgenden drei
(konvergenten) Integrale f"uhrt:
\begin{eqnarray}
 I_1&\equiv &\frac{1}{4\alpha^2}\int_{0}^{\infty}dx\,e^{-x}x^2=\frac{1}{2\alpha^2}\nonumber\\
 I_2&\equiv & -\frac{1}{\alpha}\int_{0}^{\infty}dx\,e^{-x}x\gb{\frac{x}{2\alpha}}
     = -\frac{1}{\alpha}\sum_{k=1}^{\infty} \int\limits_{(2k-1)\alpha}^{(2k+1)\alpha}dx\,e^{-x}x\gb{\frac{x}{2\alpha}}\\
 I_3&\equiv &\int_{0}^{\infty}dx\,e^{-x}\gb{\frac{x}{2\alpha}}^2
     =\sum_{k=1}^{\infty} \int\limits_{(2k-1)\alpha}^{(2k+1)\alpha}dx\,e^{-x}\gb{\frac{x}{2\alpha}}^2 \nonumber\,,
\end{eqnarray}
so dass $\textit{Cf}=I_1+I_2+I_3$.
F"ur $(2k-1)\alpha\le x\le (2k+1)\alpha$ gilt: $\gb{\frac{x}{2\alpha}}\equiv k$, d.h. die
Gau"sklammern in $I_2$ und $I_3$ k"onnen durch $k$ ersetzt werden:
\begin{eqnarray}
 I_2&=&-\frac{1}{\alpha}\sum_{k=1}^{\infty} k\int\limits_{(2k-1)\alpha}^{(2k+1)\alpha}dx\,e^{-x}x\nonumber\\
 I_3&=&\sum_{k=1}^{\infty}k^2 \int\limits_{(2k-1)\alpha}^{(2k+1)\alpha}dx\,e^{-x}\,.
\end{eqnarray}
Diese zwei einfachen Integrale liefern
\begin{eqnarray}
 \int\limits_{(2k-1)\alpha}^{(2k+1)\alpha}dx e^{-x}x&=&2e^{-2k\alpha}((1+2k\alpha)\sinh(\alpha)-\alpha\cosh(\alpha))\nonumber\\
 \int\limits_{(2k-1)\alpha}^{(2k+1)\alpha}dx e^{-x}&=&2e^{-2k\alpha}\sinh(\alpha)\,,
\end{eqnarray}
womit nur noch Summen der Form
\begin{equation}
 \sum_{k=1}^{\infty}ke^{-2k\alpha} \quad\text{und}\quad \sum_{k=1}^{\infty}k^2e^{-2k\alpha}\,,
\end{equation}
berechnet werden m"ussen. Da diese aber nur Ableitungen der \emph{geometrischen Reihe} sind,
sind diese auch einfach zu berechnen:
\begin{eqnarray}
 \sum_{k=1}^{\infty}ke^{-2k\alpha}&=&\frac{e^{-2\alpha}}{(1-e^{-2\alpha})^2}\nonumber\\
 \sum_{k=1}^{\infty}k^2e^{-2k\alpha}&=&\frac{e^{-2\alpha}+e^{-4\alpha}}{(1-e^{-2\alpha})^3}\,.
\end{eqnarray}
Man erh"alt also letztendlich f"ur $\textit{Cf}$:
\begin{equation}
 \textit{Cf}=\frac{1}{2\alpha}\left(\frac{1}{\alpha}-\frac{1}{\sinh(\alpha)}\right)\,.
\end{equation}
Unter Benutzung von (\ref{eq.csch}) erh"alt man folgende N"aherungsformel:
\begin{equation}
 \textit{Cf}=\frac{1}{12}-\frac{7}{720}\alpha^2+\frac{31}{30240}\alpha^4+{\cal O}(\alpha^6)\,.
\end{equation}

\subsection{Statistik geordneter Zufallsgr"o"sen}\label{app.ostat}
\setcounter{equation}{0}

Seien $X_1,X_2,\ldots ,X_N$ Zufallsgr"o"sen mit der kumulierten Verteilungsfunktion (cdf)
$F_i(x)=\Pr{X_i\le x}$ wobei im folgenden $F(x)=F_1(x)=\ldots=F_N(x)$ gelten soll.
$X_{(1)},X_{(2)},\ldots,X_{(N)}$ seien die geordneten Zufallsgr"o"sen ($X_{(1)}\le X_{(2)}\le\ldots\le X_{(N)}$)
mit der cdf
\begin{equation}
 P_i(x)=\Pr{X_{(i)}\le x}=\sum_{k=i}^{N}\left({N\atop k}\right)F^k(x)(1-F(x))^{N-k}
\end{equation}
was bedeutet, da"s mindestens $i$ der $X_k$ kleiner oder gleich $x$ sind.
In den Spezialf"allen $i\in 1,N$ gilt offensichtlich
\begin{eqnarray}
 P_N(x)&=&\Pr{X_{(N)}\le x}=\Pr{\mathrm{alle}\, X_i\le x}=F^N(x)\nonumber\\
 P_1(x)&=&\Pr{X_{(1)}\le x}=1-\Pr{\mathrm{alle}\, X_i > x}=1-(1-F(x))^N
\end{eqnarray}
$P_i(x)$ l"a"st sich auch schreiben als $P_i(x)=I(i,N-i+1;P_i(x))$, wobei
$I(a,b;x)$ die normierte unvollst"andige Beta-Funktion ist, definiert durch
\begin{eqnarray}
 I(a,b;x)&=&\frac{B(a,b;x)}{B(a,b)}\nonumber\\
 B(a,b;x)&=&\int_0^x t^{a-1}(1-t)^{b-1}\,dt
\end{eqnarray}
wobei $B(a,b)=B(a,b;1)$ die Beta-Funktion ist. Die
Wahrscheinlichkeitsdichte-Funktion (pdf) $p_i(x)$ f"ur $X_{(i)}$
kann nun auf zwei verschiedene Arten berechnet werden. Zum einen
l"a"st sich die Wahrscheinlichkeit ein $X_{(i)}$ in einem kleinen
Intervall $[x,x+\delta x]$ zu finden, schreiben als $p_i(x)\delta
x=P_i(x+\delta x)-P_i(x)+{\cal O}({\delta x}^2)$. F"ur $\delta
x\rightarrow 0$ ist also $p_i(x)=P_i^{\prime}(x)$, womit sofort
folgt
\begin{equation}
 p_i(x)=\frac{1}{B(i,N-i+1)}F^{i-1}(x)(1-F(x))^{N-i}f(x)
 \label{os1}
\end{equation}
wobei $f(x)=F^{\prime}(x)$ die pdf der $X_k$ ist.
Zum anderen kann man sich auch eine mehr anschauliche Herleitung f"ur $p_i(x)$ "uberlegen:
Wieder soll die Wahrscheinlichkeit f"ur $x<X_{(i)}\le x+\delta x$ bestimmt werden, d.h. die
Wahrscheinlichkeit daf"ur, da"s
\begin{itemize}
 \item f"ur $i-1$ der $X_k$ gilt: $X_k\le x$
 \item ein $X_k\in [x,x+\delta x]$ ist und
 \item f"ur $N-i$ der $X_k$ gilt: $X_k> x+\delta x$
\end{itemize}
F"ur diese Aufteilung gibt es offensichtlich $\frac{N!}{(i-1)! 1! (N-i)!}=
i\left({N \atop N-i}\right)=\frac{1}{B(i,N-i+1)}$ M"oglichkeiten und die Wahrscheinlichkeit f"ur
eine solche Aufteilung ist
\begin{eqnarray}
 &&(\Pr{X_k\le x})^{i-1}\cdot \Pr{x<X_k\le x+\delta x}\cdot (\Pr{X_k>x+\delta x})^{N-i}=\nonumber\\
 &&F^{i-1}(x)\cdot (F(x+\delta x)-F(x))\cdot (1-F(x+\delta x))^{N-i}\,.
\end{eqnarray}
Ingesamt erh"alt man also
\begin{eqnarray}
 \Pr{x<X_{(i)}\le x+\delta x}&=&\frac{1}{B(i,N-i+1)}\cdot\nonumber\\
  &&F^{i-1}(x)\cdot(F(x+\delta x)-F(x))\cdot (1-F(x+\delta x))^{N-i}
\end{eqnarray}
und f"ur $\delta x\rightarrow 0$ folgt wieder der Ausdruck (\ref{os1}) f"ur $p_i(x)$.

Um sp"ater die pdf f"ur Differenzen (oder auch anderen Funktionen) von georneten Zufallsgr"o"sen zu bestimmen,
braucht man noch die gemeinsame pdf f"ur zwei Gr"o"sen $X_{(i)}$ und $X_{(j)}$ mit $x<X_{(i)}\le x+\delta x$ und
$y<X_{(j)}\le y+\delta y$,  wobei $1\le i< j\le N$ gelten soll.
Diese sei mit $p_{ij}(x,y)$ bezeichnet. Geht man analog zu der oben beschriebenen anschaulichen Herleitung vor,
erh"alt man direkt
\begin{equation}
 p_{ij}(x,y)=\underbrace{\frac{N!}{(i-1)!(j-i-1)!(N-j)!}}_{\equiv C_{ij}}F^{i-1}(x)f(x)(F(y)-F(x))^{j-i-1}f(y)(1-F(y))^{N-j}
\end{equation}
da hier gilt
\begin{itemize}
 \item $(i-1)$ $X_k$ sind $\le x$
 \item ein $X_k$ ist in $[x,x+\delta x]$
 \item $(j-i-1)$ $X_k$ sind in $[x+\delta x,y]$
 \item ein $X_k$ ist in $[y,y+\delta y]$ und
 \item $(N-j)$ $X_k$ sind $> y+\delta y$
\end{itemize}
d.h. es gibt $\frac{N!}{(i-1)!1!(j-i-1)!1!(N-j)!}$ Verteilungsm"oglichkeiten, jede mit der Wahrscheinlichkeit
$F^{i-1}(x)(F(x+\delta x)-F(x))(F(y)-F(x+\delta x))^{j-i-1}(F(y+\delta y)-F(y))(1-F(y+\delta y))^{N-j}$.
Diese Herleitung kann nat"urlich auch auf mehr als zwei Zufallsgr"o"sen verallgemeinert werden.

Jetzt kann die pdf der Gr"o"se $\Delta_{ij}=X_{(j)}-X_{(i)}$ bestimmt werden. Dazu setzen wir $y=\delta_{ij}+x$
(d.h. $\delta_{ij}$ ist der Abstand von $x$ und $y$). Die pdf $\tilde p(\delta_{ij})$ von
$\Delta_{ij}$ ergibt sich sofort aus
\begin{equation}
 \tilde p(\delta_{ij})=\int_{\Bbb R}dx\,p_{ij}(x,\delta_{ij}+x)
\end{equation}
Jetzt berechnen wir $\tilde p(\delta_{ij})$ noch in folgendem Spezialfall:
 $f(x)=\left\{ {1\,,\quad 0\le x\le 1 \atop 0\,,\quad \mathrm{sonst}}\right.$ , d.h.
im folgenden seien alle Funktionen nur auf $[0,1]$ von Null verschieden.
F"ur die Wahrscheinlichkeitsdichte-Funktionen gilt dann
\begin{eqnarray}
 p_i(x)&=&\frac{1}{B(i,N-i+1)}x^{i-1}(1-x)^{N-i}\nonumber\\
 p_{ij}(x)&=&C_{ij}x^{i-1}(y-x)^{j-i-1}(1-y)^{N-j}\nonumber\\
 \tilde p(\delta_{ij})&=&C_{ij}\int_0^{1-\delta_{ij}}dx\,x^{i-1}\delta_{ij}^{j-i-1}(1-x-\delta_{ij})^{N-j}
\end{eqnarray}
Man erh"alt also
\begin{equation}
 \tilde p(\delta_{ij})=\frac{1}{B(j-i,N-j+i+1)}\delta_{ij}^{j-i-1}(1-\delta_{ij})^{n-j}
\end{equation}
und insbesondere im Fall $j=i+1$ ($\delta_i\equiv \delta_{i,i+1}$):
\begin{equation}
 \tilde p(\delta_i)=\frac{1}{B(1,N)}(1-\delta_i)^{N-1}=N(1-\delta_i)^{N-1}
\end{equation}


\subsection{Programm zur Simulation der Dynamik}\label{app.simprog}
\setcounter{equation}{0}

Hier ein Listing des Haupteils des Simulationsprogramms:

\begin{verbatim}
 //************ program begins ... *****************
 //open output-file

 f=fopen(output,"wb");
 fprintf(f,"N=%d, Samples=%d, iseed=%d\n",N,Nsamp,iseeda);
 fprintf(f,"time=%le, dt=%le, T=%le\n",time,dt,T);

 for(k=0;k<nE;k++) // loop over driving force
  {if(nE==1) E=(bE+vE)/2;
   else E=vE+k*(bE-vE)/(nE-1);    // electric field or driving force

   for(j=0;j<Nsamp;j++)
    {iseeds=iseeda+j*108;
     x=ran2(iseeds); //only for initialzation
     //Initialization
     for(n=0;n<N;n++)
      {phi[n]=0.0;   //initially surface is flat

       // random phase: a[n]
       a[n]=M_2PI*ran2();
      }
     Jsav=0.0; // to calculate average J_cdw
     avc=0;    //counter for "
     //TIME EVOLUTION
     for(i=0;i<nstep;i++)
      {for(n=0;n<N;n++)
        { np=n+1;if(np==N) np=0;
          nm=n-1;if(nm<0) nm=N-1;

          noise[n]=Tsdev*sqrt(-2.0*log(ran2()))*cos(M_2PI*ran2());

          //calc d(phi)/dt
          deri1[n]=(phi[np]+phi[nm]-2*phi[n])+sin(phi[n]-a[n])
                   //discrete laplacian         pinning
                        -E             +noise[n];
                   //driving force   temperature
        }  //end n
       for(n=0;n<N;n++) phit[n]=phi[n]+dt*deri1[n];

       //modified RK:
       for(n=0;n<N;n++)
        { np=n+1;if(np==N) np=0;
          nm=n-1;if(nm<0) nm=N-1;
          deri2[n]=(phit[np]+phit[nm]-2*phit[n])+sin(phit[n]-a[n])-E+noise[n];
        }  //end n
       for(n=0;n<N;n++)
        {phi[n]=(phi[n]+phit[n])/2;
         deri[n]=(deri1[n]+deri2[n])/2;
        }
       for(n=0;n<N;n++) phi[n]=phi[n]+dt*deri[n];

       if((i>=nskip) && (i%ical==0)) // calculate the average J_cdw
        {Jav=0.0;
         for(n=0;n<N;n++) Jav+=deri[n];
         Jav=Jav/N;
         Jsav+=Jav;
         avc++;
        }
      }//end time
     Jsav=Jsav/avc;
     savj[j]=Jsav;
    }//end sample-loop
   //  sample average for J_cdw
   J1=0.0; // final average
   J2=0.0;
   for(j=0;j<Nsamp;j++)
    {J1+=savj[j];J2+=savj[j]*savj[j];}
   // calculate error bars due to sample average
   if(Nsamp>1) err=sqrt((J2-J1*J1/Nsamp)/(Nsamp*(Nsamp-1)));
   else err=0.0;
   fprintf(f,"%le %le %le\n",E,J1/Nsamp,err);
  }  // end of loop over driving force
 fclose(f);
 //finish

\end{verbatim}

{\bf Bemerkungen:}
\bi
 \item Die Funktion \texttt{ran2()} liefert eine Zufallszahl
       im Intervall $[0,1]$ (entnommen aus~\cite{NUM:NR})
 \item  \texttt{Tsdev=sqrt(2.0*T);} ist die Standardabweichung
  f"ur die gau"sverteilten Zufallszahlen $\eta(T)$
\ei

\newpage


\subsection{Teilpublikation}\label{app.publ}

Andreas Glatz and Mai Suan Li, {\it Collective dynamics of
one-dimensional charge density waves}, Phys. Rev. {\bf B} 64,
184301 (2001). (Teile von Kapitel 2)

\vspace{1cm}

Andreas Glatz and Thomas Nattermann, {\it One-Dimensional
Disordered Density Waves and Superfluids: The Role of Quantum
Phase Slips and Thermal Fluctuations}, Phys. Rev. Lett. 88, 256401
(2002). (basierend auf Kapitel 3 und 4)



%% file: dank.tex
\section*{Danksagung}
\addcontentsline{toc}{section}{Danksagung}

\vspace{3cm}

Ich m"ochte allen danken, die zum Gelingen dieser Arbeit beigetragen haben,
insbesondere:

\bi
\item Prof. Dr. T. Nattermann, der die Arbeit ausgezeichnet betreut hat,
und stets, auch aus der Ferne, zu Diskussionen bereit war und mit seinen
Hinweisen neue Anst"o"se gab,

\item Dr. Mai Suan Li, der mir hilfreiche Hinweise bei den numerischen
Untersuchungen gegeben hat,

\item Dr. Stefan Scheidl, Dr. Bernd Rosenow und Simon Bogner, die
mir mit ihrem fachkundigen Rat zur Seite standen

\item und schlie"slich den derzeitigen und ehemaligen Mitbenutzern
des Zimmers 102 (Dr. Peter Arndt, Simon Bogner und Frank Kr"uger)
f"ur die gute Atmosph"are w"ahrend des vergangenen Jahres.

\ei

%% file: erklaerung.tex
\section{Erkl"arung}
\addcontentsline{toc}{section}{Erkl"arung}
\vspace{2cm}

Hiermit best"atige ich, dass ich meine Diplomarbeit selbst"andig angefertigt und keine
anderen als die angegebenen Quellen und Hilfsmittel benutzt sowie Zitate kenntlich gemacht habe.

\vspace{4cm}

K"oln, den 20. Juni 2001\hspace{.5\linewidth}\\
\parbox{\linewidth}{\raggedleft\footnotesize{Andreas Glatz}}